\documentclass[11pt]{article}
\pdfoutput=1
\usepackage{putex}
\usepackage{comment}
\usepackage{graphicx}
\usepackage{caption}
\usepackage{amsmath}
\usepackage{array}
\usepackage{subcaption}
\usepackage{epstopdf}
\usepackage{enumerate}
\usepackage{cite}
\usepackage{youngtab}
\usepackage{tensor}
\usepackage{slashed}
\usepackage[export]{adjustbox}
\usepackage[aligntableaux=center]{ytableau}
\usepackage[utf8]{inputenc}
\usepackage[
      colorlinks=true,
      linkcolor=blue,
      urlcolor=blue,
      filecolor=black,
      citecolor=red,
      ]{hyperref}

\usepackage{tikz}

\usepackage{youngtab}

\newcommand{\RNum}[1]{\uppercase\expandafter{\romannumeral #1\relax}}

\newcommand {\be} {\begin {equation}}
\newcommand {\ee} {\end {equation}}

\newcommand {\bes} {\begin {equation*}}
\newcommand {\ees} {\end {equation*}}

\newcommand{\ellK}{\mathbb{K}}
\newcommand{\ellE}{\mathbb{E}}

\newcommand{\beq}{\begin{equation}}
\newcommand{\eeq}{\end{equation}}

\def\eqref#1{(\ref{#1})}

\def\ie{\begin{equation}\begin{aligned}}
\def\fe{\end{aligned}\end{equation}}

\newcommand{\ket}[1]{|#1\rangle}
\newcommand{\braket}[1]{\langle#1\rangle}
\newcommand{\BO}[1]{\textcolor{blue}{\bf [#1 - BO]}}

\usepackage{esint}

\numberwithin{equation}{section}

\def\<{\langle}
\def\>{\rangle}

\def\comma{\,,}
\def\period{\,.}

\usepackage{cleveref}

\begin{document}
\preprint{PUPT-2638\\CERN-TH-219}
\institution{PU}{Joseph Henry Laboratories, Princeton University, Princeton, NJ 08544, USA}
\institution{CERN}{Department of Theoretical Physics, CERN, 1211 Meyrin, Switzerland
}

\title{\LARGE Chaos and the reparametrization mode on the AdS$_2$ string}
\authors{Simone Giombi\worksat{\PU}, Shota Komatsu\worksat{\CERN}, Bendeguz Offertaler\worksat{\PU}
}
\abstract{We study the holographic correlators corresponding to scattering 
of fluctuations of an open string worldsheet with AdS$_2$ geometry. 
In the out-of-time-order configuration, the correlators display a Lyapunov growth that 
saturates the 
chaos bound. We show that in a double-scaling limit interpolating between 
the Lyapunov regime and the late time exponential decay, the out-of-time-order 
correlator (OTOC) can be obtained exactly, and it has the same functional 
form found in the analogous calculation in JT gravity. The result can be understood as coming from high energy scattering near the horizon of a AdS$_2$ black hole, and is essentially controlled by the flat space worldsheet S-matrix.
While previous works on the AdS$_2$ string employed mainly a static gauge approach, 
here we focus on conformal gauge and clarify the role of boundary reparametrizations
in the calculation of the correlators. We find that the reparametrization mode 
is governed by a non-local action which is distinct from the Schwarzian action arising 
in JT gravity, and in particular leads to $SL(2,\mathbb{R})$ invariant boundary correlators. The OTOC 
in the double-scaling limit, however, has the same functional form as that obtained 
from the Schwarzian, and it can be computed using the reparametrization action and resumming 
a subset of diagrams that are expected to dominate in the limit. One application of our results is 
to the defect CFT defined by the half-BPS Wilson loop in ${\cal N}=4$ SYM. In this context, we show 
that the exact result for the OTOC in the double-scaling limit is in precise agreement with a recent 
analytic bootstrap prediction to three-loop order at strong coupling.}

\maketitle

\tableofcontents

\onehalfspacing
\section{Introduction}

One way to study low-dimensional holography is to start with a higher dimensional holographic setting and introduce a defect. A prototypical example of this is the open string in AdS$_5\times S^5$ incident on a straight line on the boundary, which is dual to the half-BPS Wilson line in $\mathcal{N}=4$ super Yang-Mills (SYM) \cite{Maldacena:1998im,Rey:1998ik}. Classically, the string worldsheet forms a surface of extremal area in AdS$_5$ with AdS$_2$ geometry, and transverse fluctuations of the worldsheet can be viewed as fields in AdS$_2$ that are governed by a tower of interactions determined by the Nambu-Goto action \cite{Giombi:2017cqn}. The fluctuations are dual to insertions of local operators along the Wilson operator on the boundary \cite{Drukker:2006xg,Cooke:2017qgm}, which defines a one-dimensional defect CFT. In recent years, this AdS$_2/$CFT$_1$ set-up has proven to be a versatile playground for studying AdS/CFT and its interplay with many non-perturbative techniques in conformal gauge theory, including supersymmetric localization \cite{Correa:2012at,Giombi:2018qox,Giombi:2018hsx}, integrability \cite{Drukker:2012de,Correa:2012hh,Kiryu:2018phb,Grabner:2020nis,Cavaglia:2021bnz,Cavaglia:2022qpg}, the numerical/analytic conformal bootstrap \cite{Liendo:2018ukf,Ferrero:2021bsb}, and the large charge limit \cite{Miwa:2006vd,Giombi:2021zfb,Giombi:2022anm}. See also \cite{Barrat:2021tpn,Barrat:2022eim}.

The AdS$_2$/CFT$_1$ correspondence of the open string/Wilson line can be viewed as an example of ``non-gravitational'' or ``rigid'' holography \cite{Aharony:2015zea}. At zero string coupling and in the limit of large string tension, the worldsheet decouples from closed string modes in the bulk and its fluctuations are suppressed. If one works in static gauge, the worldsheet theory does not contain a dynamical metric and shares many similarities with QFTs in non-dynamical AdS$_2$ \cite{Paulos:2016fap,Carmi:2018qzm,Ouyang:2019xdd,Beccaria:2019stp,Beccaria:2019mev,Beccaria:2019dju,Beccaria:2020qtk, Antunes:2021abs,Cordova:2022pbl} and conformal line defects in higher dimensional CFTs \cite{Billo:2016cpy,Mazac:2018mdx}. For instance, boundary correlators respect unitarity and the global conformal symmetry group $SL(2,\mathbb{R})$ (i.e., $SO(2,1)$, the isometry group of AdS$_2$) and boundary operators can be divided into primaries and descendents that satisfy the OPE, but the theory lacks a boundary stress tensor. The absence of a graviton on the worldsheet and a stress tensor on the boundary is somewhat trivial in $2d/1d$, but is also a feature of the worldvolume theories of higher dimensional branes (and of higher dimensional QFTs in non-dynamical AdS and conformal defects). It should be noted that the $SL(2,\mathbb{R})$ symmetry of the AdS$_2$/CFT$_1$ correspondence of the string/Wilson line distinguishes it from examples of topological AdS$_2$/topological CFT$_1$ with full Diff$(S^1)$ symmetry (see e.g., \cite{Mezei:2017kmw,Lin:2022rzw,Lin:2022zxd}), and also from the nearly-AdS$_2$/nearly-CFT$_1$ (NAdS$_2/$NCFT$_1$) of Jackiw-Teitelboim (JT) gravity \cite{Almheiri:2014cka,Maldacena:2016upp,Jensen:2016pah,Engelsoy:2016xyb}. In JT gravity, the boundary conformal symmetry is broken by the introduction of a scale quantifying the divergence of the dilaton at the boundary to $U(1)$, and $SL(2,\mathbb{R})$ symmetry is restored only in the ultraviolet.

On the other hand, the string worldsheet can also be viewed as defining a toy model of quantum gravity. One justification for this interpretation comes from the form of the scattering interaction on the worldsheet of the infinitely long non-interacting string in flat space \cite{Dubovsky:2012wk}:
\begin{align}\label{eq:S-matrix}
    \mathcal{S}=e^{i\ell_s^2 p^u p^v}\,,
\end{align}
where $\ell_s$ is the string length. 
This describes the phase shift picked up by left and right moving quanta with lightcone momenta $p^u$ and $p^v$ interacting on the worldsheet. The scattering matrix in \eqref{eq:S-matrix} gives the worldsheet theory a number of properties that are reminiscent of more realistic theories of gravity, including an absence of off-shell observables, a minimal length, a Hagedorn temperature, and integrable versions of black holes \cite{Dubovsky:2012wk}. Furthermore, \eqref{eq:S-matrix} is precisely the form of the Dray-'t Hooft scattering matrix \cite{Dray:1984ha,tHooft:1990fkf} that describes shock-wave scattering between high energy gravitating particles in $1+1$ dimensions. In JT gravity, the same shockwave S-matrix is responsible for the form of the out-of-time-ordered correlator (OTOC) \cite{Maldacena:2016upp,Lam:2018pvp}, which, like in higher dimensional theories of gravity \cite{Shenker:2013pqa,Shenker:2014cwa}, saturates the chaos bound \cite{Maldacena:2015waa} in the Lyapunov regime. This feature of JT gravity is one of the simplest illustrations of its usefulness as a model of low-dimensional quantum gravity (see \cite{Mertens:2022irh} for a recent review). 

By adapting the methods of \cite{Shenker:2013pqa,Shenker:2014cwa} to the worldsheet, the OTOC on the AdS$_2$ string can also be interpreted in terms of high-energy scattering of particles that are created and absorbed by operators on the boundary. At high enough energies (corresponding to long enough time separations between the operators on the boundary), the interaction between the bulk excitations is localized to a small region on the AdS$_2$ worldsheet, and the scattering matrix is well-approximated by the flat space answer in \eqref{eq:S-matrix}. It therefore follows that the OTOC on the AdS$_2$ string in the Lyapunov regime takes the same form as in JT gravity \cite{deBoer:2017xdk,Maldacena:2017axo,Murata:2017rbp}:
\begin{align}\label{eq:AdS2 OTOC Lyapunov}
    \frac{\braket{VW(t)VW(t)}}{\braket{VV}\braket{WW}}=1-\frac{\Delta_V \Delta_W}{4} \frac{\ell_s^2}{\ell^2}e^{\frac{2\pi t}{\beta}}+\ldots.
\end{align}
Here, $\ell$ is the AdS radius. The scrambling time is $t_s\sim \beta \log{\frac{\ell}{\ell_s}}$ and the Lyapunov exponent is $\lambda_{\rm OTOC}=\frac{2\pi}{\beta}$, which saturates the chaos bound. The above result can be checked using more standard AdS/CFT methods by computing the leading contact Witten diagram contributing to the euclidean four-point function and analytically continuing to the OTOC configuration \cite{Beccaria:2019dws}. The fact that the string OTOC saturates the chaos bound is quite natural when one views the AdS$_2$ string worldsheet theory as a toy model of gravity, but somewhat surprising when one views it as an example of rigid holography.

In this work, we extend the previous studies of chaos on the AdS$_2$ string worldsheet. The expression for the OTOC in \eqref{eq:AdS2 OTOC Lyapunov} is valid in the Lyapunov regime where the string length $\ell_s$ is small compared to the AdS radius $\ell$ and $t$ is much less than the scrambling time $t_s$. More generally, one can study the OTOC in the double scaling limit $t\to \infty$, $\ell_s\to 0$ with $\kappa\equiv \frac{\ell_s^2}{16\ell^2}e^{\frac{2\pi t}{\beta}}$ held fixed. In the first part of this paper, we will argue that because \eqref{eq:S-matrix} is the exact scattering matrix on the flat space string worldsheet, the scattering analysis in \cite{deBoer:2017xdk} can be extended to all orders in $\kappa$, with the result:
\begin{align}\label{eq:double scaled OTOC}
    \frac{\braket{VW(t)VW(t)}}{\braket{VV}\braket{WW}}=
\frac{1}{\kappa^{2\Delta_V}}U(2\Delta_V,1+2\Delta_V-2\Delta_W,\kappa^{-1}),&&\kappa=\frac{1}{16}\frac{\ell_s^2}{\ell^2} e^{\frac{2\pi t}{\beta}},
\end{align}
where $U(a,b,x)$ is the confluent hypergeometric function. As a non-trivial check of eq.~\eqref{eq:double scaled OTOC}, we show that it agrees up to order $\kappa^4$ with the analytic continuation to the OTOC configuration of the scalar four-point function in the Wilson line defect CFT that was recently computed to three loops by Ferrero and Meneghelli via the analytic conformal bootstrap \cite{Ferrero:2021bsb}. 

Eq.~\eqref{eq:double scaled OTOC} is also precisely the form of the OTOC in JT gravity in the late-time weak-coupling double scaling limit \cite{Maldacena:2016upp,Lam:2018pvp}. This equivalence of the OTOCs for both the AdS$_2$ string and JT gravity, both in the Lyapunov regime and in the double scaling limit, is interesting and warrants further investigation. One simple explanation of the equivalence is that it is a consequence of the local scattering interaction and the background geometry being the same for both the AdS$_2$ string and JT gravity. However, it is tempting to think that the equivalence of the OTOCs is evidence of a further, deeper connection between the string worldsheet theory and JT gravity. In particular, the dynamics of JT gravity coupled to matter is completely determined (at leading order in the genus expansion) by the Schwarzian boundary mode. Recall that JT gravity has a dynamical boundary curve that cuts out a patch of AdS$_2$ and regularizes the divergence of the dilaton. The boundary curve is governed by an effective Lagrangian given by the Schwarzian derivative, whose form is fixed by the pattern of spontaneous and explicit breaking of the $\text{Diff}(S^1)$ symmetry group of reparametrizations of the boundary \cite{Almheiri:2014cka,Maldacena:2016upp,Jensen:2016pah,Engelsoy:2016xyb}. The Schwarzian theory is exactly solvable \cite{Bagrets:2016cdf,Bagrets:2017pwq,Stanford:2017thb,Mertens:2017mtv,Mertens:2018fds,Blommaert:2018oro,Iliesiu:2019xuh,Kitaev:2018wpr,Yang:2018gdb,Suh:2020lco}, and in particular provides a rigorous method of deriving the OTOC in \eqref{eq:double scaled OTOC} \cite{Lam:2018pvp} that complements the scattering argument based on the shockwave interaction in \eqref{eq:S-matrix}. Furthermore, the Schwarzian mode appears in other contexts characterized by the same symmetry breaking pattern, like the SYK model \cite{KitaevTalks,Maldacena:2016hyu,Kitaev:2017awl} and 2d CFTs with large central charge \cite{Ghosh:2019rcj}, where it similarly determines the behavior of various observables including the OTOC. Given these observations, it is tempting to conjecture that the AdS$_2$ string also has an effective Schwarzian mode that dominates in a certain regime and in particular determines the OTOC in the double scaling limit. Indeed, this possibility has been explored in \cite{Banerjee:2018kwy,Banerjee:2018twd,Vegh:2019any,Gutiez:2022uof}. However, the boundary correlators on the AdS$_2$ string are expected to be symmetric under the global conformal group $SL(2,\mathbb{R})$, as required by their defect CFT interpretation, while the Schwarzian breaks even scale invariance and is symmetric only under the $U(1)$ group of translations along the boundary. Therefore, symmetry considerations alone seem to rule out the existence of a Schwarzian mode for the AdS$_2$ string, at least in the setting we consider.

Nonetheless, the possibility of understanding the double scaled OTOC on the string worldsheet in terms of a boundary mode is worth exploring, and we do so in the second half of this paper. In the computation of the string sigma model path integral, the integral over metrics after fixing the conformal gauge gives rise to $bc$ ghosts and an integral over reparametrizations of the boundary of the worldsheet. The appearance of the integral over boundary reparametrizations has been understood for a long time for the case of strings with boundary in flat space \cite{Polyakov:1981rd,Alvarez:1982zi,Fradkin:1982ge,Cohen:1985sm,polyakov1987gauge}, but it has also more recently been studied in the context of the string in AdS \cite{Polyakov:2000jg,Polyakov:2000ti,Rychkov:2002ni,Ambjorn:2011wz}. Building on those works, we study the AdS$_2$ string in conformal gauge and derive an effective action for the boundary reparametrizations. If $\alpha(\tau)$ is a reparametrization of the boundary of the hyperbolic disk, so that $\alpha(\tau+2\pi)=\alpha(\tau)+2\pi$, then at least for large tension its effective action can be written in terms of the extremization of a classical action for the two coordinates on AdS$_2$, with boundary conditions set by the reparametrization $\alpha$: 
\begin{align}\label{eq:reparametrization action implicit}
    S[\alpha(\tau)]=\underset{\substack{r(\sigma,\tau),\theta(\sigma,\tau)\\r(0,\tau)=0\\ \theta(0,\tau)=\alpha(\tau)}}{\text{extremize}}\left\{\frac{T_s \ell^2}{2}\int d\sigma d\tau \left[\frac{\partial_\beta r \partial^\beta r+\partial_\beta \theta \partial^\beta \theta}{\sinh^2{r}}-\frac{2}{\sinh^2{\sigma}}\right]\right\}.
\end{align}
This is an implicit representation of the action. The expansion of the action to quadratic order about the saddle point $\alpha(\tau)=\tau+\epsilon(\tau)$ can be determined explicitly:
\begin{align}\label{eq:reparametrization action explicit}
    S[\tau+\epsilon(\tau)]=\frac{T_s \ell^2}{2\pi}\int d\tau d\tau' \left[\frac{(\dot{\epsilon}(\tau)-\dot{\epsilon}(\tau'))^2-(\epsilon(\tau)-\epsilon(\tau'))^2}{[2\sin\left(\frac{\tau-\tau'}{2}\right)]^2}+O(\epsilon^3)\right].
\end{align}
Related expressions for the reparametrization action of the string in AdS appeared in \cite{Polyakov:2000jg,Polyakov:2000ti,Rychkov:2002ni,Ambjorn:2011wz}, and were used to study certain properties of the string partition function.

The reparametrization action in \eqref{eq:reparametrization action implicit} breaks the reparametrization symmetry Diff$(S^1)$ to an $SL(2,\mathbb{R})$ subgroup that is gauged (which is the familiar $SL(2,\mathbb{R})$ group of worldsheet transformations that preserve the conformal form of the metric). As discussed in \cite{Maldacena:2016upp} in the context of JT gravity, this symmetry breaking pattern together with an assumption of locality uniquely determines the effective action to be the Schwarzian. The reparametrization action for the string evades this argument because it is non-local. Furthermore, in accordance with our comments above and in contrast to the Schwarzian, the string reparametrization action has a physical $SL(2,\mathbb{R})$ symmetry in addition to the $SL(2,\mathbb{R})$ gauge symmetry.

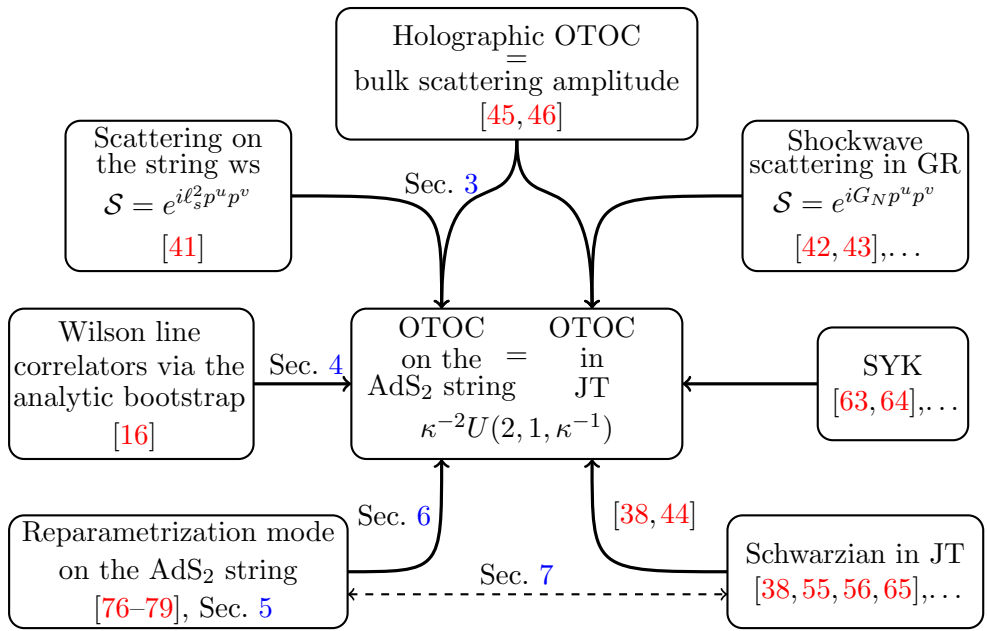
\begin{figure}[t]
    \centering
    \scalebox{1}{
        \begin{tikzpicture}
	
    	\node at (4,10.5) {\Large \bf Chaos on the AdS$_2$ string};
    	\node at (12,10.5) {\Large \bf Chaos in JT gravity};

        \draw [rounded corners = 0.2cm,thick] (5.8,4) rectangle (10.2,6);
        \node at (7,5.75) {OTOC};
        \node at (7,5.35) {on the};
        \node at (7,4.95) {AdS$_2$ string};
        \node at (8,5.35) {=};
        \node at (9,5.75) {OTOC};
        \node at (9,5.35) {in};
        \node at (9,4.95) {JT};
       \node at (8,4.4) {$\kappa^{-2}U(2,1,\kappa^{-1})$};
            
        \draw [rounded corners = 0.2cm,thick] (5.6,8.25) rectangle (10.4,10);
        \node at (8,9.6) {Holographic OTOC};
        \node at (8,9.3){=};
        \node at (8,9){bulk scattering amplitude};
        \node at (8,8.55){\cite{Shenker:2013pqa,Shenker:2014cwa}};
        
        \draw[very thick,->] (8,8.25) .. controls (8,7) and (7,8.25) ..  (7,6);       
        \draw[very thick,->] (8,8.25) .. controls (8,7) and (9,8.25) .. (9,6);

        
        \draw [rounded corners = 0.2cm,thick] (2,6.5) rectangle (5,8.5);
        \node at (3.5,8.25) {Scattering on};
        \node at (3.5,7.9) {the string ws};
        \node at (3.5,7.45) {$\mathcal{S}=e^{i\ell_s^2 p^u p^v}$};
        \node at (3.5,6.8) {\cite{Dubovsky:2012wk}};
        
        \draw[very thick,->] (5,7.5) .. controls (7,7.5) .. (7,6);     
        \node[above] at (7,7.4) {Sec.~\ref{sec:OTOC from shockwave S matrix}};
         
        
        \draw [rounded corners = 0.2cm,thick] (11,6.5) rectangle (14,8.5);
        \node at (12.5,8.25) {Shockwave};
        \node at (12.5,7.9) {scattering in GR};
        \node at (12.5,7.45) {$\mathcal{S}=e^{iG_N p^u p^v}$};
        \node at (12.5,6.8) {\cite{Dray:1984ha,tHooft:1990fkf},\ldots};        
        
        \draw[very thick,->] (11,7.5) .. controls (9,7.5) .. (9,6);
        
        
        \draw [rounded corners = 0.2cm,thick] (1.25,4) rectangle (4.5,6);
        \node at (2.85,5.7) {Wilson line };
        \node at (2.85,5.25) {correlators via the};
        \node at (2.85,4.8) {analytic bootstrap};
        \node at (2.85,4.3) {\cite{Ferrero:2021bsb}};
        
        \draw[very thick,->](4.5,5) -- (5.8,5);     
        \node[above] at (5.2,5) {Sec.~\ref{sec:OTOC on WL to 3 loops}};
        
        \draw [rounded corners = 0.2cm,thick] (12,4.25) rectangle (14,5.75);
        \node at (13,5.25) {SYK};
        \node at (13,4.75) {\cite{KitaevTalks,Maldacena:2016hyu},\ldots};

        \draw[very thick,->](12,5) -- (10.2,5);
        
        \draw [rounded corners = 0.2cm,thick] (1.25,1.75) rectangle (5.75,3.25);
        \node at (3.5,3) {Reparametrization mode};
        \node at (3.5,2.5) {on the AdS$_2$ string};
        \node at (3.5,2) {\cite{Polyakov:2000jg,Polyakov:2000ti,Rychkov:2002ni,Ambjorn:2011wz}, Sec.~\ref{sec:reparametrization mode for AdS string}};
        
        \draw[very thick,->] (5.75,2.5) .. controls (7,2.5) .. (7,4);
         \node[left] at (7,3.25) { Sec.~\ref{sec:OTOC from the string reparametrization mode}};
        
        \draw [rounded corners = 0.2cm,thick] (10.8,1.75) rectangle (14.2,3.25);
        \node at (12.5,2.75) {Schwarzian in JT};
        \node at (12.5,2.25) {\cite{Maldacena:2016upp,Kitaev:2017awl,Stanford:2017thb,Mertens:2017mtv},\ldots};
        
        \draw[very thick,->] (10.8,2.5) .. controls (9,2.5) .. (9,4);
        \node[right] at (9,3.25) {\cite{Maldacena:2016upp,Lam:2018pvp}};
        
        \draw[dashed, thick,<->] (5.75,2.2) -- (10.8,2.2);
        \node[above] at (8,2.2){Sec.~\ref{eq:AdS2 reparametrization and the Schwarzian}};
        
        \end{tikzpicture}
    }
    \caption{Outline of the paper. The OTOC on the AdS$_2$ string in the Lyapunov regime was studied in \cite{deBoer:2017xdk,Murata:2017rbp,Maldacena:2017axo,Beccaria:2019dws}. The present study of the OTOC in the double scaling limit draws on a number of previous works, including on interpreting OTOCs as holographic scattering \cite{Shenker:2013pqa,Shenker:2014cwa}, scattering on the string in flat space \cite{Dubovsky:2012wk}, four-point functions on the half-BPS Wilson loop \cite{Ferrero:2021bsb}, and the reparametrization mode of the AdS$_2$ string \cite{Polyakov:2000jg,Polyakov:2000ti,Rychkov:2002ni,Ambjorn:2011wz}. In addition, the parallel story of the OTOC in JT gravity-- and the discussions in \cite{Maldacena:2016upp,Lam:2018pvp} in particular--- served as a useful guide.}
    \label{fig:paper_outline}
\end{figure}

Using the reparametrization action in \eqref{eq:reparametrization action explicit}, we can derive the tree-level four-point functions and find agreement with the static gauge results in \cite{Giombi:2017cqn}. The computation is completely analogous to the perturbative computations in the Schwarzian theory in \cite{Maldacena:2016upp}. Technically, one finds that the contribution of the 4-point 4-derivative interaction that appears in static gauge is reproduced in conformal gauge by the reparametrization ``dressing" of free boundary-to-boundary propagators.  Furthermore, with some plausible assumptions --- i.e., that the OTOC in the double scaling limit is determined by the reparametrization action to quadratic order and does not receive corrections from the fluctuations of the matter or ghost fields in the string path integral --- we can also use the reparametrization action to reproduce the all-orders result in \eqref{eq:double scaled OTOC}.
\paragraph{Outline of the paper.} The rest of this paper is organized as follows. Section~\ref{sec:preliminaries} summarizes the definition of the boundary correlators and OTOC on the AdS$_2$ string, and reviews the static gauge computation of the OTOC in the Lyapunov regime given in eq.~\eqref{eq:AdS2 OTOC Lyapunov}. Section~\ref{sec:OTOC from shockwave S matrix} derives the all orders double-scaled OTOC given in eq.~\eqref{sec:OTOC from shockwave S matrix} using the scattering argument on the worldsheet. The analysis is a straightforward extension of the one in \cite{deBoer:2017xdk}, and is essentially equivalent to the scattering analysis in JT gravity \cite{Maldacena:2016upp,Lam:2018pvp}. Section~\ref{sec:OTOC on WL to 3 loops} presents two checks of the all orders double-scaled OTOC using results for the unit charge scalar four-point function derived in \cite{Ferrero:2021bsb}, and the large charge four-point functions derived in \cite{Giombi:2022anm}. Section~\ref{sec:reparametrization mode for AdS string} presents the conformal gauge analysis of the classical string and expresses the classical action in terms of a dynamical reparametrization mode on the string boundary. Section~\ref{sec:OTOC from the string reparametrization mode} introduces the path integral over reparametrizations and uses it to compute the tree-level four-point functions and the double scaled OTOC on the AdS$_2$ string. Section~\ref{eq:AdS2 reparametrization and the Schwarzian} summarizes the similarities and differences of the reparametrization mode on the AdS$_2$ string and the Schwarzian. Finally, Section~\ref{sec:discussion} concludes with a discussion of future directions. Several appendices are included to flesh out comments made in the body of the paper and to explain technical details. The outline of the paper is also summarized in Figure~\ref{fig:paper_outline}. 

\section{Preliminaries and warm-up}\label{sec:preliminaries}
In this section, we review the basic concepts needed in the analysis of chaos on the AdS$_2$ string. These include the definition of the boundary correlators on the string and the out-of-time-order-correlator as a diagnostic of chaos in a thermal quantum system. Then, as a warm-up to the scattering analysis and conformal gauge analysis in later sections, we review how to compute the simplest four-point function on the AdS$_2$ string at tree level--- as well as the OTOC in the Lyapunov regime--- in static gauge.

\subsection{Boundary correlators on the \texorpdfstring{AdS$_2$}{AdS2} string}

Consider an open string in AdS$_{5}\times S^5$ that is incident on a curve $\gamma$ on the AdS boundary. The dynamics of the open string is summarized by its partition function. In full generality, this is the partition function of string theory summing over all asymptotically AdS$_5\times S^5$ states that include an open string incident on $\gamma$ on the boundary of AdS. At zero string coupling, $g_s=0$, the open string decouples from the closed strings in the bulk, and the partition function is given by the sigma model path integral for a superstring in AdS$_5\times S^5$,  subject to the Dirichlet boundary conditions specified by the curve $\gamma$. To state this more precisely, let $(z,x^\mu)$ be Poincar\'e coordinates on AdS$_{5}$ with bulk coordinate $z\in [0,\infty)$, boundary coordinates $x^\mu\in \mathbb{R}$, $\mu=0,\ldots,3$, and metric $ds^2=z^{-2}(dz^2+dx^\mu dx_\mu)$. (We work in euclidean signature here). Furthermore, let $y^m$ be coordinates on $S^5$ with $y^m\in \mathbb{R}$, $m=1,\ldots,5$ with metric $ds^2=g_{mn}(y)dy^m dy^n$. In these coordinates, the open string can be represented by a map
\begin{align}
    \Sigma:(s,t)\mapsto (z(s,t),x^\mu(s,t),y^m(s,t)),
\end{align}
where $s\in [0,\infty)$ and $t\in \mathbb{R}$ are coordinates on the worldsheet. We choose the worldsheet coordinates so that the boundary is located at $s=0$. Similarly, the curve on $\mathbb{R}^4\times S^5$ that the open string is incident on can be represented by a map,
\begin{align}
    \gamma:\alpha\mapsto (\tilde{x}^\mu(\alpha), \tilde{y}^m(\alpha))
\end{align}
where $\alpha$ is the parameter along the curve. Then, we write the partition function schematically as:
\begin{align}\label{eq:string path integral}
    Z[\tilde{x}^\mu,\tilde{y}^m] &=\underset{\substack{x^\mu\rvert_{\partial}=\tilde{x}^\mu \\ y^m\rvert_{\partial}=\tilde{y}^m\\ z\rvert_{\partial}=0}}{\int}\mathcal{D}h  \mathcal{D}z \mathcal{D}x \mathcal{D}y e^{-S[h_{\alpha\beta},z,x^\mu,y^m]},
\end{align}
where we integrate over the coordinates of the string in AdS$_5\times S^5$ and the auxiliary metric $h_{\alpha\beta}$. (We have suppressed the fermionic coordinates for simplicity). The (bosonic part of the) superstring action is:
\begin{align}
    S[h_{\alpha\beta},z,x^\mu,y^m]=\frac{T_s}{2}\int d^2\sigma \sqrt{h}h^{\alpha\beta} \left[\frac{\partial_\alpha z \partial_\beta z +\partial_\alpha x^\mu \partial_\beta x_\mu}{z^2}+g_{mn}(y)\partial_\alpha y^m \partial_\beta y^n\right].
\end{align}
Here, $T_s=1/\ell_s^2$ is the string tension and $\sigma^\alpha=(s,t)$ are the worldsheet coordinates.

If we additionally restrict to the regime in which the string tension is much larger than the AdS curvature, then the path integral in \eqref{eq:string path integral} is dominated by the classical solution, and the partition function can be approximated as
\begin{align}\label{eq:classical partition function}
    Z[\tilde{x}^\mu,\tilde{y}^m]\approx e^{-S[h_{\alpha\beta},z,x^\mu,y^m]}.
\end{align}
Here, $S[h_{\alpha\beta},z,x^\mu,y^m]$ is the action of the classical string, whose worldsheet forms a surface of extremal area incident on $\gamma$. The leading quantum corrections, in $1/T_s$, come from fluctuations about the classical solution. 

The simplest (i.e., maximally symmetric) choice for $\gamma$, and the one we will consider, is a straight line in $\mathbb{R}^4$ and a point in $S^5$. The classical string incident on this contour carves out an AdS$_2$ subspace in AdS$_5$. The line on the boundary and the AdS$_2$ subspace in the bulk preserve half of the supersymmetries of AdS$_5\times S^5$ (i.e., the subgroup $OSp(4^*|4)\subset PSU(2,2|4)$), including the subgroup $SO(2,1)\times SO(3)\subset SO(5,1)$ of isometries of AdS$_5$ (where $SO(2,1)$ are isometries of AdS$_2$ and $SO(3)$ are rotations around AdS$_2$) as well as the subgroup $SO(5)\subset SO(6)$ of isometries of $S^5$. Although we can work with any parametrization of $\gamma$, it is convenient to label the points by the euclidean coordinate along the line. To be concrete, we let the contour lie along the $x^0\equiv x$ axis in $\mathbb{R}^4$, and parametrize it as $\gamma:x\mapsto (x,x^a=0,y^m=0)$ (where $a=1,2,3$ labels the three directions in $\mathbb{R}^4$ orthogonal to $x^0$). More generally, we consider the ``wavy line'' consisting of small perturbations around the straight line, which we represent as
\begin{align}\label{eq:wavy line}
    \gamma:x\mapsto (x,\tilde{x}^a(x),\tilde{y}^m(x)),
\end{align}
where $\tilde{x}^a$ and $\tilde{y}^m$ are small. Given this choice of representation of the wavy line, we denote the partition function by $Z[\tilde{x}^a,\tilde{y}^m]$ (instead of the curve reparametrization invariant expression in \eqref{eq:string path integral}) and define the boundary correlators of the AdS$_2$ string by taking derivatives of the partition function in the directions orthogonal to the line:
\begin{align}\label{eq:ads2 boundary correlators}
    \braket{x^{a_1}(x_1)y^{m_1}(x_2)\ldots}_{\text{AdS}_2}=Z^{-1}\frac{\delta}{\delta\tilde{x}^{a_1}(x_1)}\frac{\delta}{\delta\tilde{y}^{m_1}(x_2)}\ldots Z[\tilde{x}^a,\tilde{y}^m]\bigg\rvert_{\tilde{x}^a=\tilde{y}^m=0}.
\end{align}
These are the correlators we will analytically continue to the out-of-time-order configuration to study chaos on the open string.

\subsection{Defect correlators on the Wilson line}\label{sec:WL defect correlators}

One can also study the boundary correlators on the AdS$_2$ string in terms of its dual CFT description. The open string incident on the general curve $\gamma:\alpha\mapsto (\tilde{x}^\mu(\alpha),\tilde{y}^m(\alpha)$ on the AdS boundary is dual to the Wilson loop operator in $\mathcal{N}=4$ SYM that couples to both the gauge field $A_\mu$ along the path $\tilde{x}^\mu(\alpha)$ in the spacetime $\mathbb{R}^4$ and to the scalars $\Phi^I$, $I=1,\ldots,6$ along the path $\tilde{y}^m(\alpha)$ in $S^5$ \cite{Maldacena:1998im,Rey:1998ik}. Explicitly, the Wilson operator is:
\begin{align}\label{eq:WL}
    \mathcal{W}[\tilde{x}^\mu,\tilde{y}^m]=\text{Tr P}e^{\int \big(i A_\mu \dot{\tilde{x}}^\mu+ |\dot{\tilde{x}}| \theta^I \Phi^I\big) d\alpha }.
\end{align}
Here, $\theta^I(\alpha)$ are embedding coordinates on $S^5$ (satisfying $\theta^I\theta^I=1$) that can be expressed in terms of the $y^m$ coordinates--- for concreteness, we can take $y^m$ to be stereographic coordinates so that $\theta^m=\frac{\tilde{y}^m}{1+\frac{1}{4}\tilde{y}^2}$, $\theta^6=\frac{1-\frac{1}{4}\tilde{y}^2}{1+\frac{1}{4}\tilde{y}^2}$, and the metric becomes $ds^2=d\theta^Id\theta^I=(1+\frac{1}{4}y^2)^{-2}dy^m dy^m$.
The precise statement of duality is that the open string partition function is equal to the Wilson operator expectation value:
\begin{align}\label{eq:OS partition = WL VEV}
   Z[\tilde{x}^\mu,\tilde{y}^m]=\braket{\mathcal{W}[\tilde{x}^\mu,\tilde{y}^m]}_{\mathcal{N}=4\text{ SYM}}.
\end{align}
Note that the classical regime in AdS$_5\times S^5$ ($g_s=0$ and $T_s\gg1$)--- which is the regime in which the approximation in \eqref{eq:classical partition function} is valid--- corresponds to the planar limit ($N\to \infty$ with $\lambda \equiv g_{YM}^2N$ fixed) at strong coupling ($\lambda\gg1$) in $\mathcal{N}=4$ SYM. In particular, the string tension and 't Hooft coupling are related by $T_s=\frac{\sqrt{\lambda}}{2\pi}$. 

In the specific case of the AdS$_2$ string incident on the straight line on the boundary of AdS$_5$, its dual is the Wilson operator that couples to a single component of the gauge field and a single scalar: $\mathcal{W}=\text{Tr P}e^{\int (iA_0+\Phi^6)dx}$. This likewise preserves the $OSp(4^*|4)$ subgroup of the $PSU(2,2|4)$ superconformal group of $\mathcal{N}=4$ SYM and is called the half-BPS Wilson line. Given the statement of duality in \eqref{eq:OS partition = WL VEV} (and the representation of the wavy line in \eqref{eq:wavy line}), the CFT dual of the boundary correlators on the AdS$_2$ string in \eqref{eq:ads2 boundary correlators} is the half-BPS Wilson line with elementary operators in $\mathcal{N}=4$ SYM inserted along the contour:
\begin{align}\label{eq:WL correlators}
\braket{x^{a_1}(x_1)y^{m_1}(x_2)\ldots}_{\text{AdS}_2}=\braket{\braket{\mathbb{D}^{a_1}(x_1)\Phi^{m_1}(x_2)\ldots}},
\end{align}
Here, $\mathbb{D}^a=iF^{ta}+D^a\Phi^6$ are the three displacement operators, $\Phi^m$ are the five scalars orthogonal to $\Phi^6$, and the ``double bracket'' denotes correlators on the Wilson line, which are defined by:
\begin{align}
\braket{\braket{O_1(x_1)O_2(x_2)\ldots}}=\frac{\big\langle\text{Tr P}\left[O_1(x_1)O_2(x_2)\ldots e^{\int (iA_0+\Phi^6)dx}\right]\big\rangle_{\mathcal{N}=4\text{ SYM}}}{\braket{\text{Tr P}e^{\int (iA_0+\Phi^6)dx}}_{\mathcal{N}=4\text{ SYM}}}.
\end{align}
The path ordering symbol puts the operators in order on the line and connects them with the intermediate sections of the Wilson operator to make a gauge invariant object. $\mathbb{D}^a$ and $\Phi^m$ are the ``elementary operators'' on the Wilson line, but one can more generally study correlators of more general adjoint operators $O_i(x_i)$ (e.g., composites of $\mathbb{D}^a$ and $\Phi^m$). 

The operators on the Wilson line are classified by their representations under $OSp(4^*|4)$. In particular, each operator has a conformal dimension $\Delta$ specifying its behavior under the $SL(2,\mathbb{R})$ group of conformal transformations in $\mathcal{N}=4$ SYM moving around the points on the line (which correspond to the isometries of AdS$_2$) . Thus, the half-BPS Wilson line defines a 1d defect CFT whose local correlators are given by \eqref{eq:WL correlators}. As a special case, as in higher dimensions, the 1d conformal symmetry fixes the two-point function of a primary $V$ (with dimension $\Delta_V$) to be of the form:
\begin{align}\label{eq:2-pt function 1d CFT}
    \braket{\braket{V(x_1)V(x_2)}}=\frac{\mathcal{N}_V}{x_{12}^{2\Delta_V}},
\end{align}
where $x_{ij}\equiv x_i-x_j$. The three point function is also fixed up to the OPE coefficient, and the four-point function of two copies of $V$ with two copies of $W$ (with dimension $\Delta_W)$ takes the form:
\begin{align}\label{eq:4-pt function 1d CFT}
\frac{\braket{\braket{V(x_1)V(x_2)W(x_3)W(x_4)}}}{\braket{\braket{V(x_1)V(x_2)}}\braket{\braket{W(x_3)W(x_4)}}}=G(\chi),
\end{align}
where $G(\chi)$ is a general function of the conformally invariant cross-ratio, 
\begin{align}\label{eq:cross ratio line}
    \chi=\frac{x_{12}x_{34}}{x_{13}x_{24}}.
\end{align}
This is the unique independent cross-ratio in $1d$, since $1-\chi=\frac{x_{14}x_{23}}{x_{13}x_{24}}$. 

So far we have discussed the 1d CFT on the line. It is often convenient (for instance, to study thermal correlators) to study the 1d CFT on the circle instead, which is related to the line by a conformal transformation. We can label the points on the circle by an angle $\theta$. Mapping points $x_i$ on the line to points $\theta_i$ on the circle by $x=\tan{\frac{\theta}{2}}$ interchanges euclidean distances $x_{ij}$ and chordal distances $2\sin{\frac{\theta_{ij}}{2}}$ in the correlators. In particular, the two-pt function of $V$ becomes
\begin{align}
\braket{\braket{V(\theta_1)V(\theta_2)}}=\frac{\mathcal{N}_V}{[2\sin{\frac{\theta_{12}}{2}}]^{2\Delta_V}}.
\end{align}
Furthermore, the four-pt function of two copies of $V$ at $\theta_1$, $\theta_2$ and two copies of $W$ at $\theta_3$, $\theta_4$, normalized by the two point functions, is again equal to $G(\chi)$ as in \eqref{eq:4-pt function 1d CFT}, with the cross-ratio given by
\begin{align}\label{eq:cross ratio circle}
\chi=\frac{\sin{\frac{\theta_{12}}{2}}\sin{\frac{\theta_{34}}{2}}}{\sin{\frac{\theta_{13}}{2}}\sin{\frac{\theta_{24}}{2}}}.
\end{align}

\paragraph{Restricting to AdS$_2\times S^1$.} Going forward, to more cleanly separate the key concepts from technical details, we will mainly focus on an AdS$_2\times S^1$ subsector of AdS$_5\times S^5$. This means considering fluctuations of the classical AdS$_2$ along only one direction in $S^5$ in \eqref{eq:ads2 boundary correlators}, or insertions of only one scalar along the Wilson line in \eqref{eq:WL correlators}. We will mostly work with Poincar\'e coordinates $z\in[0,\infty)$, $x\in\mathbb{R}$ on AdS$_2$ and a polar angle $y\in [-\pi,\pi)$ on $S^1$, with the metric $ds^2=\frac{dx^2+dz^2}{z^2}+dy^2$. Then, we represent the string worldsheet and boundary curve as:
\begin{align}
    \Sigma&:(s,t)\mapsto (z(s,t),x(s,t),y(s,t)), &\gamma&:x\mapsto (x,\tilde{y}(x)),
\end{align}
and write the partition function as
\begin{align}\label{eq:sigma model path integral}
Z[\tilde{y}]&=\underset{\Sigma|_{\partial}=\gamma}{\int}\mathcal{D}h\mathcal{D}z \mathcal{D}x\mathcal{D}ye^{-S[h_{\alpha\beta},z,x,y]}&&\approx e^{-S[h_{\alpha\beta},z,x,y]},
\end{align}
with the action given by
\begin{align}\label{eq:xyz string polyakov action}
    S[h_{\alpha\beta},z,x,y]=\frac{T_s}{2}\int d^2\sigma \sqrt{h}h^{\alpha\beta} \left[\frac{\partial_\alpha z \partial_\beta z +\partial_\alpha x \partial_\beta x}{z^2}+\partial_\alpha y\partial_\beta y\right].
\end{align}
The four point function that we will study explicitly is:
\begin{align}\label{eq: y 4-pt function}
\braket{y(x_1)y(x_2)y(x_3)y(x_4)}_{\text{AdS}_2}=\frac{1}{Z}\frac{\delta^4Z[\tilde{y}]}{\delta \tilde{y}(x_1)\delta \tilde{y}(x_2)\delta \tilde{y}(x_3)\delta \tilde{y}(x_4)}\bigg\rvert_{\tilde{y}=0}.
\end{align}

\subsection{Static vs. conformal gauge}
The expression for the partition function in \eqref{eq:sigma model path integral} (or more generally in \eqref{eq:string path integral} and \eqref{eq:classical partition function}) is schematic. It contains a lot of redundancy due to the reparametrization symmetry of the string worldsheet, which needs to be gauge fixed in some way. Furthermore, we have not made precise the meaning of the boundary condition that $\gamma$ imposes on $\Sigma$. We address both of these points now.

One simple way to fix the gauge symmetry is to work in the static gauge, in which the worldsheet coordinates are identified with the AdS$_2$ coordinates: $z(s,t)=s$ and $x(s,t)=t$. The longitudinal coordinates $x$ and $z$ are then no longer dynamical in the path integral in \eqref{eq:sigma model path integral}. Integrating out the auxiliary metric yields the Nambu-Goto form of the action, which is now a function of only the transverse mode, $y$:
\begin{align}\label{eq:Nambu-Goto action}
    S[y]=T_s\int d^2\sigma \sqrt{\text{det}\big[g_{\alpha\beta}+\partial_\alpha y\partial_\beta y\big]}.
\end{align}
Here, $g_{\alpha\beta}=\frac{1}{s^2}\delta_{\alpha\beta}$ is the AdS$_2$ metric. The statement that the string $\Sigma$ is incident on the curve $\gamma$ in static gauge becomes simply:
\begin{align}\label{eq:static gauge bc}
    y(s=0,t)=\tilde{y}(t).
\end{align}
The main advantage of the static gauge is that it is conceptually simple, since it gets rid of the longitudinal modes and fully fixes the reparametrization gauge symmetry. The result is an effective theory for a massless scalar in AdS$_2$ governed by the Nambu-Goto action (which when expanded in powers of $y$ yields a tower of derivative interactions suppressed by powers of $T_s$) that can be studied perturbatively using Witten diagrams, as done in \cite{Giombi:2017cqn}. 

An alternative way to fix the gauge symmetry is to work in the conformal gauge, in which the worldsheet coordinates are chosen so that the auxiliary metric is conformally equivalent to the AdS$_2$ metric: $h_{\alpha\beta}=e^{2\omega}g_{\alpha\beta}$. In this case, $z$, $x$ and $y$ are all dynamical, and the string action becomes
\begin{align}\label{eq:string action conformal gauge}
    S[z,x,y]=\frac{T_s}{2}\int d^2\sigma \left[\frac{\partial_\alpha z \partial^\alpha z+\partial_\alpha x \partial^\alpha x}{z^2}+\partial_\alpha y\partial^\alpha y\right].
\end{align}
(The worldsheet indices in the above expression are contracted using $\delta_{\alpha\beta}$.) The action is supplemented by the Virasoro constraint coming from the equation of motion of the auxiliary metric. Furthermore, in the conformal gauge, the condition that the string $\Sigma$ is incident on the curve $\gamma$ can be expressed as:
\begin{align}\label{eq:conformal boundary condition}
    z(s=0,t)&=0, &x(s=0,t)&=\alpha(t), &y(s=0,t)&=\tilde{y}(\alpha(t)),
\end{align}
where $\alpha(t)$ is some reparametrization on $\mathbb{R}$.

The conformal gauge has the advantage of making the transverse mode $y$ free, but it does so at the cost of introducing new dynamical objects: the longitudinal modes $x$ and $z$ (which are governed by a non-linear action) and the reparametrization mode $\alpha$. (One can of course write the boundary condition in \eqref{eq:conformal boundary condition} without $\alpha$ as $y(0,t)=\tilde{y}(x(0,t))$, but we will see that it is useful to treat $\alpha$ as a separate dynamical object). Furthermore, the conformal gauge does not fully fix the reparametrization symmetries on the worldsheet, since it leaves behind the usual residual $SL(2,\mathbb{R})$ group of global coordinate transformations that preserve the metric up to Weyl rescaling. Nonetheless, one of the main lessons of this work is that the conformal gauge, with the boundary reparametrization mode playing the lead role, provides an interesting approach to studying the correlators and the OTOC on the AdS$_2$ string.

Although we will focus for simplicity on the motion of the string in AdS$_2\times S^1$, many of the arguments and results that follow are more general. Both the static gauge analysis reviewed in section~\ref{sec:static gauge analysis} (and treated more comprehensively in \cite{Giombi:2017cqn}) and the scattering analysis in section~\ref{sec:OTOC from shockwave S matrix} can handle arbitrary transverse fluctuations in AdS$_5$ and $S^5$. Furthermore, the conformal gauge analysis in section~\ref{sec:reparametrization mode for AdS string} can be straightforwardly extended to AdS$_2\times S^5$, 
with the only difference being that there are more transverse modes and they interact due to the curvature of $S^5$. In appendix \ref{app:string in AdS2 x Sn}, we show that the leading connected four-point function of $S^5$ fluctuations, computed in \cite{Giombi:2017cqn} using the static gauge, can be reproduced in the conformal gauge once the reparametrization mode is taken into account. By contrast, the extension to the string in AdS$_{d+1}$ for $d\geq 2$, which would involve massive transverse modes that are mixed together with the longitudinal modes, is less straightforward. It would be interesting to understand this case better, perhaps by harnessing the formalism developed in \cite{Kruczenski:2014bla}. 

Finally, most of our discussion applies equally well to open strings in more general AdS$_{d+1}\times X$ space-times with an AdS$_2\times S^1$ subsector, including the AdS$_4\times CP^3$ background dual to the ABJM theory. 

\subsection{OTOC in a 1d CFT}

Now we briefly review the out-of-time-order correlator (OTOC) as a diagnostic of chaos in thermal quantum systems \cite{Larkin1969QuasiclassicalMI,Shenker:2014cwa,Maldacena:2015waa}. The OTOC defines the quantum analog of the Lyapunov exponent of a classical chaotic system. In a classical system with Hamiltonian $H$ at inverse temperature $\beta$, the Lyapunov exponent $\lambda$ is defined by $Z^{-1}\int dq dp e^{-\beta H(p,q)}\left(\frac{\partial q(t)}{\partial q(0)}\right)^2\sim e^{2\lambda t}$, and measures the exponential rate of divergence of nearby classical trajectories. In a quantum system, one promotes the Poisson bracket $\{q(t),p(0)\}=\frac{\partial q(t)}{\partial q(0)}$ to a commutator, generalizes $q$ and $p$ to arbitrary Hermitian operators $V$ and $W$, and studies the observable
\begin{align}\label{eq:square commutator}
    Z^{-1}\text{Tr}\big(e^{-\beta H}\left[V(0),W(t)\right]^\dagger\left[V(0),W(t)\right]\big).
\end{align}
Here, $W(t)= e^{iHt}W e^{-iHt}$, $V(t)=e^{iHt}V e^{-iHt}$, and $Z=\text{Tr }e^{-\beta H}$.

In weakly coupled systems, there is typically an interval, between the ``dissipation'' time $t_d\sim \beta$ (i.e., the characteristic decay time of two-point functions) and a ``scrambling'' time $t_s\gg \beta$, in which the squared commutator grows exponentially like $\sim e^{\lambda_{\rm OTOC} t}$. The rate of exponential growth, $\lambda_{\rm OTOC}$, is identified as the quantum Lyapunov exponent. Expanding the squared commutator in \eqref{eq:square commutator} yields four four-point functions, two in time order and two out of time order. The time-ordered correlators factorize into the products of two point functions within the dissipation time, which means the exponential growth is driven by the out-of-time correlators. It is therefore standard to identify the following observable as a simple probe of quantum chaos:
\begin{align}\label{eq:OTOC definition}
    \braket{V_1W_3 V_2 W_4}\equiv Z^{-1}\text{Tr}\big(e^{-\beta H}V(t_1)W(t_3) V(t_2)W(t_4)\big),
\end{align}
where $V_i\equiv V(t_i)$, $W_i\equiv W(t_i)$. This OTOC slightly generalizes the one appearing in \eqref{eq:square commutator} since the times $t_i$ are allowed to take complex values; in particular, in order to regularize divergences arising from coincident insertions, it is useful to separate the operators in imaginary time. 

The analyticity and boundedness properties of the OTOC imply that the quantum Lyapunov exponent satisfies the ``chaos bound,'' $\lambda_{\rm OTOC}\leq \frac{2\pi}{\beta}$ \cite{Maldacena:2015waa}. This bound is saturated by Einstein gravity, and makes precise the statement that black holes are the ``fastest scramblers in nature'' \cite{Sekino:2008he}.

\begin{figure}
	\centering
	\includegraphics[height=5cm]{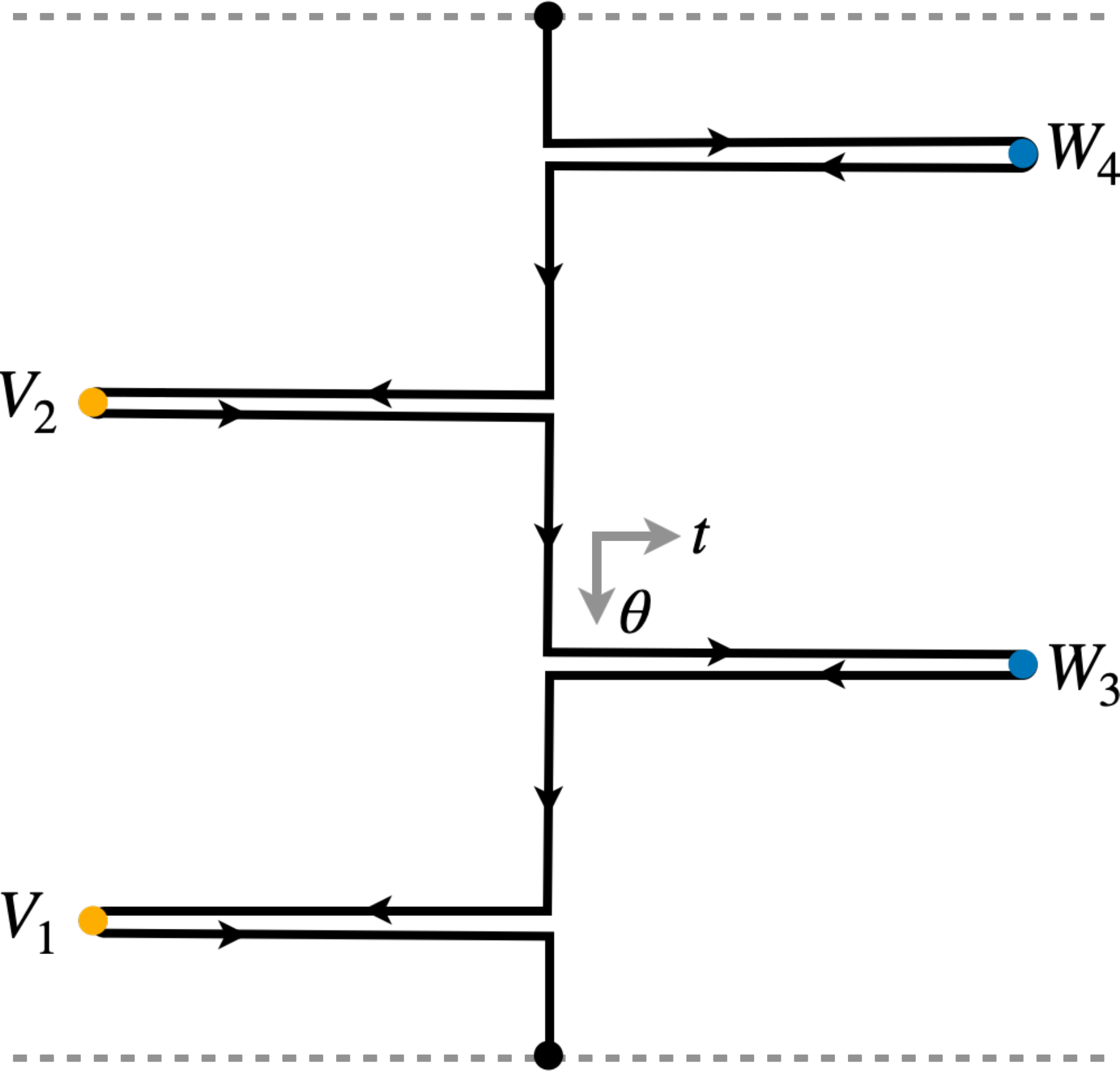}\\
    	\caption{The configuration of the operators in the OTOC on the thermal cylinder. Lorentzian time is positive to the right. Euclidean time is positive down and is periodic with period $\beta=2\pi$.}
	\label{fig:OTOC contour}
\end{figure}

In our case, we are studying the $1d$ CFT on the boundary of the AdS$_2$ string, so we can compute the OTOC by first computing the euclidean four-point function on the circle --- or on the line and then mapping to the circle --- and then analytically continuing to real time. Going forward, we work in units where the inverse temperature is $\beta=2\pi$, so the euclidean time is the angle $\theta$. We put the two copies of $V$ at $\theta_1$ and $\theta_2$ and two copies of $W$ at $\theta_3$ and $\theta_4$ such that $-\pi<\theta_4<\theta_2<\theta_3<\theta_1<\pi$, and then analytically continue $\theta_1$ and $\theta_2$ backwards in real time and $\theta_3$ and $\theta_4$ forwards in real time. For concreteness, we work with the configuration in which the four operators are spaced equally around the euclidean circle (see Figure~\ref{fig:OTOC contour}):
\begin{align}\label{eq:OTOC euclidean times}
    \theta_1&=\frac{3\pi}{4}-\frac{it}{2}, &\theta_2&=-\frac{\pi}{4}-\frac{it}{2}&\theta_3&=\frac{\pi}{4}+\frac{it}{2}, &\theta_4&=-\frac{3\pi}{4}+\frac{it}{2}.
\end{align}
From \eqref{eq:cross ratio circle}, it follows the conformally invariant cross ratio as a function of $t$ is
\begin{align}\label{eq:OTOC cross ratio}
    \chi(t)&=\frac{2}{1-i\sinh{t}}.
\end{align}
When we do the analytic continuation along this path, we should start with the expression for $G(\chi)$ that is valid on the interval $\chi>1$ (for which the operator order is $VWVW$). This is different from taking $\chi\to 0^+$ starting with the expression for $G(\chi)$ valid on the interval $0<\chi<1$ (for which the operator order is $VVWW$), which corresponds instead to the OPE limit.

\subsection{Warm-up: \texorpdfstring{AdS$_2$}{AdS2} string at tree-level in static gauge}\label{sec:static gauge analysis}

Now we will compute the four-point function in \eqref{eq: y 4-pt function} to leading order in $1/T_s$ in the static gauge, and then extract the OTOC in the Lyapunov regime. At leading order, it is sufficient to approximate the partition function by the action of the classical string, as in \eqref{eq:sigma model path integral}, because quantum corrections are suppressed by higher powers of $1/T_s$. Thus, our task is to find the solution $y$ extremizing the action in \eqref{eq:Nambu-Goto action} subject to the boundary condition in \eqref{eq:static gauge bc}. Moreover, to compute the four-point function, it is sufficient to know the action only to fourth order in $\tilde{y}(t)$, so the problem becomes a simple exercise in first order perturbation theory \cite{Giombi:2017cqn}.

We begin by expanding the action in \eqref{eq:Nambu-Goto action} in powers of $y$:
\begin{align}\label{eq:expanded action}
    S[y]&=T_s\int d^2\sigma \sqrt{g}L_{2n}(\partial_\alpha y).
\end{align}
The three lowest order Lagrangian densities are:
\begin{align}
    L_0&=1,&L_2&=\frac{1}{2}g^{\alpha\beta}\partial_\alpha y \partial_\beta y,&
    L_4&=-\frac{1}{8}(g^{\alpha\beta}\partial_\alpha y \partial_\beta y)^2.\label{eq:L_n}
\end{align}
$L_0$ gives rise to a divergent contribution to the area of the minimal surface that we drop; $L_2$ tells us that $y$ is a massless scalar in AdS$_2$; $L_4$ contains the lowest order interactions. (When we keep track of all $3+5$ transverse directions in AdS$_5\times S^5$, we find that the five transverse modes in $S^5$ are massless, the three transverse modes in AdS$_5$ have mass $m^2=2$, and all the transverse modes interact at fourth order \cite{Giombi:2017cqn}.)

The equation of motion for $y$ following from \eqref{eq:expanded action} is:
\begin{align}\label{eq:eom perturbative expansion}
    \square y&=-\sum_{n=2}^\infty \frac{1}{\sqrt{g}}\partial_\alpha \left(\sqrt{g}\frac{\partial L_{2n}}{\partial(\partial_\alpha y)}\right)\equiv j;
\end{align}
where $\square=\frac{1}{\sqrt{g}}\partial_\alpha (\sqrt{g}g^{\alpha\beta}\partial_\beta)=s^2(\frac{\partial^2}{\partial t^2}+\frac{\partial^2}{\partial s^2})$. To proceed, we can expand $y$ as $y=y_1+y_2+\ldots$ such that $y_n$ is of order $O(\tilde{y}^{2n-1})$ and expand $j$ as $j=j_2+j_3+\ldots$ such that $j_n$ is of order $O(\tilde{y}^{2n-1})$ and is composed of the $y_m$ with $m<n$. The $n$th order equation of motion following from \eqref{eq:eom perturbative expansion} is $\square y_n=j_n$. Furthermore, the boundary condition in \eqref{eq:static gauge bc} becomes $y_1(0,t)=\tilde{y}(t)$ and $y_n(0,t)=0$ for $n\geq 2$. Thus, we can solve recursively for $y_n$ for any $n$ using boundary-to-bulk and bulk-to-bulk propagators, and the resulting Witten diagrams are all tree level. This lets us determine the classical action in principle to any order in $\tilde{y}$. 

At lowest order, the equation of motion for $y_1$ is $\square y_1=0$. Given the boundary condition $y_1(0,t)=\tilde{y}(t)$, the solution is:
\begin{align}\label{eq:v1 and v2 bdy-to-bulk}
    y_1(s,t)&=\int dt' K(s,t,t') \tilde{y}(t),
\end{align}
where $K(s,t,t')$ is the boundary-to-bulk propagator for a massless field in AdS$_2$,
\begin{align}\label{eq:bdy-to-bulk propagator}
    K(s,t,t')&=\frac{1}{\pi}\frac{s}{s^2+(t-t')^2}.
\end{align}

In fact, to compute the four-point function, it suffices to determine $y_1$ only. This is because the classical action to order $O(\tilde{y}^4)$ is given by:
\begin{align}\label{eq:classical action 4th order}
    S_{\rm cl}[\tilde{y}]=T_s\int d^2\sigma \sqrt{g}\left[L_2(\partial_\alpha y_1)+L_4(\partial_\alpha y_1)\right)+O(\tilde{y}^6).
\end{align}
Note that the quadratic Lagrangian is $L_2(\partial_\alpha y) =L_2(\partial_\alpha y_1)+g^{\alpha\beta}\partial_\alpha y_1 \partial_\beta y_2+O(\tilde{y}^6)$, but the integral of the term involving $y_1$ and $y_2$ is zero (as follows from integration by parts, the equation of motion for $y_1$, and the fact that $y_2=0$ on the boundary). Substituting \eqref{eq:bdy-to-bulk propagator} into \eqref{eq:classical action 4th order} and using \eqref{eq:L_n} yields the following expression for the classical action to fourth order in $\tilde{y}$:
\begin{align}\label{eq:fourth order classical action}
   S_{\rm cl}[\tilde{y}]=-\frac{T_s}{2\pi}\int dt_1 dt_2 \frac{\tilde{y}(t_1)\tilde{y}(t_2)}{(t_1-t_2)^2}-\frac{T_s}{8}\int d^4t\text{ } \tilde{y}(t_1)\tilde{y}(t_2)\tilde{y}(t_3)\tilde{y}(t_4)F(t_1,t_2,t_3,t_4)+O(\tilde{y}^6).
\end{align}
The quadratic piece comes from integrating $\int \sqrt{g}L_2(\partial_\alpha y_1)$ by parts and using $\partial_s K(0,t,t')=\pi^{-1}(t-t')^{-2}$. We have expressed the quartic piece in terms of the function $F$, which we define by:
\begin{align}\label{eq:F(t1 t2 t3 t4)}
    F(t_1,t_2,t_3,t_4)=\int d^2\sigma \sqrt{g}[g^{\alpha\beta}\partial_\alpha K(s,t,t_1)\partial_\beta K(s,t,t_2)][g^{\gamma\delta}\partial_\gamma K(s,t,t_3)\partial_\delta K(s,t,t_4)].
\end{align}
This correponds to a four-point contact Witten diagram with a four derivative interaction. It can be evaluated in the terms of the so-called $D$ functions \cite{DHoker:1999kzh}, which in AdS$_2$ reduce to expressions involving logs and rational functions (see e.g. section $4$ of \cite{Giombi:2017cqn}). The result is:
\begin{align}\label{eq:fg5432ws24}
   F(t_1,t_2,t_3,t_4)&=\frac{1}{t_{12}^2t_{34}^2}\bar{F}(\chi),
\end{align}
where $\chi=\frac{t_{12}t_{34}}{t_{13}t_{24}}$ and $\bar{F}$ is the conformally invariant part of $F$:
\begin{align}
\bar{F}(\chi)&=\frac{1}{8\pi^3}\frac{\chi^2}{(\chi-1)^3}\biggr[-4\chi^3+12\chi^2-16\chi+8+(2\chi^4-7\chi^3+9\chi^2-4\chi+2)\log(\chi^2)\nonumber\\&+(-2\chi^4+7\chi^3-9\chi^2+5\chi-1)\log((1-\chi)^2)\biggr].
\end{align}

Finally, given the expression for the classical action in \eqref{eq:fourth order classical action}, we take the necessary variational derivatives of the partition function $Z[\tilde{y}]\approx e^{-S_{\rm cl}[\tilde{y}]}$ to get the two-point function:
\begin{align}\label{eq:2-pt static gauge}
    \braket{V_1V_2}=\frac{T_s}{\pi}\frac{1}{x_{12}^2}.
\end{align}
and the four-point function:
\begin{align}
\braket{V_1V_2V_3V_4}&=\frac{T_s^2}{\pi^2}\left[\frac{1}{x_{12}^2x_{34}^2}+\frac{1}{x_{13}^2x_{24}^2}+\frac{1}{x_{14}^2x_{23}^2}\right]\nonumber\\&+T_s\left[F(x_1,x_2,x_3,x_4)+F(x_1,x_3,x_2,x_4)+F(x_1,x_4,x_2,x_3)\right]\nonumber\\&=\frac{T_s^2}{\pi^2}\frac{1}{x_{12}^2x_{34}^2}\bigg[1+\chi^2+\frac{\chi^2}{(1-\chi)^2}+\frac{\pi^2}{T_s}\big(\chi^2 \bar{F}(\chi^{-1})+\frac{\chi^2\bar{F}(1-\chi)}{(1-\chi)^2}+\bar{F}(\chi)\big)\bigg]
\end{align}
The final result for the four-point function normalized by the two-point functions is:
\begin{align}\label{eq:4-pt static gauge}
    \frac{\braket{V_1V_2V_3V_4}}{\braket{V_1V_2}\braket{V_3V_4}}&\equiv G(\chi)=G_{\rm free}(\chi)+\frac{1}{2\pi T_s}G_{\rm tree}(\chi),
\end{align}
where the free and tree-level contributions are explicitly:
\begin{align}
    G_{\rm free}(\chi)&=1+\chi^2+\frac{\chi^2}{(1-\chi)^2},\label{eq:Gfree}\\
    G_{\rm tree}(\chi)&=-\frac{2(\chi^2-\chi+1)^2}{(1-\chi)^2}+\frac{-2+\chi+\chi^3-2\chi^4}{2\chi}\log\left((1-\chi)^2\right)\nonumber\\&+\frac{\chi^2(2-4\chi+9\chi^2-7\chi^3+2\chi^4)}{2(\chi-1)^3}\log(\chi^2).\label{eq:Gtree}
\end{align}
This reproduces the result in \cite{Giombi:2017cqn} for the four-point function of four identical scalars.\footnote{\cite{Giombi:2017cqn} used stereographic coordinates on $S^5$ while we use polar coordinates on $S^1$. The boundary correlators are independent of the choice of coordinates.} Meanwhile, quantum fluctuations about the classical solution in the path integral would correspond to Witten diagrams with loops; these give rise to corrections to \eqref{eq:2-pt static gauge} starting at order $T_s^0$  and to \eqref{eq:4-pt static gauge} starting at order $1/T_s^2$.

Given the four-point function in \eqref{eq:4-pt static gauge}, it is straightforward to analytically continue it to the OTOC configuration along the path in \eqref{eq:OTOC cross ratio}. In particular, the late time behaviors of $\chi$, $\log{\chi^2}$ and $\log((\chi-1)^2)$ along the path in \eqref{eq:OTOC cross ratio} are given by:  
\begin{align}\label{eq:analytic continuation logs}
    \chi(t)&=4ie^{-t}+O(e^{-2t}),&\log(\chi(t)^2)&= -2t+O(t^0),& \log((\chi(t)-1)^2)&=2\pi i+O(e^{-t}).
\end{align}
Thus, the exponentially growing piece at order $T_s^{-1}$ in \eqref{eq:4-pt static gauge} comes from the $\chi^{-1}\log(1-\chi)^2$ term, and we find:
\begin{align}\label{eq:AdS2 lyapunov otoc}
\frac{\braket{V_1V_2V_3V_4}}{\braket{V_1V_2}\braket{V_3V_4}}=1-\frac{1}{4T_s}e^{t}+\ldots.
\end{align}
This matches \eqref{eq:AdS2 OTOC Lyapunov} for the case $\Delta_V=\Delta_W=1$ (and working in units where $\ell=1$ and $\beta=2\pi$). 

The maximal chaos of the string worldsheet was demonstrated in this way, using the static gauge results for the correlators of the AdS$_2$ string, in \cite{Beccaria:2019dws}. It had also been demonstrated before that in \cite{Maldacena:2017axo,deBoer:2017xdk} using a scattering analysis and in \cite{Murata:2017rbp} by computing geodesic distances on the string worldsheet with shocks. To determine the OTOC beyond the Lyapunov regime using the static gauge approach, one would in principle need to compute more complicated Witten diagrams that also include loops. This seems difficult, although it is possible that the OTOC in the double scaled limit can be evaluated using some sort of eikonal approximation in which only a subset of simple diagrams survive. By contrast, it is straightforward to extend the scattering analysis used in \cite{Maldacena:2017axo,deBoer:2017xdk} to compute the full double scaled OTOC, as we show in the next section.

\section{OTOC from scattering on the \texorpdfstring{AdS$_2$}{AdS2} string}\label{sec:OTOC from shockwave S matrix}
\begin{figure}
	\centering
	\includegraphics[height=5cm]{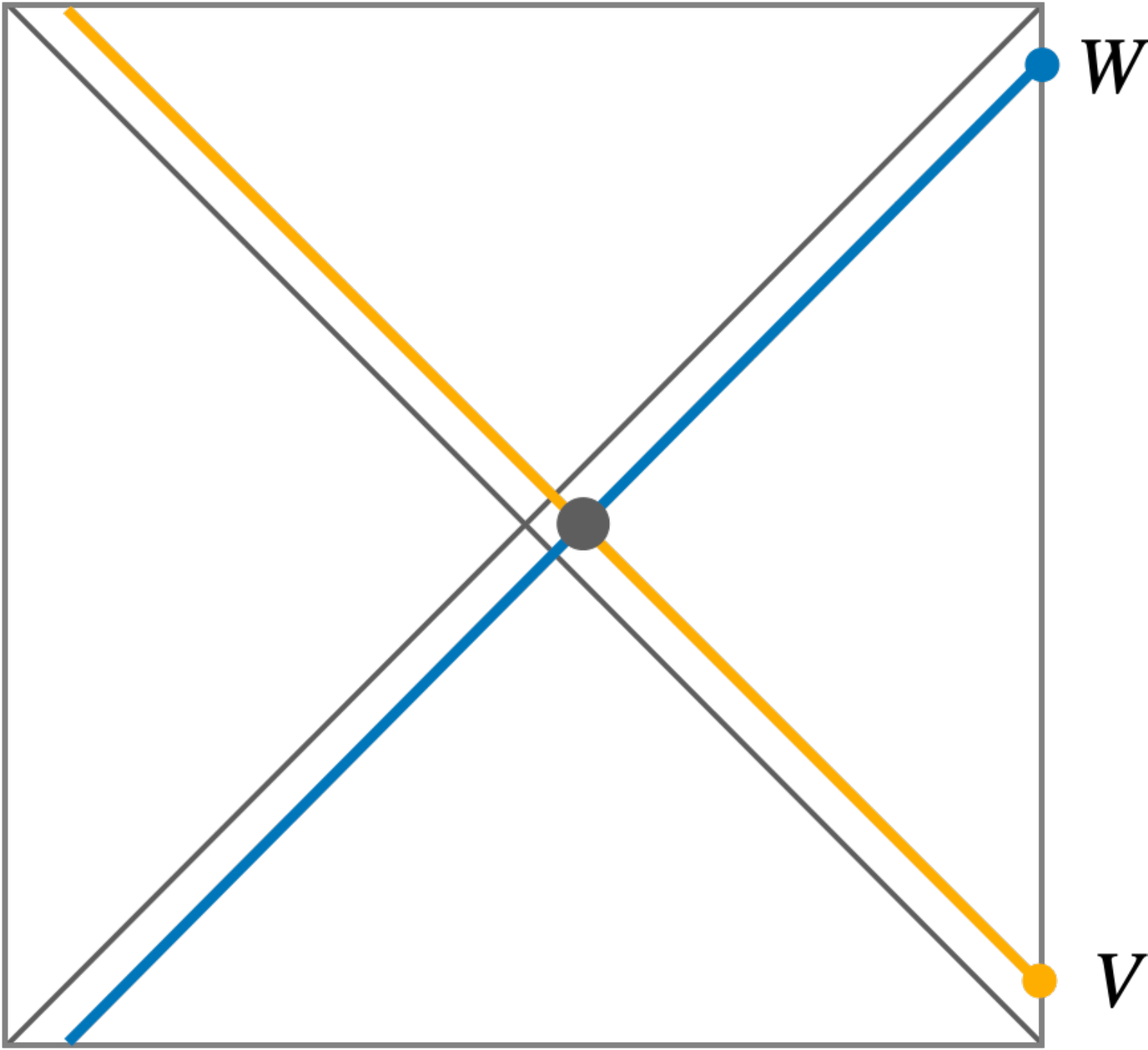}\\
    	\caption{The OTOC as a high energy scattering process on the string worldsheet.}
	\label{fig:scattering process}
\end{figure}

In this section, we derive the double-scaled OTOC in \eqref{eq:double scaled OTOC} by interpreting it as a $2\to 2$ worldsheet scattering amplitude of particles that are emitted and absorbed by $V$ and $W$ on the boundary of the string (see Figure~\ref{fig:scattering process}). This is a straightforward extension of \cite{deBoer:2017xdk} (and appendix C of \cite{Maldacena:2017axo}), which adapted the ideas of \cite{Shenker:2014cwa} to the worldsheet. The only extra input we need is the exact result for the scattering matrix on the free string in flat space that was derived in \cite{Dubovsky:2012wk}.  As we will see, the scattering process takes place in the same background (i.e., the AdS$_2$ ``black hole'') and is governed by the same local scattering interaction (i.e., the ``shockwave'' scattering matrix in two dimensions) as in JT gravity, so the analysis and final result are essentially the same as in \cite{Maldacena:2016upp,Lam:2018pvp}. Nonetheless, for completeness we provide a self-contained presentation of the scattering analysis.

Our starting point is the thermal AdS$_2$ string. One simple way to turn on a temperature is to work in AdS-Rindler coordinates covering a wedge of AdS$_{d+1}$ (see, e.g., \cite{Emparan:1999gf,Czech:2012be,Parikh:2012kg, Casini:2011kv}):
\begin{align}\label{eq:Rindler-AdS5 metric}
ds^2_{AdS_{d+1}}&=-\left(\frac{r^2}{\ell^2}-1\right)dt^2+\frac{dr^2}{\frac{r^2}{\ell^2}-1}+r^2 dH_{d-1}^2.
\end{align}
Here, $\ell$ is the radius of AdS$_{d+1}$, $r\in[\ell,\infty)$, $t\in \mathbb{R}$, and $H_{d-1}$ is the $(d-1)$-dimensional hyperbolic space with unit radius. (In this section only, we work in Lorentzian signature.) This metric has a horizon at $r=\ell$, which can be thought of as the Rindler horizon of an accelerating observer in AdS$_{d+1}$ whose trajectory starts and ends at $t=\pm \infty$ on the boundary at $r=\infty$. By the standard argument continuing \eqref{eq:Rindler-AdS5 metric} to euclidean time, the horizon has temperature $T=\frac{1}{2\pi \ell}$. The string we are interested in is extended in the $t$ and $r$ directions and sits at a point in $H_{d-1}$, with the following induced worldsheet metric:
\begin{align}\label{eq:AdS2 BH}
ds^2=-\left(\frac{r^2}{\ell^2}-1\right)dt^2+\frac{dr^2}{\frac{r^2}{\ell^2}-1}.
\end{align}
These coordinates cover a wedge of AdS$_2$, which we can think of as the exterior of an AdS$_2$ ``black hole.''

Gravity in the bulk Rindler wedge in \eqref{eq:Rindler-AdS5 metric} is dual to gauge theory on $\mathbb{R}\times \ell H_{d-1}$ at the fixed temperature $T=\frac{1}{2\pi \ell}$. (Note that at large $r$, the metric approaches $\frac{r^2}{\ell^2}(-dt^2+\ell^2 dH_{d-1}^2)$.) Thus, the string whose metric is given in \eqref{eq:AdS2 BH} is dual to a stationary quark in hyperbolic space at a special value of temperature set by the hyperbolic radius.\footnote{See \cite{Perlmutter:2016pkf} for a discussion about studying thermal OTOCs in $\mathbb{R}\times H_{d-1}$ using vacuum correlators in flat space.} Alternatively, because $\mathbb{R}\times \ell H_{d-1}$ is conformally equivalent to a Rindler wedge in Minkowski space, gravity in the bulk Rindler wedge in \eqref{eq:Rindler-AdS5 metric} is also dual to gauge theory in an accelerated frame, and the AdS$_2$ string is dual to a uniformly accelerating quark in the Minkowski vacuum.\footnote{A more direct way to see this is as follows. Let $X,Y,Z$ be Poincar\'e coordinates on a euclidean AdS$_3$ submanifold of euclidean AdS$_{d+1}$ with metric $ds^2=\ell^2 Z^{-2}(dX^2+dY^2+dZ^2)$. The AdS$_2$ string incident on the circle of radius $\ell$ is the hemisphere $X^2+Y^2+Z^2=\ell^2$, $Z\geq 0$. We analytically continue the string to Lorentzian signature by setting $Y=iT$, which yields the hemi-hyperboloid $X^2+Z^2-T^2=\ell^2$, $Z\geq 0$, in the Poincar\'e wedge of AdS$_3$ with metric $ds^2=\ell^2 Z^{-2}(-dT^2+dX^2+dZ^2)$. The string is incident on the hyperbola $X^2-T^2=\ell^2$ on the boundary at $Z=0$, whose two branches define the trajectories of a quark and anti-quark that experience uniform acceleration $a=\pm\ell^{-1}$ in the $X$ direction and measure an Unruh temperature $T=\frac{1}{2\pi \ell}$ in the Minkowski vacuum. If we parametrize the region of the hyperboloid that is accessible to the accelerating quark as
$X=\ell\big(1-\frac{\ell^2}{r^2}\big)^{1/2}\cosh(t/\ell)$, $T=\ell\big(1-\frac{\ell^2}{r^2}\big)^{1/2}\sinh(t/\ell)$, $Z=\ell^2/r$, the induced metric on the worldsheet is precisely \eqref{eq:AdS2 BH}, with $t$ being the proper time of the quark.

The accelerating quark-antiquark pair in $\mathcal{N}=4$ SYM connected by the AdS$_2$ string in the bulk is sometimes called the holographic EPR pair (see, e.g., \cite{Xiao:2008nr,Jensen:2013ora,Sonner:2013mba}), and was the version of the thermal AdS$_2$ string used to study chaos in \cite{Murata:2017rbp}. Yet another way to make the AdS$_2$ string thermal is to dangle it from the boundary into the horizon of a BTZ black hole, as in \cite{deBoer:2017xdk}.}

\begin{figure}[t]
\centering
\begin{minipage}{0.49\hsize}
\centering
\includegraphics[clip, height=5cm]{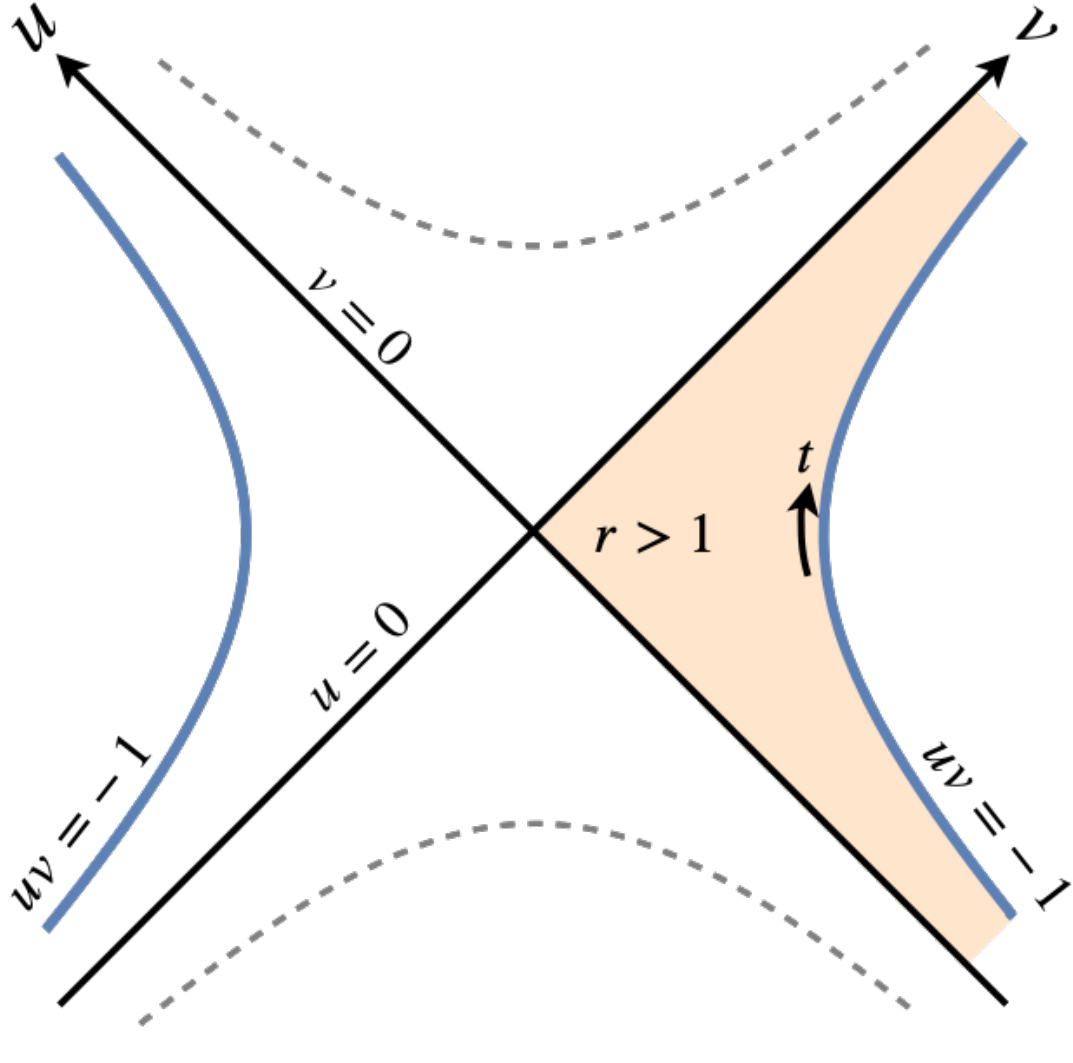}\\
{\bf a.} AdS$_2$ Kruskal diagram
\end{minipage}
\begin{minipage}{0.49\hsize}
\centering
\vspace{0.25cm}
\includegraphics[clip, height=4.5cm]{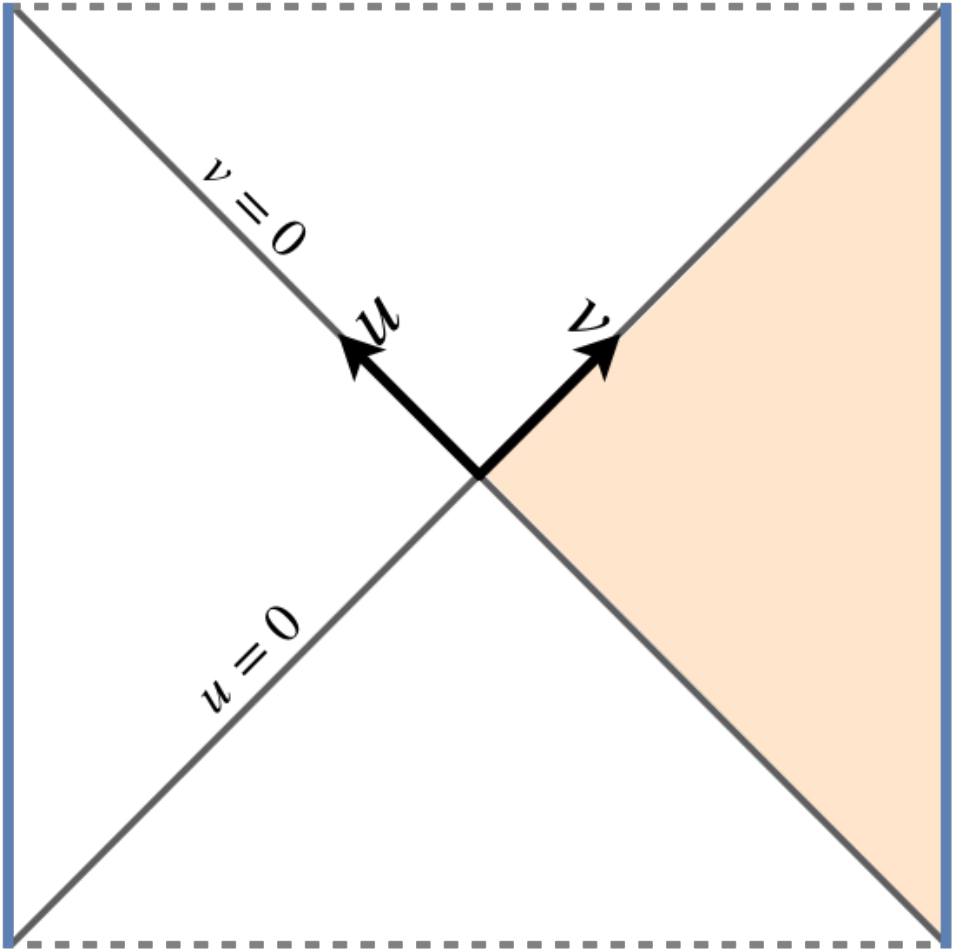}\\
\vspace{0.25cm}
{\bf b.} AdS$_2$ Penrose diagram
\end{minipage}
\caption{The Kruskal and Penrose diagrams for the AdS$_2$ worldsheet. The Kruskal coordinates in eq.~\eqref{eq:metric Kruskal} cover both sides of the $AdS_2$ black hole, while the Rindler coordinates in eq.~\eqref{eq:AdS2 BH} cover only the right wedge.}
\label{fig:AdS2 Kruskal and Penrose}
\end{figure}

The Schwarzschild coordinates in \eqref{eq:AdS2 BH} are adapted to the observer on the boundary, who has a horizon and measures the time $t$. To describe events near the horizon, it is useful to instead use Kruskal coordinates. Specifically, on the AdS$_2$ wedge, let
$u=-\ell\sqrt{\frac{r-\ell}{r+\ell}}e^{-t/\ell}$, $v=\ell \sqrt{\frac{r-\ell}{r+\ell}}e^{t/\ell}$, which produces the metric:
\begin{align}\label{eq:metric Kruskal}
    ds^2=-\frac{4dudv}{\big(1+\frac{uv}{\ell^2}\big)^2}. 
\end{align}
The Kruskal coordinates span $u\in \mathbb{R}$, $v\in \mathbb{R}$ subject to $uv>-\ell^2$, and cover both sides of the black hole patch of AdS$_2$. The two AdS$_2$ boundaries are at $uv=-\ell^2$, the future and past horizons of the boundary observer are at $u=0$ and $v=0$, and the wedge $r>\ell$ accessible to the boundary observer corresponds to $u<0$, $v>0$. The Kruskal and Penrose diagrams for the AdS$_2$ black hole are given in Figure~\ref{fig:AdS2 Kruskal and Penrose}. We also note that, at the boundary, the Kruskal coordinates are related to the boundary time by $u=-\ell e^{-t/\ell}$ and $v=\ell e^{t/\ell}$. For the remainder of this section, we will work in units where $\ell=1$ and $\beta=2\pi$.

Finally, we are ready to compute the OTOC measured by the observer in $\mathbb{R}\times H_{d-1}$ on the boundary of the AdS$_{d+1}$ wedge. We apply the general definition of the OTOC in \eqref{eq:OTOC definition}. The state in this case is the thermal bath at inverse temperature $\beta=2\pi$ plus the stationary quark; the operators $V_i$ and $W_i$ are general adjoint operators inserted along the Wilson line of the quark (e.g., they can be the displacement operators that generate transverse fluctuations of the string); and the times $t_i$ refer to the boundary or Schwarzschild time, which we will take to be $t_i=-i\theta_i$ with $\theta_i$ given concretely in \eqref{eq:OTOC euclidean times}. To make the scattering interpretation more manifest, it is convenient to write the OTOC in terms of the purification of the thermal state, which is a state living in two copies of $\mathbb{R}\times H_{d-1}$. Namely, we write \eqref{eq:OTOC definition} as
\begin{align}\label{eq:OTOC as transition amplitude}
    \braket{V_1W_3 V_2 W_4}=\braket{\rm TFD|V_1W_3V_2W_4|TFD}=\braket{\text{out}|\text{in}},
\end{align}
where
\begin{align}\label{eq:in-and-out states}
	\ket{\rm in}&\equiv V_2 W_4\ket{\rm TFD}, &\ket{\rm out}&\equiv W_3^\dagger V_1^\dagger \ket{\rm TFD}.
\end{align}
Here, $\ket{\rm TFD}$ denotes the thermofield double state (including the Wilson lines of the quark and its copy), which has the property that tracing over one (e.g., the left) copy of $\mathbb{R}\times H_{d-1}$ produces the thermal state in the other (e.g., the right) copy. In the bulk, $\ket{\rm TFD}$ is just the vacuum state of AdS$_{d+1}$ (plus the AdS$_2$ string), which appears thermal to a Rindler observer who only has access to a wedge of AdS$_{d+1}$. We take the four operators $V_1$, $V_2$ and $W_3$, $W_4$ in \eqref{eq:OTOC as transition amplitude} to act on only one (e.g., the right) copy of the CFT, so tracing out the left CFT reproduces \eqref{eq:OTOC definition}. 

\begin{figure}
    \centering
    \begin{minipage}{0.49\textwidth}
        \centering
        \includegraphics[width=0.6\textwidth]{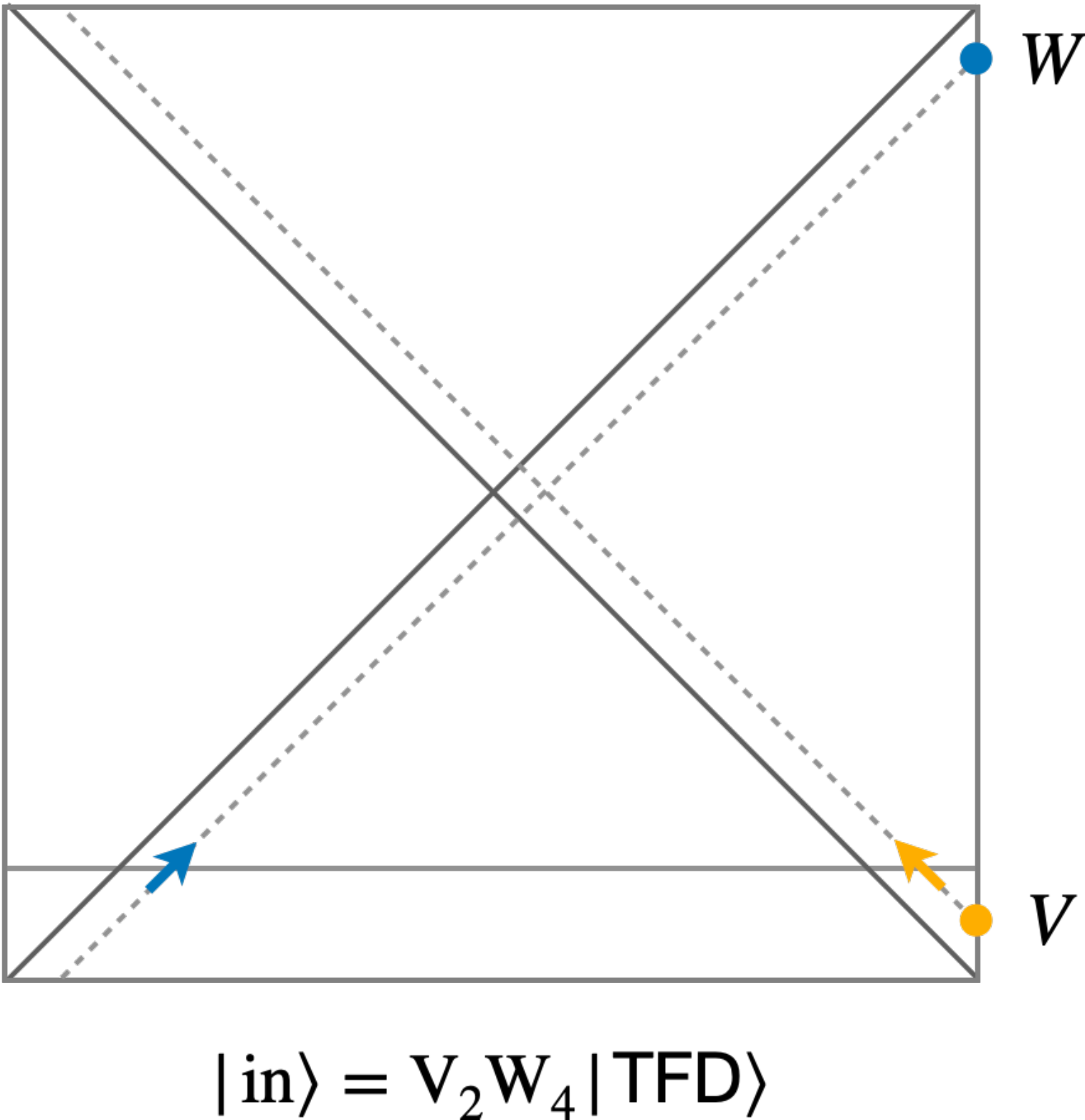}
    \end{minipage}
    \begin{minipage}{0.49\textwidth}
        \centering
        \includegraphics[width=0.6\textwidth]{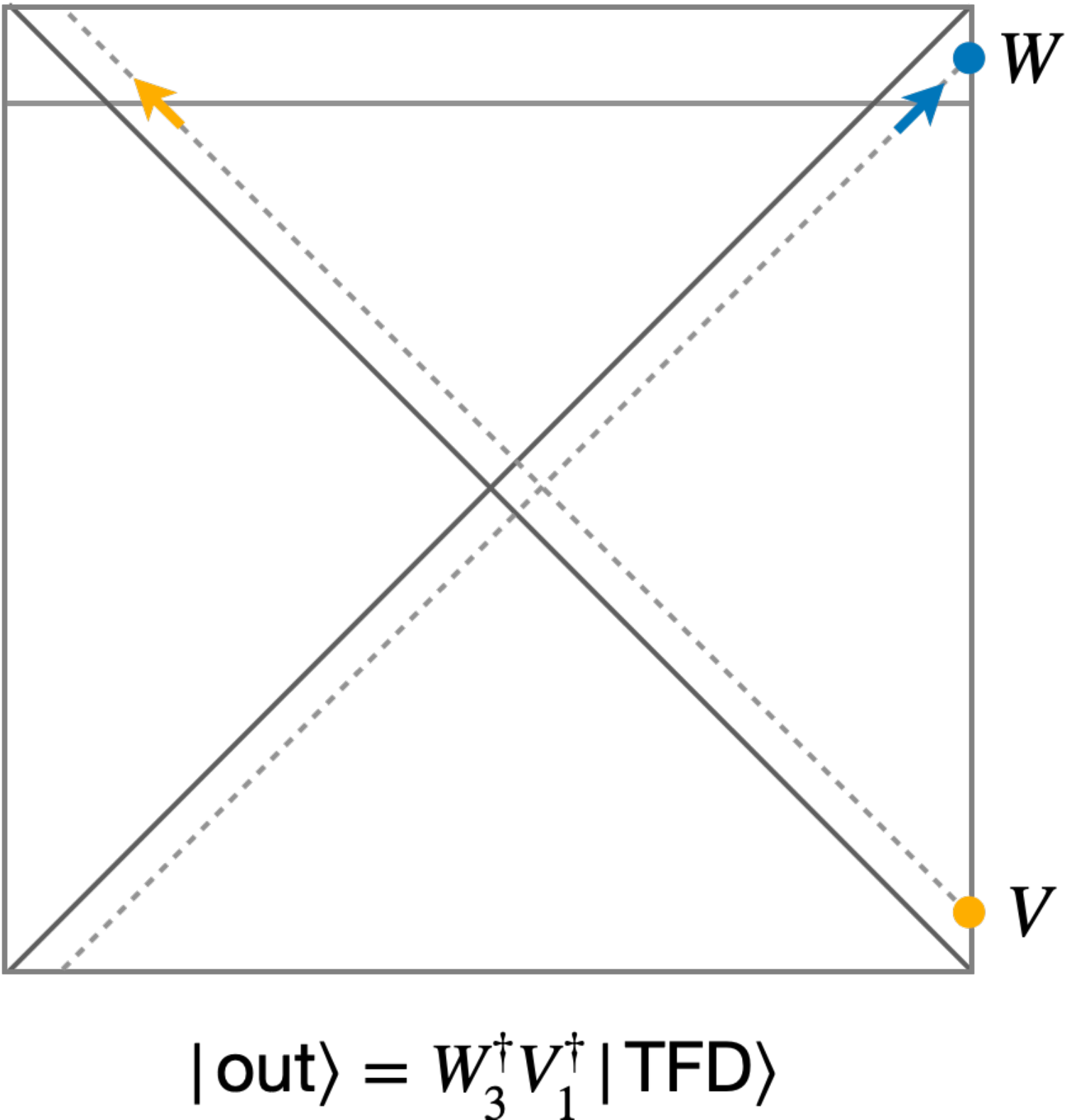}  
    \end{minipage}
    \caption{In and out scattering states.}
    \label{fig:OTOC in/out states}
\end{figure}

The overlap $\braket{\text{out}|\text{in}}$ in \eqref{eq:OTOC as transition amplitude} can be interpreted as a scattering amplitude on the worldsheet of the AdS$_2$ string. As the names imply, we can interpret $\ket{\rm in}$ as an ``in'' state and $\ket{\text{out}}$ as an ``out'' state. This is depicted in Figure~\ref{fig:OTOC in/out states} and can be understood heuristically as follows \cite{Shenker:2014cwa}. To create the state $V_2W_4\ket{\rm TFD}$, we start with the vacuum, evolve it forward to time $\frac{t}{2}$ and create a $W$ particle near the boundary in the upper right of the Penrose diagram, then evolve it backward to time $-\frac{t}{2}$ (during which the $W$ particle propagates backwards freely in the bulk and until it reaches the lower left of the Penrose diagram) and create a $V$ particle near the boundary in the bottom right of the Penrose diagram. The end result is a state with a right-moving $W$ particle in the bottom left of the Kruskal diagram and a left-moving $V$ particle in the bottom right of the Kruskal diagram. By analogous reasoning,  $W_3^\dagger V_1^\dagger \ket{\rm TFD}$ is interpreted as setting up a state with a left-moving $V$ particle in the top left of the Kruskal diagram and a right-moving $W$ particle in the top right of the Kruskal diagram.

The states in \eqref{eq:in-and-out states} can be expressed in terms of in and out states in the Kruskal momentum basis, in which the scattering interaction is simple.
Since $V$ acts on the boundary at early times, it creates a particle that travels along the $u$ horizon with positive momentum $p^u$; since $W$ acts at late times, it creates a particle that travels along the $v$ direction with positive momentum $p^v$. Thus, following \cite{Shenker:2014cwa,deBoer:2017xdk,Lam:2018pvp}, we write
\begin{align}
    \ket{\rm in}&=\int dp_2^u dp_4^v \Psi_{\Delta_V}(p_2^u,t_2)\Phi_{\Delta_W}(p_4^v,t_4)\ket{p_2^u,p_4^v}_{\rm in}.\label{eq:JMhCkQR2Wp}\\
    \ket{\rm out}&=\int dp_1^u dp_3^v \Psi_{\Delta_V}(p_1^u,t_1^*)\Phi_{\Delta_W}(p_3^v,t_3^*)\ket{p_1^u,p_3^v}_{\rm out}.\label{eq:lDYm704G1T}
\end{align}
(Note that $V_1^\dagger=V(t_1^*)$  and $W_3^\dagger=W(t_3^*)$ assuming $V$ and $W$ are Hermitian). Here, $\Psi_{\Delta_V}(p^u)$ is the wavefunction specifying the state in which $V$ creates the particle moving along the $v=0$ horizon. Likewise, $\Phi_{\Delta_W}(p^v)$ is the wavefunction specifying the state in which $W$ creates the particle moving along the $u=0$ horizon. We take the normalization of the Kruskal momentum eigenstates to be
\begin{align}\label{eq:in out states normalization}
    {}_{\rm in}\braket{p|q}_{\rm in}={}_{\rm out}\braket{p|q}_{\rm out}=p\delta(p-q).
\end{align}

It follows from \eqref{eq:JMhCkQR2Wp} and \eqref{eq:lDYm704G1T} that the scattering representation of the OTOC in the Kruskal momentum basis is \cite{Shenker:2014cwa,deBoer:2017xdk,Lam:2018pvp}:
\begin{align}\label{eq:OTOC kruskal momentum rep}
    \braket{V_1W_3V_2W_4}&=\int \underset{i}{\Pi} dp_i \Psi_{\Delta_V}(p_1^u,t_1^*)^* \Phi_{\Delta_W}(p_3^v,t_3^*)^* {}_{\rm out}\braket{p_1^u,p_3^v|p_2^u,p_4^v}_{\rm in}\Psi_{\Delta_V}(p_2^u,t_2) \Phi_{\Delta_W}(p_4^v,t_4).
\end{align}
Thus, the two ingredients we need to evaluate the above expression are the explicit forms of the wavefunctions $\Psi_\Delta$ and $\Phi_\Delta$ and the scattering matrix ${}_{\rm out}\braket{p_1^u,p_3^v|p_2^u,p_4^v}_{\rm in}$.

First, in the approximation where the particles propagate freely in AdS$_2$ until they interact near $u=v=0$, the wavefunctions are given by suitable Fourier transforms of the boundary-to-bulk propagators \cite{Shenker:2014cwa}. In Kruskal coordinates on AdS$_2$, for a scalar field of conformal dimension $\Delta$ and mass $m^2=\Delta(\Delta-1)$, the boundary-to-bulk propagator from the point $(u_i,v_i)=(-e^{-t_i},e^{t_i})$ on the boundary to the point $(u,v)$ in the bulk is given by 
\begin{align}\label{eq:boundary-to-bulk propagator}
    K_\Delta(u,v,t_i)&=c_\Delta \left(\frac{1+uv}{(1+uv_i)(1+vu_i)}\right)^\Delta, & c_\Delta &=\frac{(-1)^\Delta\Gamma(\Delta)}{\pi\sqrt{\Gamma(2\Delta)}}.
\end{align}
Then, to get the wavefunction for the particle created by $V(t_i)$ at an early time $t_i$, we take the Fourier transform of the boundary-to-bulk propagator connecting $t_i$ on the boundary to the $u=0$ horizon in the bulk. Likewise, to get the wavefunction for the particle created by $W(t_i)$ at a late time $t_i$, we take the Fourier transform of the boundary-to-bulk propagator connecting $t_i$ on the boundary to the $v=0$ horizon in the bulk. This means:
\begin{align}
    \Psi_\Delta(p^u,t_i)&=\int dv e^{2ip^u v}K_{\Delta}(0,v,t_i),&\Phi_\Delta(p^v,t_i)&=\int du e^{2ip^v u}K_{\Delta}(u,0,t_i).
\end{align}
We evaluate the integrals and arrive at the following expressions for the momentum wavefunctions:\footnote{If we want to keep track of the convergence of momentum integrals, it is convenient to assume that the imaginary parts of the four times satisfy $-\pi<\text{Im}(t_1)<\text{Im}(t_3)<0<\text{Im}(t_2)<\text{Im}(t_4)<\pi$ (as is the case in \eqref{eq:OTOC euclidean times} with $t_i=-i\theta_i$) and then analytically continue the final answer to more general values of the times.}
\begin{align}\label{eq:Kruskal wavefunctions}
    \Psi_\Delta(p^u,t_i)&=\theta(p^u)\frac{(2ip^u v_i)^\Delta }{\sqrt{\Gamma(2\Delta)}p^u}e^{2ip^u v_i},&\Phi_\Delta(p^v,t_i)&=\theta(p^v)\frac{(2ip^vu_i)^\Delta }{\sqrt{\Gamma(2\Delta)}p^v}e^{2ip^v u_i}.
\end{align}

One can check that the momentum space wavefunctions satisfy the identity:\footnote{In terms of the boundary-to-bulk propagator, \eqref{eq:wavefunction two-pt function} is 
\begin{align}
    \frac{i\pi }{2}\int du K_\Delta(u,0,t_1)\partial_u K_\Delta(u,0,t_2)=\frac{i\pi}{2}\int dv K_\Delta(0,v,t_1)\partial_v K_\Delta(0,v,t_2)=\frac{1}{[2\sin\left(\frac{it_{12}}{2}\right)]^{2\Delta}}.
\end{align}
}
\begin{align}\label{eq:wavefunction two-pt function}
    \int dp^u p^u \Psi_{\Delta}(p^u,t_1)^*\Psi_{\Delta}(p^u,t_2)=\int dp^v p^v\Phi_{\Delta}(p^v,t_1)^*\Phi_{\Delta}(p^v,t_2)=\frac{1}{\left[2\sin(\frac{it_{12}}{2})\right]^{2\Delta}}.
\end{align}
We chose the specific normalization $c_\Delta$ in \eqref{eq:boundary-to-bulk propagator} to make the r.h.s. of \eqref{eq:wavefunction two-pt function} simple. The identity in \eqref{eq:wavefunction two-pt function}, together with our choice of normalization for the momentum eigenstates in \eqref{eq:in out states normalization}, ensures that when we represent the single particle states created by $V_1^\dagger$ and $V_2$ as $\ket{V_1^\dagger}=\int dp^u \Psi_{\Delta_V}(p^u,t_1^*)\ket{p^u}$ and $\ket{V_2}=\int dp^u \Psi_{\Delta_V}(p^u,t_2)\ket{p^u}$, and the single particle states created by $W_3^\dagger$ and $W_4$ as $\ket{W_3^\dagger}=\int dp^v \Phi_{\Delta_W}(p^v,t_3^*)\ket{p^v}$ and  $\ket{W_4}=\int dp^v \Phi_{\Delta_W}(p^v,t_4)\ket{p^v}$, then the overlaps of the single-particle states reduce to unit-normalized conformal two-point functions:
\begin{align}
    \braket{V_1V_2}&=\braket{V_1^\dagger|V_2}=\frac{1}{[2\sin\left(\frac{it_{12}}{2}\right)]^{2\Delta_V}}, &  \braket{W_3W_4}=\braket{W_3^\dagger|W_4}&=\frac{1}{[2\sin\left(\frac{it_{34}}{2}\right)]^{2\Delta_W}}.
\end{align}

Next, we turn to the scattering matrix in \eqref{eq:OTOC kruskal momentum rep}. We take the scattering of excitations on the string worldsheet to be governed by:
\begin{align}\label{eq: out and in states}
    \ket{p^u,p^v}_{\rm out}&=e^{-i\ell_s^2 p^u p^v}\ket{p^u,p^v}_{\rm in}.
\end{align}
Namely, the Kruskal momentum of the two particles are individually conserved and the in and out states are related by a phase $e^{i\delta(s)}$, where the phase shift is proportional to the center-of-mass energy of the particles measured by an inertial observer at $u=v=0$: $\delta(s)=\frac{1}{4}\ell_s^2 s=\ell_s^2 p^u p^v$. We expect \eqref{eq: out and in states} to correctly capture the scattering of the two particles on the string worldsheet created by $V$ and $W$ in the limit we are considering because, when $V$ acts at early times and $W$ acts at late times, they generate particles that are localized near the $v=0$ and $u=0$ horizon and effectively interact only in a small region around $u=v=0$. The worldsheet and AdS$_{d+1}$ are approximately flat in this region and we can therefore invoke the exact result for the scattering matrix of excitations of the infinitely long free string in flat space, which we stated in \eqref{eq:S-matrix} and which was derived in \cite{Dubovsky:2012wk}. As explained in section $2$ of that paper, the exact scattering matrix can be extracted from the spectrum of excitations on the free critical string in Minkowski space with one coordinate compactified to a circle, after taking the radius to infinity to allow for asymptotic scattering states.\footnote{The scattering matrix on the string worldsheet was also derived using conformal gauge in \cite{Dubovsky:2016cog}. It would be interesting to make contact with that analysis, especially using the conformal gauge analysis in sections~\ref{sec:reparametrization mode for AdS string} and \ref{sec:OTOC from the string reparametrization mode}.}

It follows from \eqref{eq:in out states normalization} and \eqref{eq: out and in states} that 
\begin{align}\label{eq:scattering matrix}
    {}_{\rm out}\braket{p_2^u,p_4^v|p_1^u,p_3^v}_{\rm in}=e^{i\ell_s^2 p_1^u p_3^v}p_1^u p_3^v\delta(p_1^u-p_2^u) \delta(p_3^v-p_4^v).
\end{align}
Therefore, substituting \eqref{eq:Kruskal wavefunctions} and \eqref{eq:scattering matrix} into \eqref{eq:OTOC kruskal momentum rep}, we find:
\begin{align}\label{eq:ULUxWFhsH5}
    \braket{V_1W_3V_2W_4}&=(4v_1v_2)^{\Delta_V}(4u_3u_4)^{\Delta_W}\int_0^\infty dp^u dp^v \frac{(p^u)^{2\Delta_V-1}}{\Gamma(2\Delta_V)}\frac{(p^v)^{2\Delta_W-1}}{\Gamma(2\Delta_W)}e^{2ip^u(v_2-v_1)}e^{2ip^v (u_4-u_3)}e^{i\ell_s^2 p^up^v}.
\end{align}

Changing variables to $p=2ip^u(v_1-v_2)$ and $q=2ip^v(u_3-u_4)$ and rotating the contours to lie along the positive real $p$ and $q$ axes, the integral becomes
\begin{align}\label{eq:OTvzuRODWt}
    \braket{V_1W_3V_2W_4}&=\frac{(v_1v_2)^{\Delta_V}(u_3u_4)^{\Delta_W}}{[i(v_1-v_2)]^{2\Delta_V}[i(u_3-u_4)]^{2\Delta_W}}\int_0^\infty dp dq\frac{ p^{2\Delta_V-1}q^{2\Delta_W-1}}{\Gamma(2\Delta_V)\Gamma(2\Delta_W)}  e^{-p-q-\kappa pq},
\end{align}
where we have introduced
\begin{align}\label{eq:kappa}
    \kappa&=\frac{i\ell_s^2}{4(v_1-v_2)(u_3-u_4)}=\frac{i\ell_s^2 e^{(t_3+t_4-t_1-t_2)/2}}{16\sinh\left(\frac{t_{12}}{2}\right)\sinh\left(\frac{t_{34}}{2}\right)}~~\to~~\frac{\ell_s^2 e^t}{16}. 
\end{align}
In the last step, we put the times $t_i=-i\theta_i$ in the symmetric configuration specified by \eqref{eq:OTOC euclidean times}.

We identify the combination $(v_1v_2)^{\Delta_V}(i(v_1-v_2))^{-2\Delta_V}$ in \eqref{eq:OTvzuRODWt} as the two-point function $\braket{V_1V_2}$, and the combination $(u_3u_4)^{\Delta_W}(i(u_3-u_4))^{-2\Delta_W}$ as the two-point function $\braket{W_3W_4}$. Meanwhile, the double integral can be evaluated by first integrating over $q$ and then putting the integral over $p$ in the integral representation of the confluent hypergeometric function.\footnote{Namely, $U(a,b,z)=\Gamma(a)^{-1}\int_0^\infty dt e^{-zt}t^{a-1}(1+t)^{b-a-1}$.\label{fn:hypergeometric integral rep}} The final result for the OTOC is:
\begin{align}\label{eq:OTOC late time (reprise)}
    \frac{\braket{V_1W_3V_2W_4}}{\braket{V_1V_2}\braket{W_3W_4}}&=\kappa^{-2\Delta_V}U(2\Delta_V,1+2\Delta_V-2\Delta_W,\kappa^{-1}).
\end{align}
This is what was claimed in \eqref{eq:double scaled OTOC}. When expanded to leading order in $\kappa$, it also reproduces \eqref{eq:AdS2 OTOC Lyapunov}.  
 
Let us also briefly review how the time-ordered correlators are computed in the scattering picture. For instance, we can write $\braket{W_3 V_1 V_2 W_4}=\braket{V_1^\dagger W_3^\dagger|V_2W_4}$. (The discussion of $\braket{V_1 W_3 W_4 V_2}$ is similar). From the order of the operators, it follows that $\ket{V_1^\dagger W_3^\dagger}=V_1^\dagger W_3^\dagger \ket{\rm TFD}$ is an in state. It can be expressed in the momentum basis using the r.h.s. of \eqref{eq:lDYm704G1T} except with $\ket{p_1^u,p_3^v}_{\rm out}$ changed to $\ket{p_1^u,p_3^v}_{\rm in}$. In this case, in contrast to \eqref{eq:OTOC kruskal momentum rep} for the out-of-time-order correlator, it follows that the time-ordered correlator can be represented:
\begin{align}\label{eq:OBlQd5sZnJ}
    \braket{W_3V_1V_2W_4}&=\int \underset{i}{\Pi} dp_i \Psi_{\Delta_V}(p_1^u,t_1^*)^* \Phi_{\Delta_W}(p_3^v,t_3^*)^* {}_{\rm in}\braket{p_1^u,p_3^v|p_2^u,p_4^v}_{\rm in}\Psi_{\Delta_V}(p_2^u,t_2) \Phi_{\Delta_W}(p_4,t_4).
\end{align}
Given the normalization of the states in \eqref{eq:in out states normalization} and the wavefunction identity in \eqref{eq:wavefunction two-pt function}, this becomes
\begin{align}\label{eq:UZJorTs4bx}
    \braket{W_3V_1V_2W_4}&=\braket{V_1V_2}\braket{W_3W_4}.
\end{align}
Thus, the functional difference between the time-order and out-of-time-order correlators in the scattering picture is the absence or presence of the extra phase picked up by the scattering interaction. Without it, the two particles pass through each other without interacting. 

Some additional comments are in order. Firstly, our presentation of the scattering analysis on the string worldsheet is essentially equivalent to the one in \cite{deBoer:2017xdk}, which also arrived at \eqref{eq:OTvzuRODWt}. The only difference is that, because we use the exact scattering matrix on the worldsheet derived in \cite{Dubovsky:2012wk}, we believe \eqref{eq:OTvzuRODWt} is reliable in the double-scaled limit, not just to leading order in $\ell_s^2$. Secondly, our presentation is also essentially equivalent to the ones in \cite{Maldacena:2016upp,Lam:2018pvp}, which studied (among other things) the double scaled OTOC in JT gravity using the scattering analysis. The analysis is the same in both contexts because the two inputs in the scattering analysis--- the particle wavefunctions and the scattering matrix near the horizon--- are the same. The wavefunctions are the same because both the string worldsheet and the geometry of the bulk in JT gravity is AdS$_2$. Furthermore, the high energy scattering interaction in JT gravity is also given by \eqref{eq: out and in states} (with $\ell_s^2$ replaced by $16 \pi G_N/\tilde{\Phi}$ where $\tilde{\Phi}$ is a scale set by the divergence of the dilaton; see section~\ref{sec:Schwarzian from JT gravity}) \cite{Lam:2018pvp}. In gravity, this scattering matrix has a very natural interpretation in terms of the shockwave interaction between the high energy particles moving along the horizon of the black hole \cite{Dray:1984ha,tHooft:1990fkf,Shenker:2013pqa,Shenker:2014cwa}. In the frame of the first particle moving along $u$, the second particle moving along $v$ moves near the speed of light and generates a gravitational shockwave that, according to \eqref{eq:scattering matrix}, shifts the position of the first particle by $v\to v+\ell_s^2 p^u$. Likewise, in the frame of the second particle, the first particle generates a gravitational shockwave that shifts the second particle by $u\to u+\ell_s^2 p^v$. It is interesting that, as shown in \cite{Dubovsky:2012wk}, this same shockwave interaction appears to describe the scattering of particles on the string worldsheet, which gives the string worldsheet a ``gravitational flavor'' despite it not having a dynamical metric.

We can also comment on the sensitivity of the final result in \eqref{eq:OTOC late time (reprise)} on the details of the scattering process. Most of the details are washed out when we take the high energy limit. Consider the general integrable $2\to 2$ scattering matrix that is analytic, unitary and crossing symmetric \cite{Zamolodchikov:1991vx}:
\begin{align}\label{eq:riKHHU6xND}
    \mathcal{S}(s)&=\prod_i \frac{\mu_i+s}{\mu_i-s}e^{iP(s)}.
\end{align}
Here $\mu_i$ are the masses of the resonances and $P(s)$ is a scattering phase in the UV, which in our case is given by $\frac{1}{4}\ell_s^2 s$. In the scattering process above, the OTOC depends only on the $s\to \infty$ behavior of the scattering matrix, and is therefore not sensitive to the $\mu_i$. More generally, $S$-matrices of asymptotically free quantum field theory approach $1$ in $s\to \infty$ and lead to the same OTOC after the gravitional dressing by $P(s)=\frac{1}{4}\ell_s^2 s$. On the other hand, if there were $\ell_s^2$ corrections to $P(s)$ that are, for instance, of the form $P(s)=\frac{1}{4}\ell_s^2 s+b\ell_s^6 s^3+\ldots$, then these would survive in the double scaling limit. The result would differ from \eqref{eq:OTOC late time (reprise)} starting at order $\kappa^3$ in the small $\kappa$ expansion. Such corrections could arise from higher-charge analogs of the $T\bar{T}$ deformation as discussed by Smirnov and Zamolodchikov \cite{Smirnov:2016lqw} and  is expected to drastically change the UV behavior of the theory. It would be interesting to find a set up in top-down holography that gives these corrections.

\section{Two checks of the all-orders result for the double scaled OTOC}\label{sec:OTOC on WL to 3 loops}

In this section, we perform two checks of the result for the OTOC given in \eqref{eq:OTOC late time (reprise)}, using results for certain four-point functions on the Wilson line computed previously in the literature. First, the four-point function of unit charge scalars on the Wilson line was computed to fourth order in the strong coupling expansion in \cite{Ferrero:2021bsb}. Second, the four point function of two unit charge scalars and two charge $J$ scalars was computed in the double scaling limit $J\to \infty$, $\sqrt{\lambda}\to \infty$ with $J/\sqrt{\lambda}$ fixed in \cite{Giombi:2022anm}. We can continue both of these four point functions to the OTOC configuration, and check that they agree with \eqref{eq:OTOC late time (reprise)} in the appropriate regime.

\subsection{Four-point function of unit scalars at three loops}
The strong coupling expansion of the four-point function of four unit scalars on the Wilson line takes the form:
\begin{align}\label{eq:4-pt function 3-loops}
    \frac{\braket{ \Phi(x_1) \Phi(x_2)\Phi(x_3) \Phi(x_4)}}{\braket{\Phi(x_1) \Phi(x_2)}\braket{ \Phi(x_3) \Phi(x_4)}}&=G_{\rm free}(\chi)+\lambda^{-\frac{1}{2}}G_{\rm tree}(\chi)+\lambda^{-1}G_{1\text{-loop}}(\chi)\\&\hspace{3cm}+\lambda^{-\frac{3}{2}}G_{2\text{-loop}}(\chi)+\lambda^{-2}G_{3\text{-loop}}(\chi)+O(\lambda^{-\frac{5}{2}}).\nonumber
\end{align}
Here, $\Phi$ denotes one of the orthogonal scalars on the Wilson line (e.g., $\Phi=\Phi^1$), which has scaling dimension $\Delta_{\Phi}=1$ on the defect. This four-point function was computed to three-loops in \cite{Ferrero:2021bsb} by combining input from the AdS$_2$ string and the analytic bootstrap in the Wilson line defect CFT. To summarize, Ferrero and Meneghelli started with the free and tree-level results for the four-point function computed in the static gauge on the AdS$_2$ string (i.e., essentially the expressions in \eqref{eq:Gfree}-\eqref{eq:Gtree}). Then, by choosing an appropriate basis of functions (consisting of rational functions, logs and polylogs), imposing the bootstrap crossing equation, and carefully disentangling the contributions of degenerate operators to the conformal blocks, they were able to completely fix $G_{\rm 1-loop}(\chi)$, $G_{\rm 2-loop}(\chi)$,  and $G_{\rm 3-loop}(\chi)$. For instance, the $1$-loop contribution takes the form:
\begin{align}
    G_{\rm 1-loop}(\chi)&=r_1(\chi)\log(\chi-1)^2+r_2(\chi)\log(\chi-1)\log(\chi)+r_3(\chi)\log(\chi)^2\\
    &+r_4(\chi)\log(\chi-1)^2+r_5(\chi)\log(\chi)+r_6(\chi),
\end{align}
where $r_i(\chi)$, $i=1,\ldots,6$ are known rational functions of $\chi$. Similarly, $G_{\rm 2-loop}(\chi)$ and $G_{\rm 3-loop}(\chi)$ involve various combinations of rational functions, $\log(\chi)$, $\log(\chi-1)$ and also the trilogarithms $\text{Li}_3(1/\chi)$ and $\text{Li}_3(1/(1-\chi))$. The explicit expressions can be found in the supplementary Mathematica notebook provided with \cite{Ferrero:2021bsb}.\footnote{More precisely, \cite{Ferrero:2021bsb} provides the result for the four-point function of four general scalars, which depends on the conformally invariant cross-ratio $\chi$ and also on two $SO(5)$ invariant cross-ratios of the scalar polarizations, which are parametrized in the Mathematica notebook by the variables $\zeta_1$ and $\zeta_2$. For simplicity one can restrict to the case of four identical scalars by setting $\zeta_1=\zeta_2^{-1}=e^{\frac{i\pi}{3}}$, but the conclusions for the more general OTOC are the same. One should also note that the expressions in \cite{Ferrero:2021bsb} are given for the interval $0<\chi<1$. To get the expressions for $\chi>1$, one should use the relation $G(\chi)=\chi^2G(\chi^{-1})$, or, for the case of general scalars, $G(\chi,\zeta_1,\zeta_2)=\frac{\chi^2}{\zeta_1\zeta_2}G(\chi^{-1},\zeta_1^{-1},\zeta_2^{-1})$.}

When we analytically continue the explicit expression for \eqref{eq:4-pt function 3-loops} along the path in \eqref{eq:OTOC cross ratio}, using \eqref{eq:analytic continuation logs} as well as $\text{Li}_3(1/\chi(t))=-\frac{1}{6}t^3+O(t^2)$ and $\text{Li}_3(1/(1-\chi(t)))=\zeta(3)+O(e^{-t})$, we find that the contributions that survive in the double scaling limit $t\to \infty$, $\lambda\to \infty$ with $\lambda^{-\frac{1}{2}}e^{t}$ fixed come from the terms with the highest powers of $\log(\chi-1)$ at each order. More precisely, the terms relevant in the double scaling limit are:
\begin{align}
    G(\chi)&=1+\ldots+\lambda^{-\frac{1}{2}}\left[-2\chi^{-1}\log(\chi-1)+\ldots\right]+\lambda^{-1}\left[\frac{9}{2}\chi^{-2}\log(\chi-1)^2+\ldots\right]\nonumber\\&+\lambda^{-3/2}\left[-12\chi^{-3}\log(\chi-1)^3+\ldots\right]+\lambda^{-2}\left[\frac{75}{2}\chi^{-4}\log(\chi-1)^4+\ldots\right]+O(\lambda^{-5/2}),
\end{align}
and the OTOC in the double scaling limit becomes
\begin{align}\label{eq:WL OTOC 4 loops}
    \frac{\braket{\Phi_1\Phi_3 \Phi_2\Phi_4}}{\braket{\Phi_1\Phi_2}\braket{\Phi_3\Phi_4}}&=1-\frac{\pi}{2}\lambda^{-\frac{1}{2}}e^{t}+\frac{9\pi^2}{32}\lambda^{-1}e^{2t}-\frac{3\pi^3}{16}\lambda^{-\frac{3}{2}}e^{3t}+\frac{75\pi^4}{512}\lambda^{-2}e^{4t}+O(\lambda^{-\frac{5}{2}}e^{5t}).
\end{align}
This matches the expansion of \eqref{eq:OTOC late time (reprise)} to fourth order in $\kappa$ after we set $\Delta_V=\Delta_W=\Delta_{\Phi}=1$, $\kappa=\frac{\ell_s^2}{16}e^t$ (in accordance with \eqref{eq:kappa}), and remember that the AdS/CFT dictionary identifies $\ell_s^2=\frac{2\pi}{\sqrt{\lambda}}$.

As an aside, we also note the first subleading terms in the large $t$ expansion at each order in the large $\lambda$ expansion:
\begin{align}\label{eq:3-loop OTOC w/ corrections}
    \frac{\braket{\Phi_1\Phi_3 \Phi_2\Phi_4}}{\braket{\Phi_1\Phi_2}\braket{\Phi_3\Phi_4}}=1+O\big(e^{-2t}\big)&+\lambda^{-\frac{1}{2}}\bigg[-\frac{\pi}{2}e^{t}+O(t^0)\bigg]+\lambda^{-1}\bigg[\frac{9\pi^2}{32}e^{2t}+\frac{\pi t}{4}e^{t}+O\big(e^{t}\big)\bigg]\nonumber\\&+\lambda^{-\frac{3}{2}}\left[-\frac{3\pi^3}{16}e^{3t}-\frac{9\pi^2t}{32} e^{2t}+O\big(e^{2t}\big)\right]\nonumber\\&+\lambda^{-2}\left[\frac{75\pi^4}{512}e^{4t}+\frac{9\pi^3t}{32}e^{3t}+O\big(e^{3t}\big)\right]+O(\lambda^{-\frac{5}{2}}).
\end{align}
Interestingly, the first subleading terms at order $\lambda^{-1}$, $\lambda^{-\frac{3}{2}}$, and $\lambda^{-2}$ appear to be consistent with the Lyapunov exponent receiving a $1/\sqrt{\lambda}$ correction:
\begin{align}\label{eq:EhBQJYMxeT}
    \lambda_{\rm OTOC}&=1-\frac{1}{2\sqrt{\lambda}}+O(\lambda^{-1}).
\end{align}

It is unclear how to interpret this observation. When one considers stringy corrections to the Lyapunov exponent for the OTOC in the AdS$_5\times S^5$ bulk, the four-point function takes the general form $1-\frac{\#}{N}\text{exp}\left(\frac{2\pi}{\beta}(1-\frac{\#}{\sqrt{\lambda}})t+\ldots\right)+\ldots$. Thus, there are two independent parameters, $N^{-1}$ and $\lambda^{-\frac{1}{2}}$, controlling the scrambling time and the corrections to the Lyapunov exponent. By contrast, the two expansions are both controlled by $\lambda^{-\frac{1}{2}}$ in \eqref{eq:3-loop OTOC w/ corrections}. And indeed, higher order terms (e.g., the $t^2e^{t}$ term at order $\lambda^{-\frac{3}{2}}$ and the $t^2e^{2t}$ term at order $\lambda^{-2}$) are not consistent with \eqref{eq:EhBQJYMxeT}. 

\subsection{Four-point function of two heavy and two light scalars}
The scattering result for the OTOC in \eqref{eq:OTOC late time (reprise)} simplifies when one of the operators has a large conformal dimension. In particular, in the further double scaling limit $\kappa\to 0$ and $\Delta_V\to \infty$ with $\Delta_V \kappa$ held fixed, \eqref{eq:OTOC late time (reprise)} becomes\footnote{Eq.~\eqref{eq:general large charge OTOC} follows from the saddle point expansion applied to the integral representation in footnote \ref{fn:hypergeometric integral rep}. It can alternatively be derived by interpreting the OTOC as a $V$ two-point function in the state  $W\ket{\rm TFD}$. In this limit, the OTOC is approximately $ \braket{VWVW}/\braket{VV}/\braket{WW}\approx e^{-\ell}$ where $\ell$ is the renormalized length of the geodesic connecting the two insertions of $V$ on opposite boundaries of the AdS$_2$ geometry with a shockwave, where the shockwave arises due to the back-reaction to $W$. See section 3.5 of \cite{Shenker:2013pqa} and also \cite{Murata:2017rbp}.}
\begin{align}\label{eq:general large charge OTOC}
    \frac{\braket{V_1W_3V_2W_4}}{\braket{V_1V_2}\braket{W_3W_4}}&=\frac{1}{\left(1+\frac{1}{8}\Delta_V\ell_s^2e^t\right)^{2\Delta_W}}.
\end{align}
This provides an opportunity for another check of the scattering result. In the Wilson line CFT, a simple way to get operators with arbitrarily large conformal dimensions is to take composites of the scalar fields. One can define $\Phi^J\equiv (\epsilon\cdot \Phi)^J$, where $\epsilon\in \mathbb{C}^5$ is a null polarization vector ($\epsilon^2=0$) and $J$ is a positive integer. This is a chiral primary that transforms in the rank $J$ symmetric traceless representation of $SO(5)_R$ and its conformal dimension is protected and equal to its $R$ charge, $\Delta=Q_R=J$.

The four-point functions of two unit scalars and two charge $J$ scalars in the double scaling limit $\lambda\to \infty$, $J\to \infty$ with $J/\sqrt{\lambda}$ fixed were computed in \cite{Giombi:2022anm}. They can be determined from the Green's functions for transverse fluctuations of the worldsheet of an open string incident on the straight line on the boundary of AdS$_5$ and rotating with large angular momentum in $S^5$. We can analytically continue the four-point functions to the OTOC configuration, and take $t\to \infty$ and $J/\sqrt{\lambda}\to 0$ with $\frac{J}{\sqrt{\lambda}}e^{t}$ fixed. The details are given in appendix \ref{app:heavy-light OTOC}, and the result is:
\begin{align}
    \frac{\braket{\Phi(\theta_1)\Phi^J(\theta_3)\Phi(\theta_2) \Phi^J(\theta_4)}}{\braket{\Phi(\theta_1)\Phi(\theta_2)}\braket{\Phi^J(\theta_3)\Phi^J(\theta_4)}}=\frac{1}{\big(1+\frac{\pi}{4} \frac{J}{\sqrt{\lambda}}e^{t}\big)^2}\label{eq:PhPhJJ OTOC}
\end{align}
Comparing \eqref{eq:PhPhJJ OTOC} with \eqref{eq:general large charge OTOC}, we see that the large charge OTOC on the Wilson line matches the scattering picture OTOC in the light-light-heavy-heavy regime once we  set $\Delta_W=\Delta_\Phi=1$ and $\Delta_V=\Delta_{\Phi^J}=J$, and $\ell_s^2=\frac{2\pi}{\sqrt{\lambda}}$. This is a check of the scattering result for the OTOC assuming that the two ways of taking $\lambda$, $J$ and $t$ to infinity\footnote{I.e., we first take $\lambda,t\to \infty$ and then $J\to \infty$, $e^t/\sqrt{\lambda}\to 0$ to get \eqref{eq:general large charge OTOC}, and first take $J,\lambda\to \infty$ and then $t\to \infty$, $J/\sqrt{\lambda}\to 0$ to get \eqref{eq:PhPhJJ OTOC}} commute.

\section{The reparametrization mode on the \texorpdfstring{AdS$_2$}{AdS2} string}\label{sec:reparametrization mode for AdS string}

In the remainder of this work, we study the open string in AdS$_2\times S^1$ in the conformal gauge. Unlike the static gauge discussed in section~\ref{sec:static gauge analysis}, the conformal gauge features an intrinsic worldsheet metric, which might shed some light on the ``gravitational flavor'' of the string worldsheet that is hinted at by the maximal growth of the OTOC and by the discussions in \cite{Dubovsky:2012wk}. In this section, we will see that the conformal gauge analysis leads naturally to a dynamical reparametrization mode on the string boundary, which we will use in section~\ref{sec:OTOC from the string reparametrization mode} to compute the boundary correlators of the string to leading order as well as the OTOC in the double scaling limit. In this sense, the boundary reparametrization mode of the AdS$_2$ string is analogous to the Schwarzian mode in JT gravity, as we discuss in more detail in section~\ref{eq:AdS2 reparametrization and the Schwarzian}. 

There is a long history of integrals over boundary reparametrizations appearing in the study of open strings with fixed boundaries, going back to Douglas' solution to the Plateau problem \cite{douglas1931solution}. Douglas showed that the area of the minimal surface in $\mathbb{R}^d$ incident on a closed curve $\gamma:\alpha\mapsto \vec{x}(\alpha)$ is given by the following bilocal integral:
\begin{align}\label{eq:douglas integral}
    A=\underset{\alpha}{\text{minimize}}\left[\frac{1}{4\pi}\int_0^{2\pi} d\tau \int_0^{2\pi} d\tau' \frac{\big[\vec{x}(\alpha(\tau))-\vec{x}(\alpha(\tau'))\big]^2}{[2\sin\left(\frac{\tau-\tau'}{2}\right)]^2}\right].
\end{align}
Here, $\alpha(\tau)$ is a reparametrization of $\gamma$ and is minimized over. Similar expressions appear in the amplitudes of open strings propagating in flat space between fixed spacetime contours, except the reparametrizations of the boundary curve are integrated over. The integral over boundary reparametrizations is a remnant of the path integral over the worldsheet metric after gauge fixing. See \cite{Polyakov:1981rd, Alvarez:1982zi,Fradkin:1982ge, Cohen:1985sm,polyakov1987gauge} for some older references and \cite{Orland:2001rq, Makeenko:2009rf, Makeenko:2010fj, Makeenko:2010dq, Ambjorn:2014rwa,Tseytlin:2020izn} for some more recent related work.

Boundary reparametrizations have also featured in a few studies of the open string in AdS \cite{Polyakov:2000ti,Polyakov:2000jg,Rychkov:2002ni,Ambjorn:2011wz,Kruczenski:2014bla}. In particular, the area of the classical AdS string can also be represented as a non-local effective action minimized over the boundary reparametrizations, but an important difference compared to the string in flat space is that a general closed-form expression for the effective action that is analogous to eq.~\eqref{eq:douglas integral} is not easily obtained by elementary methods.\footnote{The AdS$_2$ sigma model is integrable, so one expects that it should be possible in principle to write down the general solution to the equation of motion using integrability techniques, such as the Pohlmeyer reduction (see for instance \cite{Janik:2011bd} for related calculations). We leave this to future work.} Nonetheless, one can identify a Douglas-type integral that is valid perturbatively and use it to study certain aspects of the dynamics of the AdS string (e.g., whether the AdS string satisfies the loop equations of planar Yang-Mills theory \cite{Polyakov:2000ti,Polyakov:2000jg}, and the one-loop corrections to the partition function \cite{Rychkov:2002ni,Ambjorn:2011wz}).

Our discussion of the AdS string reparametrization mode in the present section proceeds in a similar spirit to \cite{Polyakov:2000ti,Polyakov:2000jg,Rychkov:2002ni,Ambjorn:2011wz}, and our use of the reparametrization mode to compute the boundary correlators and the OTOC on the string in section~\ref{sec:OTOC from the string reparametrization mode} is guided by the example of the Schwarzian theory in \cite{Maldacena:2016upp}. 

\subsection{Conformal gauge}
We start with the Polyakov action for the string in AdS$_2\times S^1$, which is given in \eqref{eq:xyz string polyakov action}. We fix the conformal gauge by using a worldsheet coordinate transformation and Weyl rescaling to set the auxiliary metric equal to the AdS$_2$ metric: $h_{\alpha\beta}=\frac{1}{s^2}\delta_{\alpha\beta}$. The action then takes the form in \eqref{eq:string action conformal gauge}. It will be convenient to split it into three terms:
\begin{align}\label{eq:string action conformal gauge 2}
    S[x,z,y]&=S_{L}[x,z]+S_{T}[y]+T_sA_{\rm ws}.
\end{align}
The first and second terms are the actions of the longitudinal modes $x$, $z$ and the transverse mode $y$, which are:
\begin{align}
    S_{L}[x,z]&=\frac{T_s}{2}\int d^2\sigma \left[\frac{\partial^\alpha x \partial_\alpha x+\partial^\alpha z \partial_\alpha z}{z^2}-\frac{2}{s^2}\right],\label{eq:longitudinal action}\\
    S_{T}[y]&=\frac{T_s}{2}\int d^2\sigma \partial^\alpha y \partial_\alpha y.\label{eq:transverse action}
\end{align}
The third term in \eqref{eq:string action conformal gauge 2} is the (regularized) area of the worldsheet measured using the AdS$_2$ metric:
\begin{align}\label{eq:A_plane=0}
    A_{\rm ws}=\int d^2\sigma \sqrt{h}.
\end{align}
Note that $\sqrt{h}=s^{-2}$. We have separated out the area term from the longitudinal action in \eqref{eq:longitudinal action} to make the latter well behaved near the boundary, as we discuss later.

The open string is incident on a curve $\gamma$ on the boundary of AdS$_2\times S^1$ that can be represented by a map  $\gamma:\alpha\mapsto (\tilde{x}(\alpha),\tilde{y}(\alpha))$ from $\mathbb{R}$ to $\mathbb{R}\times S^1$. The general form of the boundary condition for the string is therefore:
\begin{align}\label{eq:general boundary condition}
    z(0,t)&=0,& x(0,t)&=\tilde{x}(\alpha(t)), & y(0,t)&=\tilde{y}(\alpha(t)).
\end{align}
Here, $\alpha(t)$ is a reparametrization of the boundary curve that appears because the parameter $\alpha$ along the curve $\gamma$ is not the same as the worldsheet coordinate $t$ along the boundary of the string. It is not possible to choose a parametrization of $\gamma$ for which $t$ and $\alpha$ can be identified, because putting the auxiliary metric in the conformal gauge in general requires a coordinate transformation (which is different for different auxiliary metrics) that changes $t$ at the boundary (see, e.g., the discussion in \cite{polyakov1987gauge}).

In the following, it will be more convenient in our study of boundary correlators to use the boundary condition
\begin{align}\label{eq:longitudinal boundary condition}
    z(0,t)&=0,& x(0,t)&=\alpha(t),
\end{align}
for the longitudinal modes and 
\begin{align}\label{eq:transverse boundary condition}
     y(0,t)&=\tilde{y}(\alpha(t))
\end{align}
for the transverse mode. As we saw in \eqref{eq:conformal boundary condition}, this is the boundary condition if the points on the boundary curve are labelled using the AdS$_2$ boundary coordinate: i.e., $\gamma:\alpha\to (\alpha,\tilde{y}(\alpha))$. It is also equivalent to \eqref{eq:general boundary condition} after renaming $\alpha\to \tilde{x}^{-1}\circ \alpha$ and $\tilde{y}\circ\tilde{x}\to \tilde{y}$. 

We will first study the classical string, and postpone the discussion of quantum corrections to section~\ref{sec:OTOC from the string reparametrization mode}. The equations of motion for the longitudinal modes follow from the action in \eqref{eq:longitudinal action}:
\begin{align}\label{eq:longitudinal eom}
    0&=\partial^\alpha\left(\frac{1}{z^2}\partial_\alpha x\right),&0&=\partial^\alpha\left(\frac{1}{z^2}\partial_\alpha z\right)+\frac{1}{z^3}(\partial^\alpha x \partial_\alpha x+\partial^\alpha z \partial_\alpha z),
\end{align}
The equation of motion for the transverse mode follows from \eqref{eq:transverse action}:
\begin{align}\label{eq:transverse eom}
    0=\partial^\alpha\partial_\alpha y.
\end{align}
These are supplemented by the Virasoro constraint (i.e., the equation of motion of $h_{\alpha\beta}$ following from the action in \eqref{eq:xyz string polyakov action}), which sets the stress tensor on the worldsheet equal to zero:
\begin{align} \label{eq:virasoro}
    0&=T_{\alpha\beta}^{L}+T^{T}_{\alpha\beta}.
\end{align}
We have separated the contributions to the stress tensor from the longitudinal and transverse modes:
\begin{align}
    T_{\alpha\beta}^{L}&= \frac{\partial_\alpha x\partial_\beta x + \partial_\alpha z \partial_\beta z}{z^2}-\frac{1}{2}\delta_{\alpha\beta}\frac{\partial^\gamma x\partial_\gamma x + \partial^\gamma z \partial_\gamma z}{z^2},\label{eq:longitudinal stress tensor}\\
    T_{\alpha\beta}^{T}&=\partial_\alpha y \partial_\beta y-\frac{1}{2}\delta_{\alpha\beta}\partial^\gamma y \partial_\gamma y.\label{eq:transverse stress tensor}
\end{align}

Thus, the action of the classical string can be found by first solving \eqref{eq:longitudinal eom}-\eqref{eq:virasoro} subject to the boundary conditions in \eqref{eq:longitudinal boundary condition}-\eqref{eq:transverse boundary condition} to get the classical solutions for $x(s,t)$, $z(s,t)$, and $y(s,t)$, and then evaluating the action in \eqref{eq:string action conformal gauge 2}. However, it is more enlightening to order this procedure differently. Because the longitudinal and transverse actions are decoupled, we can solve the equations of motion independently for $y(s,t)$ in terms of $\tilde{y}(\alpha(t))$ and $z(s,t)$ and $x(s,t)$ in terms of $\alpha(t)$. The only way the longitudinal and transverse modes are coupled is through their boundary conditions (which both depend on $\alpha(t)$) and the Virasoro constraint. Imposing the Virasoro constraint then fixes $\alpha(t)$ (as we will see, up to an $SL(2,\mathbb{R})$ transformation) and thus determines the classical solutions and classical action uniquely in terms of $\tilde{y}$. We can summarize this logic as:
\begin{align}
    S_{\rm cl}[\tilde{y}]&=T_sA_{\rm ws}+S_L[\alpha]+S_T[\tilde{y}\circ \alpha]\bigg\rvert_{\rm Virasoro}\label{eq:8qr0R8pczj}\\&=T_sA_{\rm ws}+S_L[\alpha]+S_T[\tilde{y}\circ \alpha]\hspace{2cm}\text{s.t. } T_{\alpha\beta}^L[\alpha]+T_{\alpha\beta}^T[\tilde{y}\circ \alpha]=0.\label{eq:bjeEswcvYB}
\end{align}

Let's explain the notation in these two equations. First, $S_{\rm cl}[\tilde{y}]$ denotes the action of the classical string, which is uniquely determined by the curve $\gamma$ (i.e., by the function $\tilde{y}$ given our parametrization of the curve). Second, $S_L[\alpha]\equiv S_L[x,z]$ and $T_{\alpha\beta}^L[\alpha]\equiv T_{\alpha\beta}^L[x,z]$ where $x(s,t)$ and $z(s,t)$ are the unique solutions to the equations of motion in \eqref{eq:longitudinal eom} with the boundary conditions in \eqref{eq:longitudinal boundary condition}. Likewise, $S_T[\tilde{y}\circ \alpha]\equiv S_T[y]$ and $T^T_{\alpha\beta}[\tilde{y}\circ \alpha]\equiv T_{\alpha\beta}^T[y]$ where $y(s,t)$ is the unique solution to the equation of motion in \eqref{eq:transverse eom} with the boundary condition in \eqref{eq:transverse boundary condition}. In other words, $S_L[x,z]$, $S_T[y]$, $T^L_{\alpha\beta}[x,z]$, $T^T_{\alpha\beta}[y]$ are the off-shell actions and stress tensors while $S_L[\alpha]$, $S_T[\tilde{y}\circ \alpha]$, $T^L_{\alpha\beta}[\alpha]$, $T^T_{\alpha\beta}[\tilde{y}\circ \alpha]$ are ``almost on-shell,'' except that they equal the longitudinal and transverse actions and stress tensors of the classical string only if we also fix $\alpha(t)$ by imposing the Virasoro constraint. To avoid cumbersome notation, we distinguish the off-shell and on-shell quantities only by their arguments; hopefully which we mean will also be clear from context.

Our goal in the remainder of this section is to implement \eqref{eq:8qr0R8pczj}. We first study the transverse and longitudinal modes separately and impose the Virasoro constraint only at the end. A crucial point in the analysis is that imposing the Virasoro constraint is equivalent to minimizing over the boundary reparametrizations, which means the classical action can also be written:
\begin{align}\label{eq:classical action in conformal gauge via extremization}
    S_{\rm cl}[\tilde{y}]=T_sA_{\rm ws}+\underset{\alpha}{\text{extremize}}\big[S_L[\alpha]+S_T[\tilde{y}\circ \alpha]\big].
\end{align}
See, e.g., \cite{Polyakov:1981rd,polyakov1987gauge,Rychkov:2002ni}. This point is also illustrated by the Douglas integral in \eqref{eq:douglas integral}. The connection between the Virasoro constraint and extremization over boundary reparametrizations is rather natural given that the auxiliary metric for a string with boundary can be put in conformal gauge only if the coordinate transformations are allowed to reparametrize the boundary, which means that varying the auxiliary metric varies the boundary reparametrization. We will also explicitly demonstrate the equivalence for the case of the string in AdS$_2\times S^1$. The main advantage of \eqref{eq:classical action in conformal gauge via extremization} over \eqref{eq:8qr0R8pczj} is that promoting the extremization to an integral over reparametrizations is a simple way to include quantum corrections to the classical result, which is the approach we will take in section~\ref{sec:OTOC from the string reparametrization mode}.

\paragraph{Complex notation.} Before proceeding, we note that the longitudinal and transverse equations of motion, stress tensors and actions can be expressed neatly using complex notation. Let $w=t+is, \bar{w}=t-is$ be complex coordinates on the upper half plane, and let $\partial\equiv \frac{1}{2}(\partial_t-i\partial_s)$ and $\bar{\partial}\equiv \frac{1}{2}(\partial_t+i\partial_s)$ be holomorphic and antiholomorphic derivatives. Furthermore, define the complex longitudinal and transverse stress tensors,
\begin{align}\label{eq:complex stress tensors}
    T^L&=T^{L}_{tt}-iT^{L}_{st},& \bar{T}^L&=T^{L}_{tt}+iT^{L}_{st},&
    T^T&=T^{T}_{tt}-iT^{T}_{st},& \bar{T}^T&=T^{T}_{tt}+iT^{T}_{st}.
\end{align}
Finally, combine the two longitudinal modes $x$ and $z$ into a single complex longitudinal mode:
\begin{align}\label{eq:complex longitudinal mode}
    X&= x+iz, &\bar{X}&= x-iz.
\end{align}

Then, the longitudinal and transverse stress tensors can be written as
\begin{align}
    T^L&= -\frac{8\partial X \partial \bar{X}}{(X-\bar{X})^2},&\bar{T}^L &=-\frac{8\bar{\partial}\bar{X}\bar{\partial}X}{(X-\bar{X})^2},&
    T^T&=2 \partial y \partial y,&\bar{T}^T&=2\bar{\partial}y \bar{\partial}y,\label{eq:stress tensor complex}
\end{align}
and the equations of motion for $X$, and $y$ simplify to
\begin{align}\label{eq:longitudinal and transverse eom complex}
    \partial \bar{\partial} X &= \frac{2\partial X \bar{\partial}X}{X-\bar{X}},&\partial \bar{\partial} \bar{X} &= \frac{2\partial \bar{X} \bar{\partial}\bar{X}}{\bar{X}-X}, &0=\partial \bar{\partial}y.
\end{align}
The equations of motion imply that the longitudinal and transverse stress tensors are conserved, which in this notation means they are holomorphic or antiholomorphic: 
\begin{align}\label{eq:holomorphicity stress tensor}
    0=\bar{\partial} T^T=\partial \bar{T}^T=\bar{\partial} T^L=\partial \bar{T}^L.
\end{align}
Finally, the transverse and longitudinal actions in complex notation are:
\begin{align}
   S_{L}[X]&=-2T_s \int dw d\bar{w}\left[\frac{\partial X \bar{\partial} \bar{X}+\partial \bar{X} \bar{\partial} X}{(X-\bar{X})^2}-\frac{1}{(w-\bar{w})^2}\right],& S_T[v]&=T_s\int dw d\bar{w} \partial y \bar{\partial} y.
\end{align}

\subsection{Transverse mode}\label{sec:transverse mode}

We now analyze the transverse mode. Because the action in \eqref{eq:transverse action} and equation of motion in \eqref{eq:transverse eom} for $y$ are those of a free massless scalar on AdS$_2$, we can immediately write down the on-shell transverse mode and the transverse action in terms of $\tilde{y}(\alpha(t))$. Indeed, these are the zeroth order results for the transverse mode in static gauge in section~\ref{sec:static gauge analysis}, except that the boundary condition now depends on the reparametrization $\alpha(t)$ and the transverse mode is exactly free in the conformal gauge. Thus, $y(s,t)$ is simply
\begin{align}\label{eq:y solns eom}
    y(s,t)&=\int dt' K(s,t,t')\tilde{y}(\alpha(t')),
\end{align}
where $K(s,t,t')=\frac{1}{\pi}\frac{s}{s^2+(t-t')^2}$. Furthermore, the action is:
\begin{align}\label{eq:transverse action divergent}
    S_{T}[\tilde{y}\circ \alpha]=-\frac{T_s}{2\pi}\int dt dt' \frac{\tilde{y}(\alpha(t))\tilde{y}(\alpha(t'))}{(t-t')^2}.
\end{align}

The bilocal integral in \eqref{eq:transverse action divergent} is technically infinite because of the divergence when $t$ and $t'$ are coincident. It is customary to think of the action as being implicitly regularized. For instance, we can replace $(t-t')^2\to (t-t')^2+s^2$ and take $s\to 0$ or $(t-t')^2\to (t-t')^{2\eta}$ and analytically continue $\eta\to 1$. For the purpose of taking variational derivatives of the transverse action (for instance, to compute correlation functions), this is perfectly satisfactory and the result is independent of the regularization scheme. However, for a massless scalar in AdS$_2$, the action is finite without need for regularization,\footnote{This is not true for massive scalars, which diverge near the AdS boundary.} and can be expressed as:
\begin{align}\label{eq:transverse action on-shell explicit}
    S_T[\tilde{y}\circ \alpha]=\frac{T_s}{4\pi}\int dt dt'\frac{[\tilde{y}(\alpha(t))-\tilde{y}(\alpha(t'))]^2}{(t-t')^2}.
\end{align}
This form of the action is manifestly finite.

It is instructive to explicitly show the steps needed to get \eqref{eq:transverse action on-shell explicit}. This is a good warm-up for when we study the longitudinal modes in the next section, where we will be interested in the value of the longitudinal action rather than its variational derivatives. We first integrate \eqref{eq:transverse action} by parts and express it as a boundary term:
\begin{align}
    S_{T}[\tilde{y}\circ \alpha]=-\frac{T_s}{2}\int dt y(t,0)\partial_s y(t,0)=-\frac{T_s}{2}\lim_{s\to 0}\int dt dt' \tilde{y}(\alpha(t))\tilde{y}(\alpha(t'))\partial_s K(s,t,t')\label{eq:tyUdUUufGI}.
\end{align}
The second equality follows from \eqref{eq:y solns eom}. Moving the limit inside the integral and using $\partial_s K(0,t,t')=\frac{1}{\pi}\frac{1}{(t-t')^2}$ formally leads to \eqref{eq:transverse action divergent}. However, the result is divergent and we can be more careful about interchanging the limit and integral. Instead, we move $\partial_s$ outside the integral, and use the fact that $\int dt K(s,t,t')=1$ to replace $\tilde{y}(\alpha(t))\tilde{y}(\alpha(t'))\to -\frac{1}{2}(\tilde{y}(\alpha(t))-\tilde{y}(\alpha(t')))^2$ without changing the value of the action. This replacement improves the convergence of the integral and makes it legitimate to now take the derivative and limit inside the integral. The final result is \eqref{eq:transverse action on-shell explicit}. 

As an aside, we note that if the string is in $\mathbb{R}^3$ instead of AdS$_2\times S^1$, then the two longitudinal modes $x$ and $z$ are also decoupled and their on-shell actions are both also given by \eqref{eq:transverse action on-shell explicit}, with the boundary value modified appropriately. The resulting expression for the action of the classical string is precisely the Douglas integral in \eqref{eq:douglas integral}, after changing the integration variable along the curve using $t=\tan{\frac{\tau}{2}}$ and invoking the fact that minimizing over the boundary reparametrization imposes the Virasoro constraint.

\subsection{Longitudinal modes}\label{sec:longitudinal modes}

It is more difficult to analyze the longitudinal modes than the transverse modes because the equations of motion in \eqref{eq:longitudinal eom} are non-linear, and their solution given the boundary condition in \eqref{eq:longitudinal boundary condition} with a general $\alpha(t)$ is not known. Therefore, we proceed in two ways. First, we will say as much as we can about the general properties of the longitudinal action without solving for it explicitly. Second, we will study the longitudinal action to leading order in perturbation theory, treating $\alpha(t)$ as a small fluctuation about the saddle point $\alpha(t)=t$. This approximation will be sufficient to evaluate the classical action to fourth order in perturbations of the boundary curve, the four-point functions to leading order in the inverse string tension, and the OTOC in the double scaled limit.

There are three important properties of the longitudinal action that we can study without solving the equations of motion: (i) the physical and gauge $SL(2,\mathbb{R})$ symmetries, (ii) the behavior near the boundary, and (iii) the equivalence of extremizing over the reparametrization mode $\alpha(t)$ and imposing the Virasoro constraint. In addition, we can completely solve for the longitudinal modes and action when the transverse modes are turned off (i.e., when $T^T=T^L=0$), which is the zeroth order step in the perturbative analysis.

\subsubsection{Two \texorpdfstring{$SL(2,\mathbb{R})$}{SL(2,R)} symmetries}\label{sec:SL(2,R) symmetries}

The string in AdS$_2\times S^1$ has two $SL(2,\mathbb{R})$ symmetries. The first moves the string around in the target space and is physical, while the second one corresponds to transformations of the worldsheet coordinates and is gauged. The physical $SL(2,\mathbb{R})$ symmetries are simply the isometries of the target space AdS$_2$. The gauge $SL(2,\mathbb{R})$ symmetries are the usual residual worldsheet coordinate transformations that leave the AdS$_2$ metric invariant up to a Weyl rescaling and that are therefore not fixed by the conformal gauge. See table \ref{tab: coordinates} for a summary of the actions of the two $SL(2,\mathbb{R})$ symmetries.

We can study the two $SL(2,\mathbb{R})$ symmetries at three levels. In addition to considering transformations of the entire string that leave the string action $S[x,y,z]$ and the Virasoro constraint invariant, we can consider transformations acting on the longitudinal modes that leave the off-shell longitudinal action $S_L[x,z]$ invariant, or transformations that act on the boundary reparametrization $\alpha(t)$ that leave the on-shell longitudinal action $S_L[\alpha]$ invariant. Studying the symmetries at the level of the string helps us interpret the two symmetries, but our primary goal is to understand the symmetries of the on-shell longitudinal action, which will become the effective action appearing in the reparametrization path integral that we will use to compute correlators in section~\ref{sec:OTOC from the string reparametrization mode}. In that context, the physical $SL(2,\mathbb{R})$ symmetries give rise to Ward identities for the correlators while the gauge $SL(2,\mathbb{R})$ symmetries need to be gauge fixed in the path integral.

\begin{table}[]
\begin{center}
\begin{tabular}{l|c}
Target space coords. & $x$, $z$               \\
Worldsheet coords.    & $t$, $s$             \\
Target space metric  & $z^{-2}(dx^2+dz^2)$.   \\
Worldsheet metric  & $s^{-2}(dt^2 + ds^2)$  \\
\hline
$SL(2,\mathbb{R})$ transformation  & $f(x)=\frac{ax+b}{cx+d}$                  \\
Physical $SL(2,\mathbb{R})$   & $x+iz\to f(x+iz)$                       \\
Gauge $SL(2,\mathbb{R})$      & $t+is\to f(t+is)$                 
\end{tabular}
\end{center}
\caption{Coordinates used to analyze the string in AdS$_2\times S^1$ in conformal gauge and two \texorpdfstring{$SL(2,\mathbb{R})$}{SL(2,R)} symmetries.}
\label{tab: coordinates}
\end{table}

We can represent a general $SL(2,\mathbb{R})$ transformation by the function
\begin{align}\label{eq:generic SL(2,R)}
    f(x)=\frac{ax+b}{cx+d},
\end{align}
where $a,b,c,d\in \mathbb{R}$, $ad - b c=1$. This is an $SL(2,\mathbb{R})$ transformation on the line if $x$ is real and on the half-plane if $x$ is complex with positive imaginary component.

First, we consider the physical $SL(2,\mathbb{R})$ symmetry. It acts as an AdS$_2$ isometry on the longitudinal modes and trivially on the transverse mode:
\begin{align}
    x(s,t)+iz(s,t)~~&\to ~~\bar{x}(s,t)+i\bar{z}(s,t)=f(x(s,t)+iz(s,t))\label{eq:physical SL(2,R) xz}\\
    y(s,t)~~&\to ~~\bar{y}(s,t)=y(s,t).\label{eq:physical SL(2,R) y}
\end{align}
Consistency with the boundary conditions in \eqref{eq:longitudinal boundary condition}-\eqref{eq:transverse boundary condition} means it also acts on the boundary reparametrization and the boundary curve as:
\begin{align}\label{eq:physical SL(2,R) bdy}
    \alpha(t)~~&\to~~ \bar{\alpha}(t)=f(\alpha(t))\\
    \tilde{y}(\alpha) ~~&\to~~ \bar{\tilde{y}}(\alpha)=\tilde{y}(f^{-1}(\alpha)).
\end{align}
This ensures that $\bar{x}(0,t)=\bar{\alpha}(t)$ and $\bar{y}(0,t)=\bar{\tilde{y}}(\bar{\alpha}(t))$ if $x(0,t)=\alpha(t)$ and $y(0,t)=\tilde{y}(\alpha(t))$.

\begin{figure}[t]
    \centering
    \includegraphics[scale=0.25]{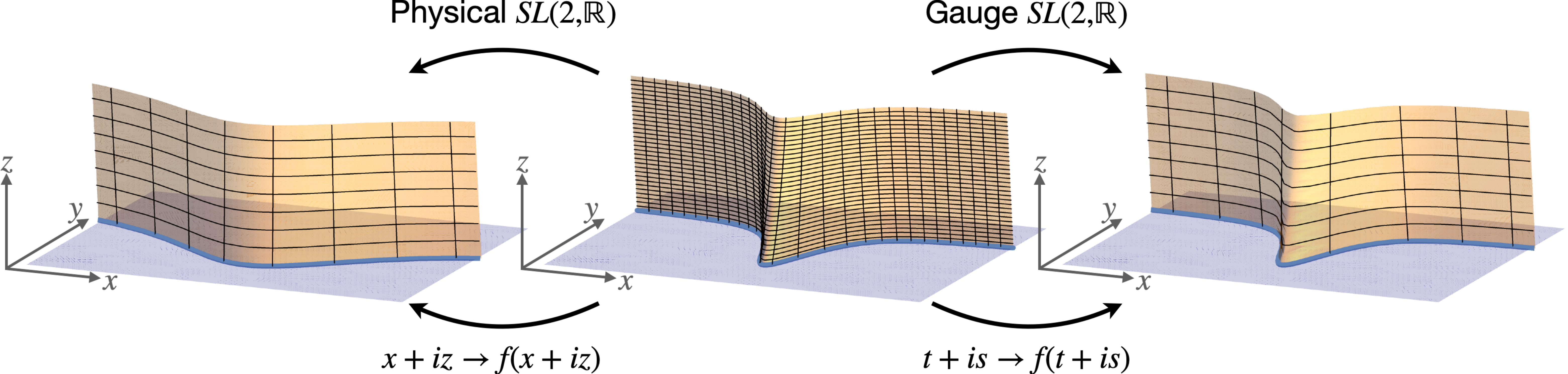}
    \caption{An example of physical and gauge $SL(2,\mathbb{R})$ transformations acting on an open string in AdS$_2\times S^1$ incident on the boundary. The physical $SL(2,\mathbb{R})$ transformation moves the string in AdS$_2\times S^1$, whereas the gauge $SL(2,\mathbb{R})$ transformation merely relabels the worldsheet coordinates (which is represented by the rescaling of the mesh lines on the string).}
    \label{fig:physicalvsgaugeSL2R}
\end{figure}

Because the transformation is an isometry of the target space AdS$_2$, it leaves the longitudinal action invariant in \eqref{eq:longitudinal action}: i.e., $S_L[\bar{x},\bar{z}]=S_L[x,z]$ off-shell and $S_L[\alpha]=S_L[\bar{\alpha}]$ on-shell. Similarly, the longitudinal stress tensor in \eqref{eq:longitudinal stress tensor} satisfies $T^L[\bar{x},\bar{z}]=T^L[x,z]$ off-shell and $T^L[\bar{\alpha}]=T^L[\alpha]$ on shell. Because the transformation does nothing to $y(s,t)$, it trivially leaves the transverse action and stress tensor invariant. The total action and the total stress tensor are therefore also both invariant. Finally, we see that this $SL(2,\mathbb{R})$ transformation is physical because it actually moves the string in the target space AdS$_2\times S^1$ and the curve on the boundary $\mathbb{R}\times S^1$--- namely, $(z(s,t), x(s,t), y(s,t))$ and $(\bar{z}(s,t),\bar{x}(s,t),\bar{y}(s,t))$ represent different strings and $(\alpha,\tilde{y}(\alpha))$ and $(\alpha,\bar{\tilde{y}}(\alpha))$ represent different boundary curves. This is illustrated in the left half of Figure~\ref{fig:physicalvsgaugeSL2R}. 

Next, we consider the gauge $SL(2,\mathbb{R})$ symmetry. It acts on the longitudinal and transverse modes through the worldsheet coordinates as:
\begin{align}
    x(s,t)+iz(s,t) ~~&\to~~\bar{x}(s,t)+i\bar{z}(s,t)=x(\bar{s},\bar{t})+iz(\bar{s},\bar{t}),\label{eq:gauge SL(2,R) xz}\\
    y(s,t) ~~&\to~~\bar{y}(s,t)=y(\bar{s},\bar{t}), \label{eq:gauge SL(2,R) y}
\end{align}
where
\begin{align}\label{eq:worldsheet SL(2,R)}
    \bar{t}+i\bar{s}=f(t+is).
\end{align}
Consistency with \eqref{eq:longitudinal boundary condition} means it also transforms the boundary reparametrization but not the boundary curve:
\begin{align}\label{eq:gauge SL(2,R) bdy}
    \alpha(t) ~~&\to ~~\bar{\alpha}(t)=\alpha(f(t))\\
    \tilde{y}(\alpha) ~~&\to ~~\bar{\tilde{y}}(\alpha)=\tilde{y}(\alpha).
\end{align}

To see how this transformation changes the actions and stress tensors, we note that the Jacobian for the worldsheet change of coordinates in \eqref{eq:worldsheet SL(2,R)}:
\begin{align}
    \frac{\partial \bar{\sigma}^\alpha}{\partial \sigma^\beta}=\frac{1}{((d+ct)^2+c^2s^2)^2}\left(\begin{array}{cc}(d+ct)^2-s^2c^2 & - 2cs (d+ct) \\ 2 c s (d+ct) & (d+ct)^2-s^2c^2\end{array}\right).
\end{align}
It satisfies $\delta^{\gamma\delta}\frac{\partial\bar{\sigma}^\alpha}{\partial \sigma^\gamma}\frac{\partial \bar{\sigma}^\beta}{\partial \sigma^\delta}=|\frac{\partial \bar{\sigma}}{\partial \sigma}|\delta^{\alpha\beta}$ and $\delta_{\alpha\beta}\frac{\partial\bar{\sigma}^\alpha}{\partial \sigma^\gamma}\frac{\partial \bar{\sigma}^\beta}{\partial \sigma^\delta}=|\frac{\partial \bar{\sigma}}{\partial \sigma}|\delta_{\gamma\delta}$ where $|\frac{\partial \bar{\sigma}}{\partial \sigma}|=(c^2 s^2+(d+ct)^2)^{-2}$. We note also that $s^{-2}=\bar{s}(s)^{-2}|\frac{\partial \bar{\sigma}}{\partial \sigma}|$ because $\sqrt{h}=s^{-2}$ is a tensor density. 

It follows that, under the transformation in \eqref{eq:gauge SL(2,R) xz}, the longitudinal action satisfies
\begin{align}\label{eq:longitudinal action under SL(2,R) gauge transformation}
    S_L[\bar{x},\bar{z}]=\frac{T_s}{2}\int d^2\sigma \bigg|\frac{\partial \bar{\sigma}}{\partial \sigma}\bigg|\left[\frac{\partial_\alpha x (\bar{\sigma})\partial^\alpha x (\bar{\sigma})+\partial_\alpha z (\bar{\sigma})\partial^\alpha z (\bar{\sigma})}{z(\bar{\sigma})^2}-\frac{2}{\bar{s}^2}\right]=S_L[x,z].
\end{align}
Likewise, under the transformation in \eqref{eq:gauge SL(2,R) y}, the transverse action satisfies $S_T[\bar{y}]=S_T[y]$. These also imply $S_L[\alpha]=S_L[\bar{\alpha}]$ and $S_T[\tilde{y}\circ \bar{\alpha}]=S_T[\tilde{y}\circ \alpha]$. Thus, both the longitudinal and transverse Lagrangians transform as worldsheet tensor densities and the actions are invariant both off-shell and on-shell. Meanwhile, the stress tensors transform as worldsheet tensors:
\begin{align}\label{eq:longitudinal stress tensor under SL(2,R) gauge transformation}
    T_{\alpha\beta}^L[\bar{x},\bar{z}]\bigg\rvert_{\sigma}&=\frac{\partial \bar{\sigma}^\gamma}{\partial \sigma^\alpha}\frac{\partial \bar{\sigma}^\delta}{\partial \sigma^\beta}T_{\gamma\delta}^L[x,z]\bigg\rvert_{\bar{\sigma}(\sigma)},
    &T_{\alpha\beta}^T[\bar{y}]\bigg\rvert_{\sigma}&=\frac{\partial \bar{\sigma}^\gamma}{\partial \sigma^\alpha}\frac{\partial \bar{\sigma}^\delta}{\partial \sigma^\beta}T_{\gamma\delta}^T[y]\bigg\rvert_{\bar{\sigma}(\sigma)}.
\end{align}
This is why it is necessary to transform both the longitudinal and transverse modes simultaneously in \eqref{eq:gauge SL(2,R) xz}-\eqref{eq:gauge SL(2,R) y} (in contrast with \eqref{eq:physical SL(2,R) xz}-\eqref{eq:physical SL(2,R) y}) because the Virasoro constraint is otherwise not preserved. Finally, we see that this $SL(2,\mathbb{R})$ transformation is gauge because it simply relabels the worldsheet coordinates without actually moving the string in target space or the curve on the boundary. This is illustrated in the right half of Figure~\ref{fig:physicalvsgaugeSL2R}.

\subsubsection{Behavior near the boundary}\label{sec:behavior near bdy}
The divergence of the AdS metric at the boundary may lead us to ask whether \eqref{eq:longitudinal action} is well defined. We now show that the longitudinal action is finite if $x$ and $z$ satisfy the equations of motion.

Since the divergence of the AdS$_2$ string comes from the region near the boundary, we want to study the general behavior of $x$ and $z$ at small $s$. Following \cite{Polyakov:2000ti,Polyakov:2000jg}, we expand $x(s,t)$ and $z(s,t)$ as
\begin{align}
    x(s,t)&=\alpha(t)+\sum_{n=1}^\infty a_n(t)s^n,&z(s,t)&=\sum_{n=1}^\infty b_n(t)s^n.
\end{align}
We used the boundary conditions in \eqref{eq:longitudinal boundary condition} to fix the zeroth order coefficients. We can fix the higher order coefficients by substituting these series expansions into the two equations of motion in \eqref{eq:longitudinal eom} and setting the coefficients of each power of $s$ in each equation to zero. The first and second order terms imply $a_1=0$, $b_1=\dot{\alpha}$ and $b_2=0$, $a_2=-\frac{1}{2}\ddot{\alpha}$. The next order terms identically vanish, which means that the equations of motion near $s=0$ do not constrain $a_3(t)$ and $b_3(t)$. For convenience, we can rewrite $a_3(t)=\frac{1}{3}g(t)$ and $b_3(t)=\frac{1}{3}\left[h(t)-\frac{1}{2}\dddot{\alpha}(t)\right]$, where $g(t)$ and $h(t)$ are as yet undetermined arbitrary functions. Thus, to order $s^3$, the longitudinal modes are given by:
\begin{align}
    x(s,t)&=\alpha(t)-\frac{\ddot{\alpha}(t)}{2}s^2+\frac{g(t)}{3}s^3+\ldots\label{eq:x series expansion}\\
    z(s,t)&=\dot{\alpha}(t)s+\bigg[\frac{h(t)}{3}-\frac{\dddot{\alpha}(t)}{6}\bigg]s^3+\ldots\label{eq:z series expansion}
\end{align}
Up to relabelling, this is equivalent to eq. (4.3) of \cite{Polyakov:2000ti}. It is easy to check that all higher order coefficients, $a_n$ and $b_n$ for $n\geq 4$, are fixed in terms of $\alpha$, $g$ and $h$. For our purposes knowing the terms in \eqref{eq:x series expansion}-\eqref{eq:z series expansion} will be sufficient.

Although the expansion of the equations of motion near the boundary does not constrain $g(t)$ and $h(t)$, we expect based on uniqueness that they should be determined by $\alpha(t)$. As pointed out in \cite{Polyakov:2000ti}, this can in principle be done by requiring $x(s,t)$ and $z(s,t)$ to be well behaved as $s\to \infty$, but is not easy to implement because the series converge only when $s$ is sufficiently small. Thus, while the series expansions are useful for understanding the behavior of the longitudinal modes near the boundary, they do not provide a viable method to solve the longitudinal action exactly.

Given the series expansions in \eqref{eq:x series expansion}-\eqref{eq:z series expansion} and the longitudinal action in \eqref{eq:longitudinal action}, we find that the longitudinal Lagrangian near $s=0$ takes the form:
\begin{align}\label{eq:longitudinal lagrangian near s=0}
    \frac{\partial_\alpha x \partial^\alpha x + \partial_\alpha z \partial^\alpha z }{z^2}-\frac{2}{s^2}=O(s^0).
\end{align}
This means the on-shell longitudinal action is finite, as desired.

Finally, although it affects neither the longitudinal nor the transverse modes, we should also comment on the area contribution to the string action in \eqref{eq:string action conformal gauge 2}. The area of AdS$_2$ is infinite and requires some sort of regularization. One method of regularization is to introduce a cut-off curve near the boundary of the worldsheet, remove the divergent contribution that is proportional to its length (which can be interpreted as a renormalization of the mass of the dual particle propagating on the boundary) and then send the cut-off to the boundary \cite{Maldacena:1998im,Rey:1998ik}.\footnote{Alternatively, one can add a boundary term to the string action that implements the Legendre transform with respect to the AdS bulk coordinate \cite{Drukker:1999zq}. The resulting action is finite and also yields \eqref{eq:regularized are hyperbolic plane} for the hyperbolic plane.
} The regularization does not depend on the specific way in which the cut-off curve is sent to the boundary, and yields the following well-known result for the regularized area of the hyperbolic plane:
\begin{align}\label{eq:regularized are hyperbolic plane}
    A_{\rm ws}=0.
\end{align}
We illustrate this procedure for both the hyperbolic plane and the hyperbolic disk in appendix~\ref{app: AdS2 area with cutoff}. An interesting feature of this analysis is that it leads to a Schwarzian term in the expression for the area--- but, importantly, the Schwarzian term decouples as the curve is sent to the boundary of AdS. We discuss this also in section~\ref{eq:AdS2 reparametrization and the Schwarzian}.

\subsubsection{Extremizing over reparametrizations and the Virasoro constraint}\label{sec:extremization and Virasoro}
Next, we demonstrate the equivalence of the Virasoro constraint and the extremization over boundary reparametrizations for the string in AdS$_2\times S^1$, which is summarized by \eqref{eq:8qr0R8pczj} and \eqref{eq:classical action in conformal gauge via extremization}, without explicitly solving for the longitudinal modes. The equivalence follows from two facts: (i) extremizing over reparametrizations $\alpha(t)$ sets one component of the sum of the longitudinal and stress tensors to be zero on the boundary of the worldsheet, and (ii) the on-shell longitudinal and transverse stress tensors are holomorphic on the worldsheet.

First, we write the variations of the on-shell transverse and longitudinal actions as boundary terms. For the transverse action in \eqref{eq:longitudinal action}, we have:
\begin{align}\label{eq:variation transverse action}
    \delta S_T&=T_s \int d^2\sigma \left[\partial_\mu ( \partial^\mu y \delta y)+ (y\text{ e.o.m.})\delta y \right]=-T_s\int dt \big[y'(0,t)\delta y(0,t)\big].
\end{align}
For the longitudinal action in \eqref{eq:transverse action}, we have:
\begin{align}\label{eq:variation longitudinal action}
    \delta S_{L}&=T_s \int d^2\sigma\left[\partial_\mu\left(\frac{\partial^\mu x \delta x}{z^2}\right)+\partial_\mu\left(\frac{\partial^\mu z \delta z}{z^2}\right)+(x\text{ e.o.m.})\delta x + (z\text{ e.o.m.})\delta z\right]\nonumber\\&=-T_s\lim_{s_0\to 0}\int dt \left[\frac{x'(s,t)\delta x(s,t)}{z(s,t)^2}+\frac{z'(s,t)\delta z(s,t)}{z(s,t)^2}\right]_{s=s_0}.
\end{align}
Here, $f'\equiv \partial_sf$ and $\dot{f}\equiv \partial_tf$ denote partial derivatives of functions on the worldsheet. In the second line, we introduced a cut-off at $s=s_0$ that we send to zero at the end. The cut-off is not strictly necessary, because eq.~\eqref{eq:longitudinal lagrangian near s=0} showed that the on-shell longitudinal action is finite near the boundary, but it is a convenient way to handle different terms in the action that may individually be singular.\footnote{For instance, introducing the cut-off at $s=s_0$ lets us drop the total $t$ derivatives when going from the first line to the second line in \eqref{eq:variation longitudinal action} without worrying about their behavior at $s=0$.}

To extremize the total action in \eqref{eq:string action conformal gauge 2} over the space of boundary reparametrization, we consider the variation of the action under $\alpha(t)\to \alpha(t)+\delta \alpha(t)$, which induces a variation in the on-shell transverse and longitudinal modes through the boundary conditions. In particular, given that $y(0,t)=\tilde{y}(\alpha(t))$, the variation induced in the transverse mode on the boundary is:
\begin{align}\label{eq:variation transverse modes}
    \delta y(0,t)&=\dot{\tilde{y}}(\alpha(t))\delta \alpha(t)=\dot{y}(0,t)\frac{\delta \alpha(t)}{\dot{\alpha}(t)}.
\end{align}
Furthermore, from \eqref{eq:x series expansion} and \eqref{eq:z series expansion}, the variation induced in the longitudinal modes near the boundary is
\begin{align}\label{eq:variation longitudinal modes}
    \delta x(s,t)&=\delta \alpha(t)+O(s^2), &\delta z(s,t)&=\delta \dot{\alpha}(t)s+O(s^3)
\end{align}

Now we can rewrite the variations of the transverse and longitudinal actions in terms of $\delta \alpha(t)$ and the stress tensors. First, given \eqref{eq:variation transverse modes}, we can rewrite the variation in the transverse action in \eqref{eq:variation transverse action} as:
\begin{align}\label{eq:Bo74FAqNmH}
    \delta S_T&=-T_s \int dt \dot{y}(0,t)y'(0,t) \frac{\delta \alpha(t)}{\dot{\alpha}(t)}=-T_s\int dt T_{ts}^{T}(0,t) \frac{\delta \alpha(t)}{\dot{\alpha}(t)}.
\end{align}
To get the second equality, we used the expression for the stress tensor in \eqref{eq:transverse stress tensor} evaluated at $s=0$. 

Second, using the leading behavior of the longitudinal modes near the boundary, given in \eqref{eq:x series expansion}-\eqref{eq:z series expansion}, and of their variation, given in \eqref{eq:variation longitudinal modes}, we can rewrite the variation in the longitudinal action in \eqref{eq:variation longitudinal action} as 
\begin{align}
    \delta S_L&=-T_s\lim_{s_0\to 0}\int dt \biggr[\frac{\left(-\ddot{\alpha}(t)s+g(t)s^2+O(s^3)\right)\left(\delta \alpha(t)+O(s^2)\right)}{s^2\dot{\alpha}(t)^2+O(s^4)}\nonumber\\&\hspace{5cm}+\frac{(\dot{\alpha}(t)+O(s^2))(s \delta \dot{\alpha}(t)+O(s^3))}{s^2\dot{\alpha}(t)^2+O(s^4)}\biggr]_{s=s_0}\nonumber\\&=-T_s\lim_{s_0\to 0}\int dt \left[\frac{1}{s_0}\left(-\frac{\ddot{\alpha}(t)}{\dot{\alpha}(t)^2}\delta \alpha(t)+\frac{1}{\dot{\alpha}(t)}\delta \dot{\alpha}(t)\right)+\frac{g(t)}{\dot{\alpha}(t)^2}\delta \alpha(t)\right].\label{eq:Nl16SfCFov}
\end{align}
The first term in the second line integrates to zero because it is a total derivative. Meanwhile, the remaining term can be written in terms of the longitudinal stress tensor. To see this, we substitute \eqref{eq:x series expansion} and \eqref{eq:z series expansion} into \eqref{eq:longitudinal stress tensor} and find that both components of the on-shell longitudinal stress tensor are finite at the boundary and are given by:
\begin{align}\label{eq:x,z stress tensor series expansion}
    T_{tt}^{L}(0,t)&=-\frac{h(t)}{\dot{\alpha}(t)},&T_{ts}^{L}(0,t)&=\frac{g(t)}{\dot{\alpha}(t)}.
\end{align}
Thus,
\begin{align}\label{eq:wQbrprkLwp}
    \delta S_L&=-T_s\int dt T_{ts}^L(0,t)\frac{\delta \alpha(t)}{\dot{\alpha}(t)}.
\end{align}

Combining \eqref{eq:Bo74FAqNmH} and \eqref{eq:wQbrprkLwp}, it follows that the change in the total action under a variation in the boundary reparametrization is:
\begin{align}
    \delta S = \delta S_T+\delta S_{L}=-T_s \int dt \left[T_{ts}^{T}(0,t)+T_{ts}^{L}(0,t)\right]\frac{\delta \alpha(t)}{\dot{\alpha}(t)}.
\end{align}
Therefore, extremizing over the reparametrizations fixes the $ts$ component of the total stress tensor to be zero on the boundary of the worldsheet:
\begin{align}\label{eq:Virasoro on the bdy}
    T_{ts}^{T}(0,t)+T_{ts}^{L}(0,t)=0.
\end{align}
This is actually equivalent to imposing the full Virasoro constraint on the entire worldsheet, because the on-shell longitudinal and transverse stress tensors are holomorphic (see \eqref{eq:holomorphicity stress tensor}). A result from complex analysis states that if either the real or imaginary components of two holomorphic functions on the upper half of the complex plane that vanish everywhere at infinity are equal, then the two functions are equal everywhere in the upper half plane. We can also make this concrete by writing the full stress tensor $T_{\alpha\beta}(s,t)$ on the worldsheet explicitly in terms of $T_{st}(0,t)$ on the boundary:
\begin{align}
    T_{st}(s,t)&=\frac{1}{\pi}\int dt' \frac{s}{s^2+(t-t')^2}T_{st}(0,t'),
    &T_{tt}(s,t)&=\frac{1}{\pi}\int dt' \frac{t-t'}{s^2+(t-t')^2}T_{st}(0,t').\label{eq:kramersKronig stress tensor 2}
\end{align}
Thus, $T_{st}(0,t)=0$ implies $T_{st}(s,t)=T_{tt}(s,t)=0$. 

\subsubsection{Longitudinal modes without transverse modes}

When the transverse mode is turned off, we can completely solve the longitudinal dynamics. This will be the starting point of the perturbative analysis that we turn to next. The transverse mode being turned off means the curve on the boundary is fixed at a point in $S^1$: e.g., $\tilde{y}(\alpha)=0$. The solution to the transverse equations of motion is then simply $y(s,t)=0$ and the transverse stress-tensor and action are zero: $T^T=S_T=0$. To study the longitudinal modes, it is convenient in this case to impose the Virasoro constraint before the equations of motion. It follows that $T^L=0$ and therefore, from \eqref{eq:stress tensor complex}, that $\partial X =0$ or $\bar{\partial}X=0$. In other words, $X=x+iz$ is either a holomorphic or an antiholomorphic function of $t+is$. It is clear from \eqref{eq:longitudinal and transverse eom complex} that in both cases the longitudinal equations of motion are automatically satisfied.

Because $z(s,t)\geq 0 $ with $z(0,t)=0$, it follows that $t+is\mapsto x+iz$ is a holomorphic map of the upper half of the complex plane into itself. Furthermore, we require that this map be a bijection, so that $s$ and $t$ are good coordinates on the string worldsheet. This allows us to invoke the result from complex analysis that the (anti)holomorphic bijections on the upper half of the complex plane are the $\rm{SL}(2,\mathbb{R})$ transformations. Thus, the general form of $x(s,t)$ and $z(s,t)$ consistent with the Virasoro constraint (and also automatically consistent with the equations of motion) is:
\begin{align}\label{eq:x+iz w/o transverse}
    x(s,t)+iz(s,t)&=\pm \frac{a(t\pm is )+b}{c(t\pm is)+d}
\end{align}
where $a,d,b,c\in \mathbb{R}$ and $ad-bc=1$. The choice of sign (with the $+$ sign for the holomorphic solutions and the $-$ sign for the antiholomorphic solutions) corresponds to a choice of orientation. Restricting \eqref{eq:x+iz w/o transverse} to the real axis determines the boundary reparametrization $\alpha(t)$:
\begin{align}\label{eq:x(t) w/o transverse}
    \alpha(t)=\pm \frac{at+b}{ct+d}.
\end{align}
We will focus on the holomorphic solution, in which case $\dot{\alpha}>0$.

The longitudinal action in \eqref{eq:longitudinal action} evaluated on the solution in \eqref{eq:x+iz w/o transverse} is $S_L=0$, and the total action in \eqref{eq:string action conformal gauge 2} reduces to the (regularized) area of the AdS$_2$ worldsheet. This simple case also concretely illustrates how imposing the Virasoro constraint picks out an $SL(2,\mathbb{R})$ subset of reparametrizations from the space of all possible reparametrizations. This is the smallest subset consistent with the gauge $SL(2,\mathbb{R})$ symmetry of the string. We might summarize this effect by saying that the imposing the Virasoro constraint in the conformal gauge breaks the reparametrization symmetry of the string from $\text{Diff}(S^1)$ to $\rm{SL}(2,\mathbb{R})$. The reparametrization mode should behave analogously when the transverse modes are turned on.\footnote{The boundary reparametrizations of the AdS$_2$ string for the case without transverse modes was studied recently in \cite{Gutiez:2022uof}. That analysis finds more non-trivial behavior for the reparametrizations than what is given in \eqref{eq:x(t) w/o transverse}. We believe the different conclusion in \cite{Gutiez:2022uof} is a result of allowing $t+is\mapsto x+iz$ to be any holomorphic function on the upper half plane. We believe one should restrict to biholomorphic bijections. }
 
\subsubsection{Perturbative analysis of the longitudinal modes}\label{sec:longitudinal modes perturbative}

Finally, we turn to the classical AdS string in the conformal gauge in the case where the transverse modes are turned on. We will work perturbatively, treating the transverse fluctuations as being small. This is the regime where we can most easily make progress and that is relevant for the computation of the boundary correlators.

When the transverse modes are turned off, we saw that the boundary mode, $\alpha(t)$, takes the form in \eqref{eq:x(t) w/o transverse} and the longitudinal modes, $x(s,t)$ and $z(s,t)$, take the form in \eqref{eq:x+iz w/o transverse}. We will expand perturbatively around the ``simplest'' solution, $\alpha(t)=t$, $x(s,t)=t$ and $z(s,t)=s$, which is a convenient choice of $SL(2,\mathbb{R})$ gauge. We write
\begin{align}\label{eq:perturbative expansion longitudinal modes}
    \alpha(t)&=t+\epsilon(t),&x(s,t)&=t+\xi(s,t),&z(s,t)&=s+\zeta(s,t),
\end{align}
and treat $\epsilon$, $\xi$ and $\zeta$ as small perturbations. The boundary conditions for $\xi$ and $\zeta$ are:
\begin{align}
    \xi(0,t)&=\epsilon(t),&\zeta(0,t)&=0.
\end{align}
We will assume that the perturbation of the boundary curve is localized: i.e., $\epsilon(t)\to 0$ as $t\to \pm \infty$, which implies that $\xi$ and $\zeta$ are localized on the worldsheet: $\xi(s,t),\zeta(s,t)\to 0$ as $t\to \pm \infty$ or $s\to \infty$.

Recall that to compute the boundary four-point function in \eqref{eq: y 4-pt function} in the classical regime, it is sufficient to compute the classical action to quartic order in $\tilde{y}$. It is clear from \eqref{eq:y solns eom} and \eqref{eq:transverse stress tensor} that $T^T$ is quadratic in $\tilde{y}$. Meanwhile, if we substitute \eqref{eq:perturbative expansion longitudinal modes} into \eqref{eq:longitudinal stress tensor} and expand in $\xi$ and $\zeta$, we see that $T^L$ is linear in $\xi$ and $\zeta$. The Virasoro constraint therefore implies that $\xi$ and $\zeta$ are quadratic in $\tilde{y}$. This means we only need to study the longitudinal action to quadratic order in $\xi$ and $\zeta$, which is a convenient simplification.

We can use the linear order equations of motion for the longitudinal modes to write the quadratic order longitudinal action as a boundary term. First, substituting \eqref{eq:perturbative expansion longitudinal modes} into \eqref{eq:longitudinal eom} and expanding, we find that the linear order equations of motion for $\xi$ and $\zeta$ are:
\begin{align}\label{eq:longitudinal linear eom}
    0&=s(\ddot{\xi}+\xi'')-2(\xi'+\dot{\zeta}),&0&=s(\ddot{\zeta}+\zeta'')+2(\dot{\xi}-\zeta').
\end{align}
Second, substituting \eqref{eq:perturbative expansion longitudinal modes} into \eqref{eq:longitudinal action} and expanding, we find that the longitudinal action to quadratic order in $\xi$ and $\zeta$ is:
\begin{align}\label{eq:expanding longitudinal action}
    S_{L}&=T_s\int d^2 \sigma\bigg[\frac{\dot{\xi}}{s^2}+\frac{\zeta'}{s^2}-\frac{2\zeta}{s^3}-\frac{2\zeta\zeta'}{s^3}+\frac{3\zeta^2}{s^4}-\frac{2\zeta\dot{\xi}}{s^3}+\frac{\partial^\alpha \xi\partial_\alpha \xi +\partial^\alpha \zeta \partial_\alpha \zeta}{2s^2}+\ldots\bigg]\\&=T_s\int d^2 \sigma \bigg[\partial_t\left(\frac{\xi}{s^2}\right)+\partial_s\left(\frac{\zeta}{s^2}\right)-\partial_s\left(\frac{\zeta^2}{s^3}\right)-\partial_t\left(\frac{\xi\zeta}{s^3}\right)+\partial_\alpha\left(\frac{\xi\partial^\alpha \xi+ \zeta\partial^\alpha \zeta}{2s^2}\right)+\ldots\bigg].\nonumber
\end{align}
We used \eqref{eq:longitudinal linear eom} to get to the second line.

All of the terms in the second line in \eqref{eq:expanding longitudinal action} are total derivatives. Now, as in \eqref{eq:variation longitudinal action}, it is convenient to introduce a cut-off at $s=s_0$, write the action as an integral over the boundary at $s=s_0$, and take $s_0\to 0$ at the end. If we apply the small $s$ expansion of $\xi$ and $\zeta$, which are essentially given by \eqref{eq:x series expansion}-\eqref{eq:z series expansion}, we see that all the terms that are singular at $s=0$ cancel, and the finite contribution to the action is:
\begin{align}\label{eq:longitudinal action bdy term}
    S_L&=-\frac{T_s}{2}\int dt \xi(0,t)g(t) =-\frac{T_s}{4}\int dt \xi(0,t)\xi'''(0,t).
\end{align}
In the second equality we used $\xi'''(0,t)=2g(t)$, which follows from \eqref{eq:x series expansion}. This action has corrections that are of third order in $\xi$ and $\zeta$.

If we can solve the equations of motion in \eqref{eq:longitudinal linear eom} for $\xi$ and $\zeta$ in terms of $\epsilon$, then \eqref{eq:longitudinal action bdy term} gives us the on-shell longitudinal action to quadratic order in $\epsilon$. Because the equations of motion are linear, the solutions can be written in terms of boundary-to-bulk integrals:
\begin{align}
    \xi(s,t)&=\int dt' K_x(s,t,t')\epsilon(t'),&\zeta(s,t)&=\int dt' K_z(s,t,t')\epsilon(t').\label{eq:longitudinal bdy-to-blk integrals}
\end{align}
It turns out that the two boundary-to-bulk propagators $K_x$ and $K_z$ are given explicitly by
\begin{align}
    K_x(s,t,t')&=\frac{4}{\pi}\frac{s^3(s^2-(t-t')^2)}{(s^2+(t-t')^2)^3},&
    K_z(s,t,t')&=-\frac{8}{\pi}\frac{s^4 (t-t')}{(s^2+(t-t')^2)^3}.\label{eq:longitudinal bdy-to-blk propagators}
\end{align}
It is easy to check that $K_x(s,t,t')$ and $K_z(s,t,t')$ solve \eqref{eq:longitudinal linear eom}, become sharply peaked at $t=t'$ as $s\to 0$ and satisfy $\int dt' K_x(s,t,t')=1$ and $\int dt' K_z(s,t,t')=0$ for any $s$. These properties together with \eqref{eq:longitudinal bdy-to-blk integrals} mean that $\xi(s,t)\to \epsilon(t)$ and  $\zeta(s,t)\to 0$ as $s\to 0$, as desired. 

Combining \eqref{eq:longitudinal action bdy term} with \eqref{eq:longitudinal bdy-to-blk integrals}-\eqref{eq:longitudinal bdy-to-blk propagators} yields the following expression for the on-shell longitudinal action to quadratic order in $\epsilon$:
\begin{align}\label{eq:DpvUsKXIcZ}
    S_{L}[t+\epsilon(t)]&=-\frac{T_s}{4}\lim_{s\to 0}\int dt  dt'  \partial_s^3 K_x(s,t,t')\epsilon(t) \epsilon(t').
\end{align}
If we substitute the explicit expression for $\partial_s^3 K_x$ and treat $s$ as a short distance regulator, the resulting quadratic action appears similar to the one in eq. (15) in \cite{Ambjorn:2011wz}. We prefer to put the action in a manifestly finite form that does not involve $s$. If we naively take the limit inside the integral in \eqref{eq:DpvUsKXIcZ}, the resulting integral is divergent. Instead, we follow a sequence of steps analogous to the ones we took to get from \eqref{eq:tyUdUUufGI} to \eqref{eq:transverse action on-shell explicit} when studying the transverse action. We start by introducing the function
\begin{align}
    J_x(s,t,t')=\frac{4}{\pi}\left[\frac{3(t-t')^2}{(s^2+(t-t')^2)^2}-\frac{14(t-t')^4}{(s^2+(t-t')^2)^3}+\frac{12 (t-t')^6}{(s^2+(t-t')^2)^4}\right],
\end{align}
which has three useful properties: (i) $\partial_s^3 K_x(s,t,t')=\partial_t \partial_{t'} J_x(s,t,t')$, (ii) $\int dt' J_x(s,t,t')=0$, and (iii) $\lim_{s\to 0}\int dt dt' J_x(s,t,t') (f(t)-f(t'))^2$ is finite as $t'\to t$ and evaluates to $\frac{4}{\pi} \int dt dt' \frac{(f(t)-f(t'))^2}{(t-t')^2}$ for any function $f(t)$ that is smooth and decays sufficiently quickly at $t\to \pm \infty$. Thus, writing $\partial_s^3 K_x$ as $\partial_t \partial_{t'}J_x$ in \eqref{eq:DpvUsKXIcZ}, we can integrate by parts to transfer the derivatives to $\epsilon(t)$ and $\epsilon(t')$, then replace $\dot{\epsilon}(t)\dot{\epsilon}(t')\to -\frac{1}{2}(\dot{\epsilon}(t)-\dot{\epsilon}(t'))^2$ without changing the value of the integral, and finally safely take the $s\to 0$ limit. The final expression for the on-shell longitudinal action to quadratic order in $\epsilon$ is:
\begin{align}\label{eq:longitudinal action quadratic line}
    S_{L}[t+\epsilon(t)]&=\frac{T_s}{2\pi}\int dt dt' \frac{(\dot{\epsilon}(t)-\dot{\epsilon}(t'))^2}{(t-t')^2}.
\end{align}
This, together with \eqref{eq:longitudinal action quadratic circle body}, is the main result of this section.

We can also get from \eqref{eq:DpvUsKXIcZ} to  \eqref{eq:longitudinal action quadratic line} more formally by noting that $\partial_s^3 K_x(0,t,t')=-\frac{24}{\pi}\frac{1}{(t-t')^4}$ and taking the limit in \eqref{eq:DpvUsKXIcZ} inside the integral to get the expression 
\begin{align}
    S_L[t+\epsilon(t)]&= \frac{6T_s}{\pi}\int dt dt' \frac{\epsilon(t)\epsilon(t')}{(t-t')^4}.
\end{align}
This yields a second representation of the quadratic action that is sometimes be useful but also only formal because it is divergent. We can think of it as being defined by letting $(t-t')^4\to (t-t')^{4\eta}$ in the denominator and then analytically continuing from $\eta<1/4$ to $\eta\to 1$. Then, writing $\frac{1}{(t-t')^4}=\partial_t \partial_{t'} \frac{1}{6}\frac{1}{(t-t')^2}$, transferring the derivatives to $\epsilon(t)$ and $\epsilon(t')$ and invoking the identity $\int dt \frac{1}{(t-t')^2}=0$ in analytic regularization to replace $\dot{\epsilon}(t)\dot{\epsilon}(t')\to -\frac{1}{2}(\dot{\epsilon}(t)-\dot{\epsilon}(t'))^2$, we again arrive at \eqref{eq:longitudinal action quadratic line}.

We can also express the quadratic longitudinal action in Fourier space. Writing $\epsilon(t)$ as its Fourier integral, $\epsilon(t)=\int\frac{d\omega}{2\pi}e^{-i \omega t}\epsilon(\omega)$ with $\epsilon(\omega)^*=\epsilon(-\omega)$, we find that the longitudinal action is\footnote{It is convenient when computing the Fourier representation of the action to replace $(t-t')^2\to (t-t')^2+s^2$ in \eqref{eq:longitudinal action quadratic line} and take $s\to 0$ at the end.}
\begin{align}\label{eq:quadratic longitudinal action fourier}
    S_L[t+\epsilon(t)]=\frac{T_s}{2\pi}\int d\omega \epsilon(\omega)\epsilon(-\omega) |\omega|^3.
\end{align}
This form of the longitudinal action appeared in \cite{Rychkov:2002ni}.

\subsubsection{Hyperbolic disk coordinates}
We can also perform the conformal gauge analysis of the string in AdS$_2\times S^1$ in hyperbolic disk coordinates instead of hyperbolic half-plane coordinates. One reason to do so is that many questions are better posed on the hyperbolic disk because the boundary coordinate is compact. We summarize the set-up and the results needed for section~\ref{sec:OTOC from the string reparametrization mode} here, and supply the details in appendix~\ref{app:conformal gauge hyperbolic disk coordinates}. 

We start with the following metric on AdS$_2\times S^1$:
\begin{align}\label{eq:AdS2 hyperbolic disk coordinates}
    ds^2=\frac{d\theta^2+dr^2}{\sinh^2{r}}+dy^2,
\end{align}
where $\theta\in [0,2\pi]$ is the boundary angular coordinate on AdS$_2$, $r\in [0,\infty)$ is the bulk coordinate (with $r=0$ labelling the boundary of AdS$_2$), and $y$ is an angular coordinate on $S^1$. Now the boundary is $S^1\times S^1$ and the curve that the string is incident on can be represented as $\gamma:\alpha\mapsto (\alpha,\tilde{y}(\alpha))$. Furthermore, the string can be represented as $\Sigma:(\sigma,\tau)\mapsto (r(\sigma,\tau),\theta(\sigma,\tau),y(\sigma,\tau))$ where $\sigma\in[0,\infty)$ and $\tau\in [0,2\pi]$ are worldsheet coordinates (with $\sigma=0$ labelling the boundary of the worldsheet).

In analogy with \eqref{eq:string action conformal gauge 2}-\eqref{eq:A_plane=0}, the string action in conformal gauge can be split into three terms,
\begin{align}\label{eq:conformal gauge action disk}
    S[\theta,r,y]&=S_{L}[\theta,r]+S_{T}[y]+T_sA_{\rm ws},
\end{align}
where the first term is the longitudinal action,
\begin{align}\label{eq:longitudinal action disk}
    S_{L}[\theta,r]&=\frac{T_s}{2}\int d^2\sigma \left[\frac{\partial^\alpha \theta \partial_\alpha \theta+\partial^\alpha r \partial_\alpha r}{\sinh^2{r}}-\frac{2}{\sinh^2{\sigma}}\right],
\end{align}
the second term is the transverse action again given by \eqref{eq:transverse action}, and the third term is the regularized area of the worldsheet. This is $A_{\rm ws}=\int d^2\sigma \sqrt{h}=-2\pi$, as is reviewed in appendix~\ref{app: AdS2 area with cutoff}.

The boundary condition of the string in the disk coordinates is:
\begin{align}\label{eq:string bc disk coordinates}
    r(0,\tau)&=0, &\theta(0,\tau)&=\alpha(\tau),& y(0,\tau)&=\tilde{y}(\alpha(\tau)),
\end{align}
where $\alpha(\tau)$ is a reparametrization of the boundary. Since $\tau$ is an angular coordinate, $r$ and $y$ are periodic in $\tau$ (i.e., $r(\tau+2\pi)=r(\tau)$, etc.) and $\theta$ is periodic in $\tau$ up to a shift by $2\pi$ (i.e., $\theta(\sigma,\tau+2\pi)=2\pi+\theta(\sigma,\tau)$ and $\alpha(\tau+2\pi)=2\pi+\alpha(\tau)$).  

As in the analysis in the hyperbolic plane coordinates, the classical string action as a function of $\tilde{y}$ can be expressed as the sum of the on-shell longitudinal and transverse actions, $S_L[\alpha]$ and $S_T[\tilde{y}\circ \alpha]$, subject to the Virasoro constraint or extremization over the boundary reparametrization:
\begin{align}
    S_{\rm cl}[\tilde{y}]&=-2\pi T_s+S_L[\alpha]+S_T[\tilde{y}\circ \alpha]\bigg\rvert_{\rm Virasoro}\label{eq:classical action with Virasoro constraint body}\\&=-2\pi T_s+\underset{\alpha}{\rm extremize}\big\{S_L[\alpha]+S_T[\tilde{y}\circ \alpha]\big\}.\label{eq:classical action with extremization over reparametrizations body}
\end{align}

The on-shell transverse action as a functional of $\tilde{y}(\alpha(\tau))$ is:
\begin{align}\label{eq:transverse action circle body}
    S_T[\tilde{y}\circ \alpha]=\frac{T_s}{4\pi}\int d\tau d\tau' \bigg[\frac{\big[\tilde{y}(\alpha(\tau))-\tilde{y}(\alpha(\tau'))\big]^2}{\bigr[2\sin\big(\frac{\tau-\tau'}{2}\big)\bigr]^2}\bigg].
\end{align}
This follows from \eqref{eq:transverse action on-shell explicit} if we change the integration variable over the boundary using $t=\tan{\frac{\tau}{2}}$, which replaces the euclidean distance $t-t'$ by the chordal distance $2\sin(\frac{\tau-\tau'}{2})$.

Meanwhile, the general solution of the on-shell longitudinal action as a functional of $\alpha(\tau)$ is not known. But if $\alpha(\tau)=\tau+\epsilon(\tau)$ where $\epsilon$ is small, then the on-shell longitudinal action to quadratic order is:
\begin{align}\label{eq:longitudinal action quadratic circle body}
    S_L[\tau+\epsilon(\tau)]&=\frac{T_s}{2\pi}\int d\tau d\tau' \frac{(\dot{\epsilon}(\tau)-\dot{\epsilon}(\tau'))^2-(\epsilon(\tau)-\epsilon(\tau'))^2}{[2\sin\left(\frac{\tau-\tau'}{2}\right)]^2}.
\end{align}
The steps needed to derive \eqref{eq:classical action with Virasoro constraint body}-\eqref{eq:longitudinal action quadratic circle body} are essentially the same as in the analysis in hyperbolic plane coordinates; see appendix~\ref{app:conformal gauge hyperbolic disk coordinates}.

The longitudinal action on the circle given in \eqref{eq:longitudinal action quadratic circle body} is sometimes nicer to work with than the longitudinal action on the line given in \eqref{eq:longitudinal action quadratic line} because the periodicity of the boundary allows us to write the bounday reparametrization as a Fourier series instead of a Fourier integral. In particular, the perturbation $\epsilon(\tau)$ about the saddle point $\alpha(\tau)=\tau$ is periodic in $\tau$ and can be expressed as:
\begin{align}\label{eq:epsilon Fourier series}
    \epsilon(\tau)=\sum_{n\in \mathbb{Z}}\epsilon_n e^{in\tau},
\end{align}
where $\epsilon_n^*=\epsilon_{-n}$. We can then evaluate the action in terms of the modes Fourier $\epsilon_n$, using the following orthogonality relation:
\begin{align}\label{eq:orthogonality relation}
    \int_0^{2\pi} \frac{d\tau}{2\pi} \frac{d\tau'}{2\pi} \frac{(e^{in \tau}-e^{in\tau'})(e^{im \tau}-e^{im\tau'})}{\big[2\sin\left(\frac{\tau-\tau'}{2}\right)\big]^2}= |n|\delta_{n,-m}.
\end{align}
The different modes decouple when we substitute \eqref{eq:epsilon Fourier series} into \eqref{eq:longitudinal action quadratic circle body}, and the action becomes:
\begin{align}\label{eq:Ov0gv3LFHW}
    S_L[\tau+\epsilon(\tau)]&=4\pi T_s \sum_{n=2}^\infty |n|
    (n^2-1)\epsilon_n\epsilon_{-n}.
\end{align}
The $|n|(n^2-1)$ dispersion relation on the circle is the discrete version of the $|\omega|^3$ dispersion relation in \eqref{eq:quadratic longitudinal action fourier} on the line. Note that $\epsilon_0$, $\epsilon_1$ and $\epsilon_{-1}$ are zero modes of the action; they correspond to the three generators of the $SL(2,\mathbb{R})$ gauge symmetry of the string.

Finally, to close this section, we return to the introductory remarks we made at its beginning. Adopting the language of \cite{Ambjorn:2011wz}, one can view \eqref{eq:longitudinal action quadratic circle body} together with \eqref{eq:transverse action circle body} (or, equivalently, \eqref{eq:longitudinal action quadratic line} together with \eqref{eq:transverse action on-shell explicit}) as defining a perturbative ``Douglas integral'' for the classical string in AdS$_2\times S^1$. Namely, the analysis in this section has shown that the regularized area of a minimal surface in AdS$_2\times S^1$ incident on the boundary curve $\gamma:\alpha\to (\alpha,\tilde{y}(\alpha))$, to fourth order in $\tilde{y}$, is given by:
\begin{align}\label{eq:AdS douglas integral}
    A&=-2\pi+\underset{\epsilon(\tau)}{\text{extremize}}\biggr[\frac{1}{4\pi}\int_{0}^{2\pi}d\tau \int_0^{2\pi} d\tau'\biggr( \frac{2(\dot{\epsilon}(\tau)-\dot{\epsilon}(\tau'))^2-2(\epsilon(\tau)-\epsilon(\tau'))^2}{[2\sin\left(\frac{\tau-\tau'}{2}\right)]^2}\\&\hspace{9cm}+\frac{[\tilde{y}(\tau+\epsilon(\tau))-\tilde{y}(\tau'+\epsilon(\tau'))]^2}{[2\sin\left(\frac{\tau-\tau'}{2}\right)]^2}\biggr)\biggr].\nonumber
\end{align}
This is less elegant than the Douglas integral for the minimal surface in $\mathbb{R}^3$ because it is only perturbative. For a more direct comparison, we should choose the boundary curve in $\mathbb{R}^3$ to be a nearly circular curve (e.g., in \eqref{eq:douglas integral} let $\vec{x}(\alpha)=(\cos{\alpha},\sin{\alpha},\tilde{y}(\alpha))$, set $\alpha(\tau)=\tau+\epsilon(\tau)$ and expand in $\epsilon$), which is the flat space analog of a nearly circular curve on the boundary of AdS$_2\times S^1$.

The equivalence the classical string action in conformal gauge given in \eqref{eq:AdS douglas integral} and in static gauge given in \eqref{eq:fourth order classical action} is demonstrated explicitly (to fourth order in $\tilde{y}$) in appendix~\ref{sec:static = conformal gauge correlators}. 

\section{The string reparametrization integral: correlators and OTOC}\label{sec:OTOC from the string reparametrization mode}

The upshot of the previous section is that the partition function of the open string in AdS$_2\times S^1$ in the classical approximation can be written as:
\begin{align}\label{eq:bVAXENVfDy}
    Z[\tilde{y}]\approx e^{-S_{\rm cl}[\tilde{y}]}=\underset{\alpha}{\text{extremize}}\bigg\{e^{-T_sA_{\rm ws}-S_L[\alpha]-S_T[\tilde{y}\circ \alpha]}\bigg\}.
\end{align}
Going forward, we drop the constant $T_sA_{\rm ws}$ term. We now explore some of the implications of promoting the extremization in \eqref{eq:bVAXENVfDy} to a path integral over the boundary reparametrizations:
\begin{align}\label{eq:string partition function reparametrization integral}
    Z[\tilde{y}]=\underset{\mathcal{M}_R}{\int} \mathcal{D}\alpha\text{ }e^{-S_L[\alpha]-S_T[\tilde{y}\circ \alpha]}.
\end{align}
Such a reparametrization path integral appeared in \cite{Rychkov:2002ni,Ambjorn:2011wz} in the context of the string in AdS and in, e.g., \cite{Polyakov:1981rd, Alvarez:1982zi,Fradkin:1982ge, Cohen:1985sm,polyakov1987gauge} in the context of the string in flat space. As discussed in those references, the integral over reparametrizations is what remains of the path integral over the worldsheet metric $h_{\alpha\beta}$ in eq.~\eqref{eq:sigma model path integral} in the string sigma model after the conformal gauge is fixed.

In the limit of large string tension, eq.~\eqref{eq:string partition function reparametrization integral} is dominated by its saddle point and reduces to \eqref{eq:bVAXENVfDy}. When the string tension is finite, \eqref{eq:string partition function reparametrization integral} includes some but not all of the quantum corrections to the classical result in \eqref{eq:bVAXENVfDy}. This is because it does not include the path integral over the matter fields (which include the longitudinal modes $x$ and $z$, the transverse mode $y$, as well as the other bosonic and fermionic fields that we have suppressed in \eqref{eq:sigma model path integral} by focusing on the motion of the string in an AdS$_2\times S^1$ subspace of AdS$_5\times S^5$) or the path integral over the $bc$ ghosts that arise when fixing the conformal gauge. Nonetheless, one might hope that the reparametrization path integral in eq.~\eqref{eq:string partition function reparametrization integral} captures at least an interesting subset of the quantum corrections to at least some interesting observables on the string. This is the attitude that we take in this section.

Furthermore, we will be somewhat heuristic about the meaning of the reparametrization path integral in \eqref{eq:string partition function reparametrization integral}. In particular, to compute the four-point function to leading order in $1/T_s$ and the OTOC in the double scaling limit, it will be sufficient to expand around the saddle point and approximate the path integral as a Gaussian whose action is the longitudinal action at quadratic order in the perturbation $\epsilon$ that is given in \eqref{eq:longitudinal action quadratic circle body}. The only ingredient needed for these computations is the propagator for $\epsilon$, which can be determined using only some general properties of the integral in \eqref{eq:string partition function reparametrization integral}. The following analysis will therefore involve some educated guesswork that is guided by the properties of the string action in conformal gauge that were discussed in section~\ref{sec:reparametrization mode for AdS string}, as well as by the example of the Schwarzian path integral in, e.g.,\cite{Maldacena:2016upp,Stanford:2017thb,Mertens:2017mtv}. 

In section \ref{sec:reparametrization mode for AdS string}, we found explicit expressions for the transverse action $S_T[\tilde{y}\circ \alpha]$ in \eqref{eq:transverse action circle body} and the longitudinal action $S_L[\alpha]$ (perturbatively) in \eqref{eq:longitudinal action quadratic circle body}. The two components of \eqref{eq:string partition function reparametrization integral} that we have not made precise--- and will not need to make too precise--- are the domain of integration $\mathcal{M}_R$ and the measure $\mathcal{D}\alpha$. Let's first consider the domain of integration. We will work with the string in hyperbolic disk coordinates, so that $\alpha(\tau)$ in the string boundary condition in \eqref{eq:string bc disk coordinates} is a reparametrization of a circle. A first guess, therefore, is that the domain of integration in \eqref{eq:string partition function reparametrization integral} should include all reparametrizations of the circle, $\text{Diff}(S^1)$. However, we recall from section~\ref{sec:SL(2,R) symmetries} that the string in AdS$_2\times S^1$ has two $SL(2,\mathbb{R})$ symmetries, one physical and one gauge. The action of these symmetries on the boundary reparametrization $\alpha(\tau)$ are given in \eqref{eq:jR9ruCAwrT} and \eqref{eq:s10Y7PjWQZ} for the case of the circle (and in \eqref{eq:physical SL(2,R) bdy} and \eqref{eq:gauge SL(2,R) bdy} for the case of the line). Another common way to represent an $SL(2,\mathbb{R})$ transformation on a circle with angular coordinate $\phi$ is
\begin{align}\label{eq:SL(2,R) on circle}
    f:\phi\mapsto f(\phi), &&\tan{\frac{f(\phi)}{2}}=\frac{a\tan{\frac{\phi}{2}}+b}{c\tan{\frac{\phi}{2}}+d},
\end{align}
where $a,b,c,d\in \mathbb{R}$ and $ad-bc=1$. Then, the physical transformation $SL(2,\mathbb{R})$ transformation sends $\alpha(\tau)\to f(\alpha(\tau))$ (i.e., it acts on the ``left'' or, equivalently, on the target space boundary coordinate) while the gauge transformation sends $\alpha(\tau)\to \alpha(f(\tau))$ (i.e., it acts on the ``right'' or, equivalently, on the worldsheet boundary coordinate). Therefore, an updated guess is that the domain of integration in \eqref{eq:string partition function reparametrization integral} should include all diffeomorphisms of the circle modulo the gauge $SL(2,\mathbb{R})$ transformations, which we can write as:
\begin{align}\label{eq:M_R}
    \mathcal{M}_R=\text{Diff}(S^1)/SL(2,\mathbb{R})_R.
\end{align}
The subscript $R$ indicates that we identify two reparametrizations if they are related by an $SL(2,\mathbb{R})$ transformation acting on the right: i.e., $\alpha\sim \alpha\circ f$.

Eq.~\eqref{eq:M_R} is also the domain of integration that appears in the Schwarzian theory \cite{Maldacena:2016upp,Stanford:2017thb,Mertens:2017mtv}, which also has a gauge $SL(2,\mathbb{R})$ symmetry. Consequently, we will have to gauge fix the $SL(2,\mathbb{R})$ modes in the same way when computing correlators. Furthermore, as discussed in \cite{Stanford:2017thb}, Diff($S^1$)/$SL(2,\mathbb{R})$ is a symplectic manifold with a natural measure, which therefore provides a precise definition of $\mathcal{D}\alpha$ in \eqref{eq:string partition function reparametrization integral}. However, for the purpose of computing correlators perturbatively, it will be sufficient --- and equivalent to working with the Schwarzian measure--- to use a naive form for the measure.

\subsection{Correlators in the reparametrization path integral}
From \eqref{eq:string partition function reparametrization integral}, we can derive the representation of the string boundary correlators in the reparametrization path integral. Recall from section~\ref{sec:preliminaries} that the correlators on the string are defined by taking orthogonal variational derivatives of the string partition function with respect to the boundary curve. In particular, we need to take derivatives of \eqref{eq:string partition function reparametrization integral} with respect to $\tilde{y}$, and label the positions of the operators by their angular coordinates, $\theta$, on the AdS$_2$ boundary. For instance, the two and four-point functions are:
\begin{align}
    \braket{y_1y_2}_{\text{AdS}_2}&=\frac{1}{Z}\frac{\delta^2 Z[\tilde{y}]}{\delta \tilde{y}(\theta_1)\delta \tilde{y}(\theta_2)},&\braket{y_1y_2y_3y_4}_{\text{AdS}_2}&=\frac{1}{Z}\frac{\delta^4 Z[\tilde{y}]}{\delta \tilde{y}(\theta_1)\delta \tilde{y}(\theta_2)\delta \tilde{y}(\theta_3)\delta \tilde{y}(\theta_4)}.
\end{align}
In \eqref{eq:string partition function reparametrization integral}, $\tilde{y}$ appears only in the transverse action. To take variational derivatives, it is useful to rewrite the transverse action in \eqref{eq:transverse action circle body} by changing the integration variable from $\tau$ to $\theta=\alpha(\tau)$. This gives:
\begin{align}\label{eq:transverse action c.o.var}
    S_T[\tilde{y}\circ \alpha]=\frac{T_s}{4\pi}\int d\theta d\theta' \frac{\dot{\tau}(\theta)\dot{\tau}(\theta')}{\big[2\sin(\frac{1}{2}[\tau(\theta)-\tau(\theta')])\big]^2}(\tilde{y}(\theta)-\tilde{y}(\theta'))^2.
\end{align}
In the above expression, $\tau(\cdot)$ denotes the inverse of $\alpha(\cdot)$. Taking two variational derivatives of \eqref{eq:transverse action c.o.var} yields
\begin{align}\label{eq:CSl6ubNNjr}
    B_\tau (\theta_1,\theta_2)\equiv -\frac{\pi}{T_s}\frac{\delta^2S_T[\tilde{y}\circ \alpha]}{\delta \tilde{y}(\theta_1)\delta\tilde{y}(\theta_2)}=\frac{\dot{\tau}(\theta_1)\dot{\tau}(\theta_2)}{\big[2\sin(\frac{1}{2}[\tau(\theta_1)-\tau(\theta_2)])\big]^2}.
\end{align}
We have introduced the notation $B_\tau(\theta_1,\theta_2)$ for this bilocal object, which looks like a conformal two-point function of unit scaling dimension ``dressed'' by the reparametrization $\tau$. This object is familiar from the study of correlators in the Schwarzian theory in JT gravity. We will typically drop the explicit dependence of $B_\tau$ on $\tau$.

The fact that the inverse $\tau(\cdot)$ appears in \eqref{eq:CSl6ubNNjr} instead of $\alpha(\cdot)$ suggests that it is the more natural way of representing the reparametrization in the path integral. This is related to the fact that $\alpha(\cdot)$ maps the worldsheet boundary coordinate $\tau$ to the AdS$_2$ boundary coordinate $\theta$ while $\tau(\cdot)$ maps $\theta$ to $\tau$, and the locations of the insertions in the boundary correlators should be labelled by $\theta$ instead of $\tau$.\footnote{Similar comments were made in the analysis of the scattering matrix on the long bosonic string in conformal gauge in section 6 of \cite{Dubovsky:2016cog}.} We therefore rewrite \eqref{eq:string partition function reparametrization integral} as
\begin{align}\label{eq:EM43CBaVED}
    Z[\tilde{y}]=\underset{\mathcal{M}_L}{\int}\mathcal{D}\tau \text{ }e^{-S_R[\tau]-S_T[\tilde{y}\circ \tau^{-1}]}.
\end{align}
Here, we have introduced the effective action for the reparametrizations,
\begin{align}\label{eq:SR = SL}
    S_R[\tau]=S_L[\alpha],
\end{align}
where $\alpha(\cdot)$ is the inverse of $\tau(\cdot)$. It is a happy accident that the subcripts $R$ and $L$ in \eqref{eq:SR = SL} stand equally well for ``reparametrization'' and ``longitudinal'' or ``right'' and ``left.'' This encapsulates the two observations that the effective action governing the reparametrization path integral in \eqref{eq:EM43CBaVED} ultimately derives from the action for the longitudinal modes in \eqref{eq:longitudinal action disk} in the conformal gauge analysis, and that $S_R$ and $S_L$ in \eqref{eq:SR = SL} are mirror images of the same object. Indeed, because $\tau(\cdot)$ and $\alpha(\cdot)$ are inverses and because the physical and gauge $SL(2,\mathbb{R})$ transformations act on the left and right, respectively, of $\alpha$, it follows that the physical and gauge transformations act on the right and left, respectively, of $\tau$. Namely, the physical transformation sends $\tau(\theta)\to \tau(f(\theta))$) and the gauge transformations sends $\tau(\theta)\to f(\tau(\theta))$. Thus, instead of \eqref{eq:M_R}, the domain of integration in \eqref{eq:EM43CBaVED} should be
\begin{align}\label{eq:VBoPjO1JbC}
    \mathcal{M}_L=\text{Diff}(S^1)/SL(2,\mathbb{R})_L,
\end{align}
which is the space of all diffeomorphisms of the circle modulo gauge $SL(2,\mathbb{R})$ transformations acting on the left (i.e., we identify $\tau\sim f\circ \tau$ if $f$ is an $SL(2,\mathbb{R})$ transformation on the circle).

Finally, when we take variational derivatives of \eqref{eq:EM43CBaVED} with respect to $\tilde{y}$, we again simply pull down insertions of the bilocal object defined in \eqref{eq:CSl6ubNNjr}. Thus, in our final formulation of correlators in the reparametrization path integral, the two-point and four-point string correlators are
\begin{align}
    \braket{y(\theta_1)y(\theta_2)}_{\text{AdS}_2}&=\frac{T_s}{\pi}\braket{B_\tau(\theta_1,\theta_2)}\label{eq:yy 2-pt function}\\
    \braket{y(\theta_1)y(\theta_2)y(\theta_3)y(\theta_4)}_{\text{AdS}_2}&=\frac{T_s^2}{\pi^2}\big[\braket{ B_\tau(\theta_1,\theta_2)B_\tau(\theta_3,\theta_4)}+\braket{B_\tau(\theta_1,\theta_3)B_\tau(\theta_2,\theta_4)}\label{eq:yyyy 2-pt function}\\&\hspace{4.3cm}+\braket{B_\tau(\theta_1,\theta_4)B_\tau(\theta_2,\theta_3)}\big].\nonumber
\end{align}
On the right hand side, the angle brackets indicate expectation values in the reparametrization path integral:
\begin{align}\label{eq:reparametrization expectation value}
    \braket{\ldots}\equiv \frac{1}{Z_R}\underset{\mathcal{M}_L}{\int} \mathcal{D}\tau e^{-S_R[\tau]}(\ldots),
\end{align}
where the integral is normalized by the partition function $Z_R$ so that $\braket{1}=1$.

For greater generality, we can instead study the following two-point and four-point functions between ``operators'' $V$ and $W$ with conformal dimension $\Delta_V$ and $\Delta_W$:
\begin{align}
    \braket{V_1V_2}\equiv \braket{V(\theta_1)V(\theta_2)}_{\text{AdS}_2}&=\braket{B_\tau(\theta_1,\theta_2)^{\Delta_V}}\label{eq:gen 2pt}\\
    \braket{V_1V_2W_3W_4}\equiv\braket{V(\theta_1)V(\theta_2)W(\theta_3)W(\theta_4)}_{\text{AdS}_2} &=\braket{B_\tau(\theta_1,\theta_2)^{\Delta_V}B_\tau(\theta_3,\theta_4)^{\Delta_W}}.\label{eq:gen 4pt}
\end{align}

Before we study \eqref{eq:gen 2pt} and \eqref{eq:gen 4pt} perturbatively, we make some additional comments about the manifestation of the physical and gauge $SL(2,\mathbb{R})$ symmetries in the reparametrization path integral. First, the gauge symmetry sends $\tau\mapsto f\circ \tau$ where $f$ is given in \eqref{eq:SL(2,R) on circle}. This leaves both the reparametrization action in \eqref{eq:EM43CBaVED} and the bilocal operator in \eqref{eq:CSl6ubNNjr} invariant under left transformations (i.e., $S_R[f\circ \tau]=S_R[\tau]$ and $B_{f\circ \tau}(\theta_i,\theta_j)=B_\tau(\theta_i,\theta_j)$). This means that the integrands in \eqref{eq:yy 2-pt function}-\eqref{eq:gen 4pt} are indeed independent of the choice of representative for each gauge orbit in $\mathcal{M}_L$. It also means that we will need to gauge fix the three $SL(2,\mathbb{R})$ modes when computing the $\epsilon$ propagator. Second, the physical $SL(2,\mathbb{R})$ transformation sends $\tau\to \tau\circ f$, which leaves the action invariant (i.e., $S_R[\tau\circ f]=S_R[\tau]$) but transforms the the bilocal operator as $B_{\tau\circ f}(\theta_1,\theta_2)=\dot{f}(\theta_1)\dot{f}(\theta_2)B_\tau(f(\theta_1),f(\theta_2))$. These properties, and also assuming the right invariance of the measure, imply Ward identities for the correlators. For example, for the two-point function in \eqref{eq:gen 2pt}, for any $f$ that is an $SL(2,\mathbb{R})$ transformation on the circle, we have
\begin{align}\label{eq:LfEU3SQMRf}
    \braket{V(\theta_1)V(\theta_2)}&=\dot{f}(\theta_1)^{\Delta_V}\dot{f}(\theta_2)^{\Delta_V}\braket{V(f(\theta_1))V(f(\theta_2))}.
\end{align}
This generalizes to $n$-point functions of the bilocal operators. These Ward identities in particular fix the two-point and three-point functions up to normalization and the four-point functions up to a function of the $SL(2,\mathbb{R})$ invariant cross-ratio, $\chi$.

\subsection{The propagator and perturbation theory}\label{sec:epsilon propagator and 4pt function}
We will now compute the four-point function to first order in $1/T_s$, and then the OTOC in the double scaling limit, from the reparametrization path integral. For these observables, it is sufficient to expand around the saddle point of the reparametrization action by writing $\tau(\theta)=\theta+\epsilon(\theta)$ and working to quadratic order in $\epsilon$. 

We can compute the correlators perturbatively to the desired order using the propagator for $\epsilon(\theta)$, which is simplest to derive in Fourier space. Thus, we write $\epsilon(\theta)=\sum_n \epsilon_n e^{in\theta}$, and the quadratic reparametrization action as:
\begin{align}\label{eq:reparametrization quadratic action fourier}
    S_R[\theta+\epsilon(\theta)]&=4\pi T_s \sum_{n=2}^\infty |n|(n^2-1)\epsilon_n \epsilon_{-n}+O(\epsilon^3)
\end{align}
We used the fact that the inverse of $\tau(\theta)=\theta+\epsilon(\theta)$ is $\alpha(\tau)=\tau-\epsilon(\tau)$ to leading order in $\epsilon$. This means that at quadratic order the reparametrization action $S_R$ is the same as the longitudinal action $S_L$, which is given in \eqref{eq:longitudinal action quadratic circle body} in $\theta$ space and in \eqref{eq:Ov0gv3LFHW} in Fourier space.

Because the action is multiplied by the string tension, expanding in powers of $1/T_s$ means expanding in powers of $\epsilon$. And at the order that we are working, we will assume that the integration measure $\mathcal{D}\tau=\mathcal{D}\epsilon$ is given in Fourier space by:
\begin{align}\label{eq:xxzVJeoMlh}
    \mathcal{D}\epsilon=Nd\epsilon_0 d\epsilon_1 d\epsilon_{-1}\prod_{n=2}^\infty d\epsilon_n d\epsilon_{-n}
\end{align}
Here, $N$ is a normalization constant that does not affect the correlators. Eq.~\eqref{eq:xxzVJeoMlh} is also the form of the measure in the Schwarzian theory, which to leading order about the saddle point $\tau=\theta$ is $\mathcal{D}\epsilon\propto d\epsilon_0 d\epsilon_1 d\epsilon_{-1}\prod_{n=2}^\infty n(n^2-1)d\epsilon_n d\epsilon_{-n}$ \cite{Stanford:2017thb}. Factoring out the infinite constant $N=\prod_{n=2}^\infty n(n^2-1)$ leads to \eqref{eq:xxzVJeoMlh}. 

There are three zero modes of the quadratic action in \eqref{eq:reparametrization quadratic action fourier}: $\epsilon_0$, $\epsilon_1$, and $\epsilon_{-1}$, which correspond to $\epsilon(\theta)=1$, $\epsilon(\theta)=e^{i\theta}$, and $\epsilon(\theta)=e^{-i\theta}$. These are the three infinitesimal $SL(2,\mathbb{R})$ gauge transformations that we need to mod out in the reparametrization integral. The simplest way to gauge fix these modes is to set $\epsilon_0=\epsilon_{\pm 1}=0$. Namely, we write $\epsilon(\theta)=\sum_{n\neq 0,\pm 1}\epsilon_n e^{in\theta}$ and do not include $\epsilon_0$, $\epsilon_{\pm 1}$ in the measure in \eqref{eq:xxzVJeoMlh}. 

The two-point function of the Fourier modes follows from \eqref{eq:reparametrization quadratic action fourier}:
\begin{align}
    \braket{\epsilon_n\epsilon_m}&=\frac{1}{4\pi T_s|n|(n^2-1)}\delta_{n,-m},&&n,m\neq 0,\pm 1.
\end{align}
The $\epsilon$ propagator is therefore 
\begin{align}
    \braket{\epsilon(\theta)\epsilon(0)}&=\frac{1}{4\pi T_s}\sum_{n\neq 0,\pm 1}\frac{e^{in \theta}}{|n|(n^2-1)}=\frac{1}{4\pi T_s}\left[-1+\frac{3}{2}\cos{\theta}+2\sin^2{\frac{\theta}{2}}\log\left(4\sin^2{\frac{\theta}{2}}\right)\right].\label{eq:epsilon propagator 0}
\end{align}
If we choose instead a more general gauge fixing condition that preserves translation invariance,\footnote{One way to get \eqref{eq:sOtAmXClR9} is to add the gauge fixing term $S_{\rm gf}=T_s\left[\frac{1}{2a}\epsilon_0^2+\frac{1}{b}\epsilon_1\epsilon_{-1}\right]$ to the quadratic action in \eqref{eq:reparametrization quadratic action fourier} and absorb the first two terms in \eqref{eq:epsilon propagator 0} into $a$ and $b$.} the propagator becomes
\begin{align}\label{eq:sOtAmXClR9}
    \braket{\epsilon(\theta)\epsilon(0)}=\frac{1}{T_s}\left[a+b\cos{\theta}+\frac{1}{2\pi}\sin^2{\frac{\theta}{2}}\log\left(4\sin^2{\frac{\theta}{2}}\right)\right].
\end{align}
The coefficients of the first two terms are gauge dependent.

To compute the four-point function in \eqref{eq:gen 4pt} using the $\epsilon$ propagator, we write the bilocal operator in \eqref{eq:CSl6ubNNjr} in terms of $\epsilon$ and expand to linear order. It is convenient to introduce the bilocal operator $\mathcal{B}$ which is normalized by the conformal two-point function:
\begin{align}\label{eq:iu6D22cnWY}
    B(\theta_i,\theta_j)=\frac{1}{[2\sin{\frac{\theta_{ij}}{2}}]^{2}}\mathcal{B}(\theta_i,\theta_j).
\end{align}
Then, we have:
\begin{align}\label{eq:dressed two-point function}
    \mathcal{B}(\theta_i,\theta_j)&=\frac{\sin^2{\frac{\theta_{ij}}{2}}}{\sin^2\big(\frac{1}{2}[\theta_{ij}+\epsilon_{ij}
    ]\big)}(1+\dot{\epsilon}_i)(1+\dot{\epsilon}_j)=1+\dot{\epsilon}_i+\dot{\epsilon}_j-\epsilon_{ij}\cot{\frac{\theta_{12}}{2}}+O(\epsilon^2).
\end{align}
where we use the shorthand $\epsilon_{ij}\equiv \epsilon(\theta_i)-\epsilon(\theta_j)$ and $\dot{\epsilon}_i\equiv \dot{\epsilon}(\theta_i)$.

\begin{figure}[t]
\centering
\begin{minipage}{0.49\hsize}\label{fig:four-pt}
\centering
\includegraphics[clip, height=4cm]{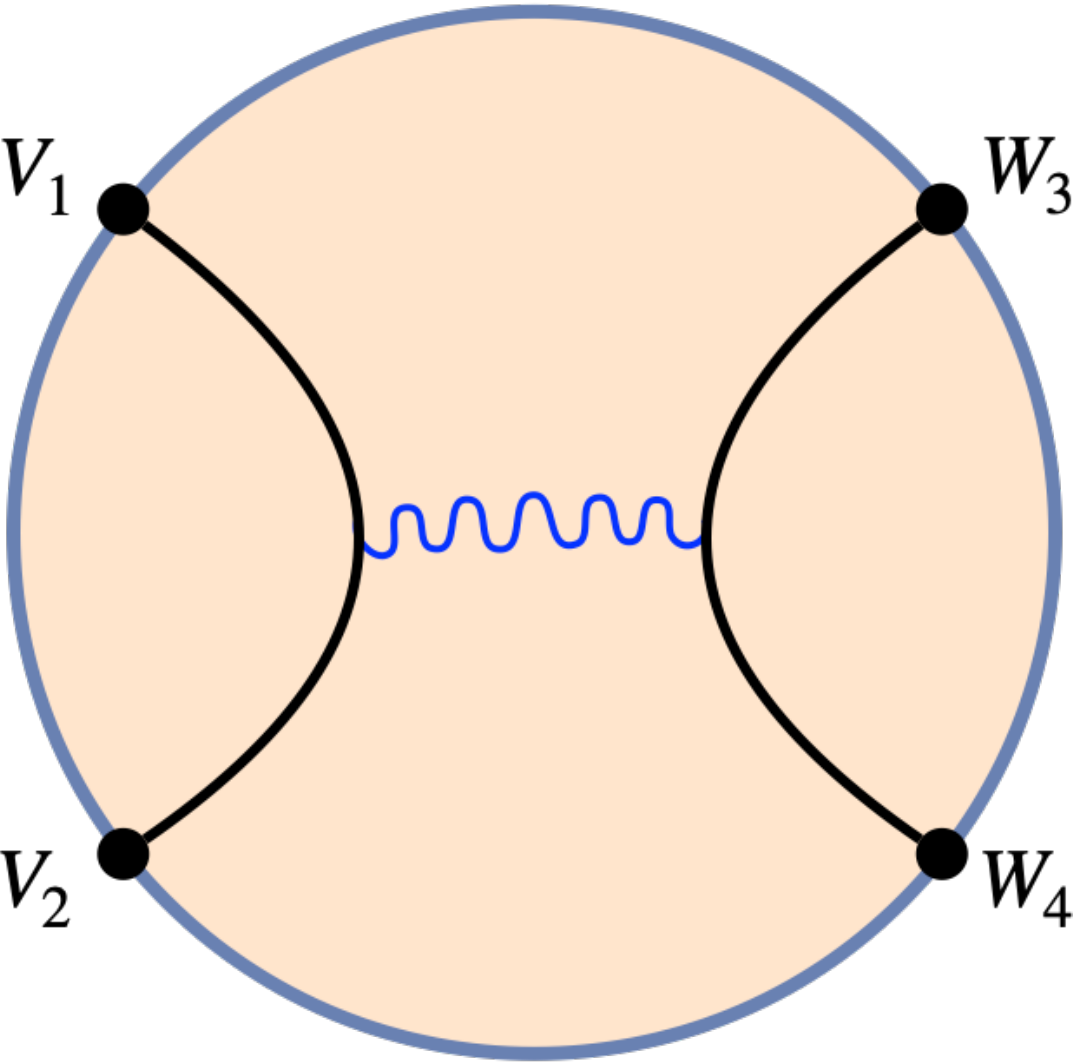}\\
{\bf a.} four-point function
\end{minipage}
\begin{minipage}{0.49\hsize}\label{fig:DS OTOC}
\centering
\includegraphics[clip, height=4cm]{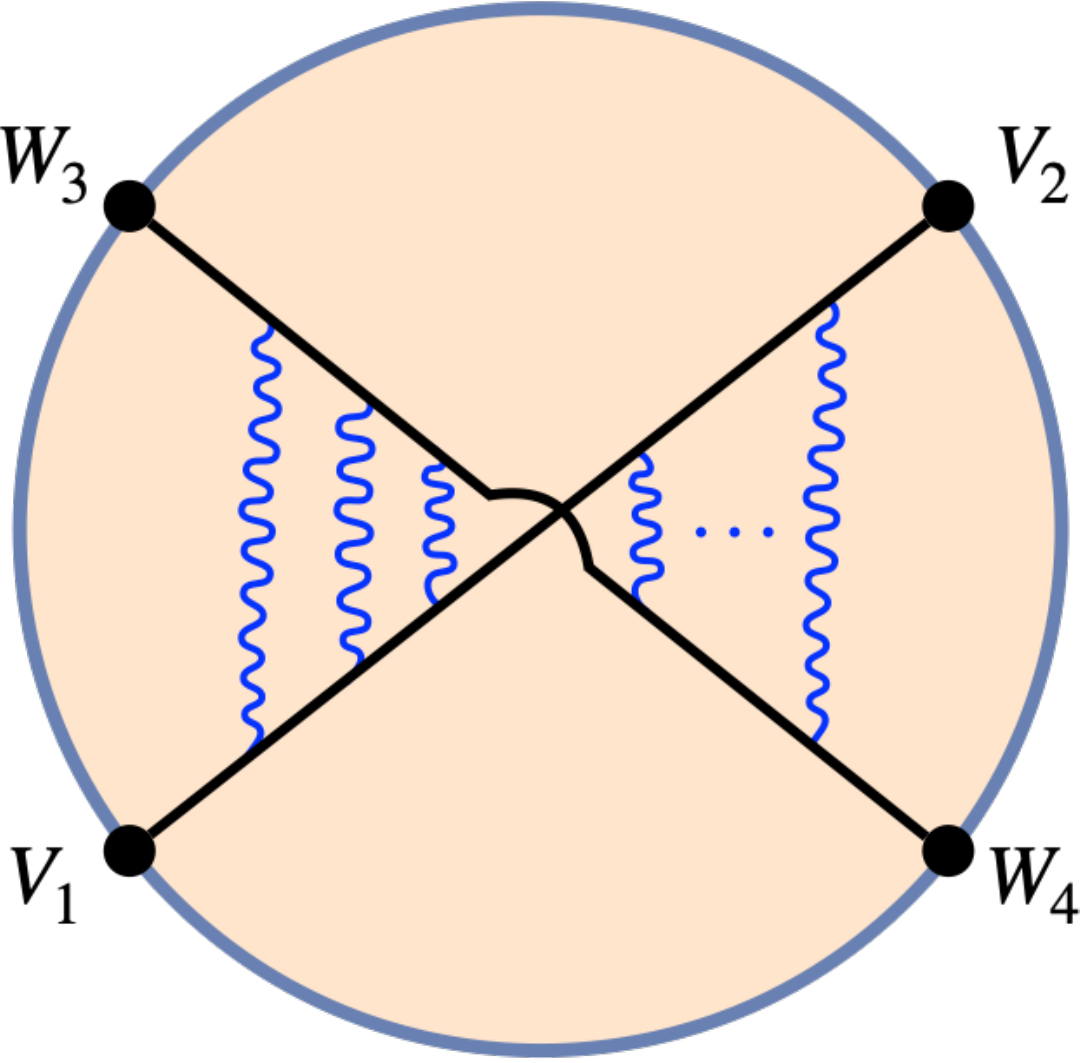}\\
{\bf b.} double scaled OTOC
\end{minipage}
\caption{Representation of the \textbf{a.} four-point function at leading order and \textbf{b.} OTOC in the double scaling limit, as computed in the reparametrization path integral. The curved black lines represent the dressed two-point functions defined in \eqref{eq:dressed two-point function}. The wavy blue lines represent contractions between the $\epsilon$'s appearing in the expansion of the dressed two-point functions using the propagator in \eqref{eq:sOtAmXClR9}. In the double scaled OTOC, the $\epsilon$ propagators connect the two dressed two-point functions directly, in analogy with the eikonal approximation in high energy gravitational scattering. There are no interactions between the $\epsilon$'s because we keep only the quadratic part of the reparametrization action in section~\ref{sec:double scaled OTOC from reparam integral}. These diagrams are only meant to be schematic and do not correspond precisely to any Feynman rules.}
\label{fig:four-pt and DS OTOC}
\end{figure}

The leading contribution to the four-point function normalized by the two-point functions is:
\begin{align}\label{eq:TLmeIfkMCn}
    \frac{\braket{V_1V_2W_3W_4}}{\braket{V_1V_2}\braket{W_3W_4}}&=\frac{\braket{\mathcal{B}(\theta_1,\theta_2)^{\Delta_V}\mathcal{B}(\theta_3,\theta_4)^{\Delta_W}}}{\braket{\mathcal{B}(\theta_1,\theta_2)^{\Delta_V}}\braket{\mathcal{B}(\theta_3,\theta_4)^{\Delta_W}}}\nonumber\\&=1+\Delta_V\Delta_W \braket{\mathcal{B}(\theta_1,\theta_2)\mathcal{B}(\theta_3,\theta_4)}_{\text{conn}}+O(1/T_s^2)
\end{align}
where the connected component of the correlator of two bilocal operators is:
\begin{align}\label{eq:f7JS8ArQyv}
    \braket{\mathcal{B}(\theta_1,\theta_2)\mathcal{B}(\theta_3,\theta_4)}_{\text{conn}}&=\braket{(\dot{\epsilon}_1+\dot{\epsilon}_2)(\dot{\epsilon}_3+\dot{\epsilon}_4)}-\cot{\frac{\theta_{12}}{2}}\braket{(\dot{\epsilon}_3+\dot{\epsilon}_4)\epsilon_{12}}\nonumber\\&-\cot{\frac{\theta_{34}}{2}}\braket{(\dot{\epsilon}_1+\dot{\epsilon}_2)\epsilon_{34}}+\cot{\frac{\theta_{12}}{2}}\cot{\frac{\theta_{34}}{2}}\braket{\epsilon_{12}\epsilon_{34}}+O(\epsilon^3).
\end{align}
We use the propagator in \eqref{eq:sOtAmXClR9} to evaluate the various terms above. The final result simplifies to:
\begin{align}\label{eq:four-pt function leading order}
    \frac{\braket{V_1V_2W_3W_4}}{\braket{V_1V_2}\braket{W_3W_4}}&=1-\frac{\Delta_V\Delta_W}{4\pi T_s}\left[4+\frac{2-\chi}{\chi}\log((1-\chi)^2)\right]+O(1/T_s^2).
\end{align}
This computation is represented graphically in Figure~\ref{fig:four-pt and DS OTOC}.

A few comments about \eqref{eq:four-pt function leading order}: Firstly, it is independent of the coefficients $a$ and $b$ of the zero mode contributions to the propagator in \eqref{eq:sOtAmXClR9}, which reflects the gauge $SL(2,\mathbb{R})$ symmetry of the string boundary correlators. Secondly, the normalized four-point function depends on the positions of the operator insertions only through the conformal cross-ratio, $\chi$, which reflects the physical $SL(2,\mathbb{R})$ symmetry of the boundary correlators. Thirdly, the four-point function of four identical operators in \eqref{eq:yyyy 2-pt function} follows from \eqref{eq:four-pt function leading order} if we set $\Delta_V=\Delta_W=1$ and sum over the three distinct pairings of the four operators.
Namely,
\begin{align}
    \frac{\braket{y_1y_2y_3y_4}}{\braket{y_1y_2}\braket{y_3y_4}}&=1+\chi^2+\frac{\chi^2}{(1-\chi)^2}-\frac{1}{4\pi T_s}\bigg[4+\frac{2-\chi}{\chi}\log((1-\chi)^2)+\frac{\chi^2}{(1-\chi)^2}\left(4+\frac{1+\chi}{1-\chi}\log(\chi^2)\right)\nonumber\\&\hspace{2cm}+\chi^2\left(4+(2\chi-1)\log\left(\frac{(1-\chi)^2}{\chi^2}\right)\right)\bigg]+O(1/T_s^2),
\end{align}
which precisely matches \eqref{eq:4-pt static gauge}. It is worth noting that doing the $\epsilon$ contractions in \eqref{eq:f7JS8ArQyv} seems technically easier than computing $D$-functions in static gauge (e.g., in going from \eqref{eq:F(t1 t2 t3 t4)} \eqref{eq:fg5432ws24}). Finally, analytically continuing \eqref{eq:four-pt function leading order} to the OTOC configuration using \eqref{eq:OTOC cross ratio} and \eqref{eq:analytic continuation logs} yields:
\begin{align}\label{eq:chEkwMOIlb}
    \frac{\braket{V_1V_2W_3W_4}}{\braket{V_1V_2}\braket{W_3W_4}}&=1-\frac{\Delta_V\Delta_W}{4T_s}e^t+\ldots
\end{align}
Thus, the maximal chaos of the AdS$_2$ string can be derived from the reparametrization mode in conformal gauge.

\paragraph{On the line.} We can also study the perturbative correlators on the line instead of the circle. The analysis is slightly more tricky because the diffeomorphisms on the line are not as well-behaved as on the circle. A concrete manifestation of this is that the three infinitesimal $SL(2,\mathbb{R})$ gauge transformations, $1$, $t$, and $t^2$, are not normalizable.\footnote{Note also that the action in \eqref{eq:longitudinal action quadratic line} is indeed zero for $\epsilon(t)=1$ and $\epsilon(t)=t$, but infinite for $\epsilon(t)=t^2$.} Nonetheless, by using a slightly more heuristic approach, we can still deduce the $\epsilon$ propagator and compute the four-point function.

We write the perturbation about the saddle point, $t(x)=x+\epsilon(x)$, as a Fourier integral: $\epsilon(x)=\int \frac{d\omega}{2\pi}e^{-i\omega x}\epsilon(\omega)$. The action in Fourier space was evaluated in \eqref{eq:quadratic longitudinal action fourier}. It follows that the two-point function of the Fourier modes is:
\begin{align}
    \braket{\epsilon(\omega)\epsilon(\nu)}=\frac{\pi}{T_s}\delta(\omega+\nu)\frac{1}{|\omega|^3}.
\end{align}
Naively, the $\epsilon$ propagator is therefore
\begin{align}\label{eq:BiyOKdFweq}
    \braket{\epsilon(x)\epsilon(0)}&=\frac{1}{4\pi T_s}\int d\omega\frac{e^{-i\omega x}}{|\omega|^3},
\end{align}
but this integral is divergent at $\omega=0$. We can regularize it by integrating over the interval $\mathbb{R}\setminus (-\delta,\delta)$ and taking $\delta\to 0$. The result is:
\begin{align}\label{eq:Pdq5ZKUGWh}
    \underset{\mathbb{R}\setminus (-\delta,\delta)}{\int} d\omega \frac{e^{-i\omega x}}{|\omega|^3}=\frac{1}{\delta^2}+\left[-\frac{3}{2}+\gamma_E+\frac{1}{2}\log{\delta^2}\right]x^2+\frac{1}{2}x^2\log{x^2}+O(\delta^2).
\end{align}
The terms in the above expression that diverge as $\delta\to 0$ are either constant in $x$ or multiply $x^2$. Since $1$ and $x^2$ are two of the infinitesimal gauge $SL(2,\mathbb{R})$ modes, their coefficients in the propagator are gauge dependent and do not affect any of the observables. We therefore use the gauge freedom to absorb all the constants, and arrive at the following expression for the propagator:
\begin{align}\label{eq:KXC6hIJayQ}
    \braket{\epsilon(x)\epsilon(0)}=\frac{1}{T_s}\left[a+bx^2+\frac{1}{8\pi }x^2\log(x^2)\right].
\end{align}
This agrees with what we get if we replace the chordal distance $2\sin{\frac{\theta}{2}}$ in \eqref{eq:sOtAmXClR9} by the euclidean distance $x$.

In analogy with \eqref{eq:CSl6ubNNjr} and \eqref{eq:iu6D22cnWY}, or by taking two variational derivatives of the transverse action in \eqref{eq:transverse action on-shell explicit}, we introduce the bilocal operator on the line:
\begin{align}
    B(x_1,x_2)\equiv\frac{\dot{t}(x_1)\dot{t}(x_2)}{[t(x_1)-t(x_2)]^2}=\frac{1}{x_{12}^2}\mathcal{B}(x_1,x_2),
\end{align}
where $\mathcal{B}$ is again the bilocal operator normalized by the conformal two point function. For small $\epsilon$,
\begin{align}\label{eq:bilocal line normalized}
    \mathcal{B}(x_1,x_2)&=\frac{(1+\dot{\epsilon}(x_1))(1+\dot{\epsilon}(x_2))}{\big[1+\frac{\epsilon_{12}}{x_{12}}\big]^2}=1+\dot{\epsilon}_1+\dot{\epsilon}_2-2\frac{\epsilon_{12}}{x_{12}}+O(\epsilon^2).
\end{align}

The four-point function on the line is then
\begin{align}\label{eq:0WilC3IFHt}
    \frac{\braket{V_1V_2W_3W_4}}{\braket{V_1V_2}\braket{W_3W_4}}&=1+\Delta_V\Delta_W \braket{\mathcal{B}(x_1,x_2)\mathcal{B}(x_3,x_4)}_{\rm conn}+O(1/T_s^2),
\end{align}
where
\begin{align}\label{eq:cVmT9ySxIQ}
    \braket{\mathcal{B}(x_1,x_2)\mathcal{B}(x_3,x_4)}_{\rm conn}&=\braket{(\dot{\epsilon}_1+\dot{\epsilon}_2)(\dot{\epsilon}_3+\dot{\epsilon}_4)}-\frac{2}{x_{12}}\braket{(\dot{\epsilon}_3+\dot{\epsilon}_4)\epsilon_{12}}\\&\hspace{2cm}-\frac{2}{x_{34}}\braket{(\dot{\epsilon}_1+\dot{\epsilon}_2)\epsilon_{34}}+\frac{4}{x_{12}x_{34}}\braket{\epsilon_{12}\epsilon_{34}}+O(\epsilon^3).\nonumber
\end{align}
When we evaluate the above correlators using the propagator in \eqref{eq:KXC6hIJayQ}, the result reproduces \eqref{eq:four-pt function leading order}.

\subsection{Double-scaled OTOC}\label{sec:double scaled OTOC from reparam integral}

Finally, we will use the reparametrization path integral to derive the OTOC in the double scaling limit $T_s\to \infty$ and $t\to \infty$ with $\kappa=\frac{e^{t}}{16 T_s}$ held fixed. To make progress, we assume that only the quadratic part of the reparametrization action is relevant when computing the OTOC. This is a simplifying assumption made out of necessity (because we do not know the higher order corrections to the reparametrization action), but it has a plausible interpretation in terms of the eikonal approximation used in high energy gravitational scattering. Moreover, it reproduces the scattering result for the OTOC that was derived in section~\ref{sec:OTOC from shockwave S matrix} and checked to fourth order in section \ref{sec:OTOC on WL to 3 loops}. 

We take advantage of the conformal symmetry of the boundary correlators to work with the reparametrization integral on the line rather than on the circle. The numerators and the denominators of the bilocal operator in \eqref{eq:bilocal line normalized} can be written as exponentials using differentiation and Schwinger parameters:
\begin{align}\label{eq:schwinger parameter + differentiation}
     (1+\dot{\epsilon}_i)^\Delta &=\left(\frac{\partial}{\partial \alpha_i}\right)^\Delta e^{\alpha_i (1+\dot{\epsilon}_i)}\bigg\rvert_{\alpha_i=0}, &\frac{1}{(1+\frac{\epsilon_{ij}}{x_{ij}})^{2\Delta}}&=\frac{1}{\Gamma(2\Delta)}\int_0^\infty dp p^{2\Delta -1}e^{-p(1+\frac{\epsilon_{ij}}{x_{ij}})}.
\end{align}
Note that the exponents are linear in $\epsilon$, which would not be true if we worked with the reparametrization path integral on the circle because of the denominator in the bilocal operator in \eqref{eq:dressed two-point function}. Using \eqref{eq:schwinger parameter + differentiation}, the four-point function in the reparametrization path integral can be written as:
\begin{align}
    \frac{\braket{V_1V_2W_3W_4}}{\braket{V_1V_2}\braket{W_3W_4}}&=\frac{\braket{\mathcal{B}(x_1,x_2)^{\Delta_V}\mathcal{B}(x_3,x_4)^{\Delta_W}}}{\braket{\mathcal{B}(x_1,x_2)^{\Delta_V}}\braket{\mathcal{B}(x_3,x_4)^{\Delta_W}}}\nonumber\\&=\left(\frac{\partial}{\partial \alpha_1}\frac{\partial}{\partial \alpha_2}\right)^{\Delta_V}\left(\frac{\partial}{\partial \alpha_3}\frac{\partial}{\partial \alpha_4}\right)^{\Delta_W}\bigg[\left(\prod_{i=1}^4 e^{\alpha_i}\right) \frac{1}{\Gamma(2\Delta_V)}\frac{1}{\Gamma(2\Delta_W)} \nonumber\\&\times \int dp p^{2\Delta_V-1} e^{-p} \int dq q^{2\Delta_W-1} e^{-q} \bigg\langle \text{exp}\bigg(-p\frac{\epsilon_{12}}{x_{12}}-q\frac{\epsilon_{34}}{x_{34}}+\sum_{i=1}^4 \alpha_i \dot{\epsilon}_i\bigg)\bigg\rangle\bigg].\label{eq:ftrdsww345tf}
\end{align}
To get the second line, we kept only the zeroth order terms in the two-point functions, $\braket{\mathcal{B}(x_i,x_j)^{\Delta}}=1+O(1/T_s)$. When we truncate the action to quadratic order, the $\epsilon$'s obey Wick's theorem and therefore the exponent in the expectation value in \eqref{eq:ftrdsww345tf} being linear in $\epsilon$ implies:
\begin{align}\label{eq:gb345tfds}
    \bigg\langle \text{exp}\bigg(-p\frac{\epsilon_{12}}{x_{12}}-q\frac{\epsilon_{34}}{x_{34}}+\sum_{i=1}^4 \alpha_i \dot{\epsilon}_i\bigg)\bigg\rangle= e^X,
\end{align}
where
\begin{align}\label{eq:X}
	X=\bigg\langle\frac{1}{2}\bigg(-p\frac{\epsilon_{12}}{x_{12}}-q\frac{\epsilon_{34}}{x_{34}}+\sum_{i=1}^4 \alpha_i \dot{\epsilon}_i\bigg)^2\bigg\rangle.
\end{align}
Making this approximation is equivalent to summing up only those contributions to the four-point function in the reparametrization path integral where the $\epsilon$'s in the bilocal operators are contracted directly, without any interactions between the $\epsilon$'s from the higher orders terms in the reparametrization action. Such contractions are represented schematically in Figure~\ref{fig:four-pt and DS OTOC}. 

The fully expanded expression for $X$ involves many Wick contractions and is rather unwieldy.\footnote{Some terms involve self-contractions of the $\epsilon$'s, which are given by $\braket{\epsilon(0)^2}=a$, $\braket{\epsilon(0)\dot{\epsilon}(0)}=0$, $\braket{\dot{\epsilon}(0)^2}=-\frac{1}{4\pi T_s}\log{\delta^2}$, where we introduce $\delta$ as a short-distance regulator into which a constant and a gauge-dependent piece have been absorbed. These self-contractions also appear in the leading correction to the two-point function, and the logarithmic divergences indicate that the external operators have anomalous dimensions of order $1/T_s$. The self-contractions are not relevant in the present analysis, since they are subleading in the double scaling limit.} Moreover, it is invariant under neither the gauge $SL(2,\mathbb{R})$ transformations (i.e., $X$ depends non-trivially on the gauge-dependent coefficients $a$ and $b$) nor the physical $SL(2,\mathbb{R})$ transformations (i.e., the various $x_i$-dependent terms in $X$ cannot be recombined into a function only the conformal cross-ratio $\chi$ in \eqref{eq:xyz string polyakov action}). This is a consequence of trying to include $1/T_s$ corrections to the four-point function while keeping only the quadratic part of the reparametrization action, and is in contrast with the computation of the leading correction to the four-point function in \eqref{eq:0WilC3IFHt}-\eqref{eq:cVmT9ySxIQ}. However, we are interested in the OTOC in the double scaling limit, in which case only the terms in \eqref{eq:X} that grow exponentially in $t$ survive; other terms, including the gauge dependent terms, drop out.

To be concrete, we put the four operators in the configuration specified in \eqref{eq:OTOC euclidean times} and map the points from the euclidean circle to the line using $x_i=\tan{\frac{\theta_i}{2}}$. We then take $T_s,t\to \infty$ with $\kappa=\frac{e^t}{16 T_s}$ held fixed. In this limit $x_{12},x_{34}\propto e^{-t/2}$ become exponentially small in $t$, while all the other distances remain finite.\footnote{Specifically, the distances between the four points at late times obey $x_{13},x_{14},x_{23},x_{24}\to -2i$, $x_{12},x_{34}\sim 4e^{-\frac{i\pi}{4}}e^{-\frac{t}{2}}$. Furthermore, the analytic continuation of the logs to late times gives $\log{x_{13}^2},\log{x_{14}^2},\log{x_{24}^2}\to 2\log{2}-i\pi$, $\log{x_{23}^2}\to 2\log{2}+i\pi$, $\log{x_{12}^2},\log{x_{34}^2}\sim -t + 4\log{2} -\frac{i\pi}{2}$.} In this case, the only term that contributes to \eqref{eq:X} is
\begin{align}\label{eq:iIpkPJrMdN}
    \frac{pq}{x_{12}x_{34}}\braket{\epsilon_{12}\epsilon_{34}}&=\frac{1}{T_s}\frac{pq}{x_{12}x_{34}}\biggr[\frac{1}{8\pi}\left(x_{13}^2\log(x_{13}^2)+x_{24}^2\log(x_{24}^2)-x_{14}^2 \log(x_{14}^2)-x_{23}^2 \log(x_{23}^2)\right)\nonumber\\&\hspace{3cm}+b\left(x_{13}^2+x_{24}^2-x_{14}^2-x_{23}^2\right)\biggr],\nonumber\\&=-\frac{pq}{8\pi T_s}\left[\chi^{-1}\log((1-\chi)^2)+\ldots\right],
 \end{align}
 where the ``$\ldots$'' in the second line denotes terms that are subleading in the double scaling limit. Analytically continuing the cross-ratio $\chi$ according to \eqref{eq:cross ratio line} in the second line in \eqref{eq:iIpkPJrMdN} (or, equivalently, continuing the individual points $x_i$ in the first line in \eqref{eq:iIpkPJrMdN}) leads to
\begin{align}
	X&=\frac{pq}{x_{12}x_{34}}\braket{\epsilon_{12}\epsilon_{34}}+\ldots\to -\kappa pq,
\end{align}
where $\kappa=\frac{e^t}{16T_s}$. After substituting this into \eqref{eq:ftrdsww345tf}, the derivatives with respect to the $\alpha_i$ are trivial and the integrals reduce to:
\begin{align}\label{eq:double scaled OTOC integral}
    \frac{\braket{V_1V_2W_3W_4}}{\braket{V_1V_2}\braket{W_3W_4}}&=\int \frac{dp\text{ }p^{2\Delta_V-1}}{\Gamma(2\Delta_V)}e^{-p}\int \frac{dq\text{ }q^{2\Delta_W-1}}{\Gamma(2\Delta_W)}e^{-q}e^{-\kappa pq},
\end{align}
which precisely matches \eqref{eq:OTvzuRODWt} and reproduces \eqref{eq:OTOC late time (reprise)}.

This derivation of the double scaled OTOC on the AdS$_2$ string from the reparametrization integral relies on two important assumptions. Firstly, we assumed that for the purpose of computing the OTOC, it is valid to approximate the string partition function by the reparametrization integral in \eqref{eq:EM43CBaVED} (or, equivalently, in \eqref{eq:string partition function reparametrization integral}) without worrying about the fluctuations of the matter fields and ghosts in the string sigma model. Secondly, we assumed that it is valid to keep only the quadratic part of the reparametrization action. Both of these assumptions have a natural interpretation in terms of the eikonal approximation in high energy scattering between between gravitating particles, where one resums the contributions of only graviton exchanges (see, e.g., \cite{Amati:1992zb,Kabat:1992tb,Cornalba:2006xk,Cornalba:2006xm,Cornalba:2007zb}). Integrating over only the reparametrizations is analogous to including only virtual gravitons, and keeping only the quadratic contribution to the reparametrization action is analogous to including only processes in which the gravitons are exchanged directly between the external particles without interacting among themselves. However, this interpretation is only heuristic because there are no gravitons on the worldsheet. Finally, it was also important in getting to \eqref{eq:double scaled OTOC integral} that we simultaneously sent the $V$ operators to the past and the $W$ operators to the future: i.e., in accordance with \eqref{eq:OTOC euclidean times}, we send the Lorentzian times $\text{Im}(\theta_1),\text{Im}(\theta_2)\to -\infty$ at the same time that we send $\text{Im}(\theta_3),\text{Im}(\theta_4)\to \infty$, but the relative rate at which they are sent to $ \pm \infty$ is unimportant.

The discussion in this section --- the honest derivation of the OTOC in the Lyapunov regime in \eqref{eq:chEkwMOIlb} and the more heuristic derivation of the OTOC in the double scaling limit in \eqref{eq:double scaled OTOC integral}--- demonstrates that maximal chaos can arise from other reparametrization actions besides the Schwarzian. However, in our derivation of the OTOC above, it is not very transparent what common property of the Schwarzian and the AdS$_2$ reparametrization actions makes the OTOC maximally chaotic. Although we will not explore this point much further, the key ingredient should be the $SL(2,\mathbb{R})$ gauge symmetry that is common to both JT gravity and the AdS$_2$ string.\footnote{We thank M\'ark Mezei and Juan Maldacena for discussions on this point.} In particular, in \cite{Maldacena:2016upp} (see also section 3.1 of \cite{Choi:2023mab}), the OTOC in the double scaling limit was computed starting from the path integral over reparametrizations of the AdS$_2$ boundary in JT gravity by identifying a set of nearly-zero modes that dominate the path integral. These nearly-zero modes consist of turning on and turning off $SL(2,\mathbb{R})$ gauge transformations that are exponentially growing in time along sections of the integration contour between the operators $V$ and $W$, and their existence and role in determining the double scaled OTOC follows entirely from the $SL(2,\mathbb{R})$ gauge symmetry of the Schwarzian theory. In particular, the exponentially growing modes $e^{\pm t}$ are the Wick rotations of the $SL(2,\mathbb{R})$ zero modes $e^{\pm i \theta}$ on the circle (see the paragraph after eq.~\eqref{eq:xxzVJeoMlh}). Essentially the same analysis should therefore be applicable in the case of the reparametrization mode of the AdS$_2$ string, and should provide an alternative way to derive \eqref{eq:double scaled OTOC integral}.

\section{The string reparametrization mode and the Schwarzian}\label{eq:AdS2 reparametrization and the Schwarzian}

In the first half of this paper, we saw that the OTOCs on the AdS$_2$ string and in JT gravity are the same both in the Lyapunov regime and in the double scaled limit. In the second half, we studied the reparametrization mode that emerges in the conformal gauge analysis of the AdS$_2$ string and used it to rederive both the leading order four-point function and the double scaled OTOC. We noted that the role of the reparametrization mode in those computations is analogous to the role of the Schwarzian in JT gravity. In this section we compare the string reparametrization mode and the Schwarzian mode in more detail.

\subsection{Review: the Schwarzian mode in JT gravity}\label{sec:Schwarzian from JT gravity}

We begin by reviewing how the Schwarzian mode arises in JT gravity, following the discussion in \cite{Maldacena:2016upp} (see also \cite{Jensen:2016pah,Engelsoy:2016xyb} and \cite{Mertens:2022irh} for a recent review). JT gravity is a toy model of gravity in two dimensions consisting of the dilaton $\Phi$, metric $h$, and matter fields \cite{Jackiw:1984je,Teitelboim:1983ux}. We consider the simplest case where there is one matter field, $y$, minimally coupled to the metric. The action is:
\begin{align}\label{eq:JT+matter}
    S_{\text{JT}+\text{matter}}[\Phi,h,y]=S_{\rm JT}[\Phi,h]+S_{\rm matter}[y,h],
\end{align}
where 
\begin{align}
    S_{\rm JT}[\Phi,h]&=S_{\rm topological}-\frac{1}{16\pi G_N}\left[\int_{\mathcal{M}} d^2\sigma \sqrt{h}\Phi(R+2)+2\int_{B}d\theta \sqrt{h_{\theta\theta}}\Phi (K-1)\right],\label{eq:JT action}\\
    S_{\rm matter}[y,h]&=\int_{\mathcal{M}} d^2\sigma \frac{\sqrt{h}}{2}\left(h^{\alpha\beta}\partial_\alpha y \partial_\beta y +m^2 y^2\right).\label{eq:matter action}
\end{align}
In \eqref{eq:JT action}, $\sigma^\alpha=(\sigma,\tau)$ are spacetime coordinates on $\mathcal{M}$ and $\theta$ is a coordinate along the boundary $B=\partial \mathcal{M}$. The $R$ in the bulk term in \eqref{eq:JT action} denotes the Ricci curvature. The $K$ in the the boundary term denotes the extrinsic curvature (and the ``$-1$'' is a counterterm to make the action finite). Finally, the topological term in \eqref{eq:JT action} is proportional to the Euler characteristic and does not affect the dynamics other than to pick out the disk topology as the leading contribution in the genus expansion.

The dilaton equation of motion is $R=-2$. Thus, the spacetime $\mathcal{M}$ is a patch of hyperbolic space. Meanwhile, the metric equation of motion implies that the dilaton diverges at the boundary, so it is necessary to regularize the theory by cutting off the boundary along a curve. In hyperbolic disk coordinates with metric $ds^2=\sinh^{-2}{\sigma}(d\sigma^2+d\tau^2)$, the cut-off curve can be parametrized as $\theta\mapsto (\sigma(\theta),\tau(\theta))$, where $\theta\in [0,2\pi]$ is a rescaled proper length coordinate satisfying:
\begin{align}\label{eq:tdsrfde4e}
    h_{\theta\theta}=\frac{\dot{\sigma}(\theta)^2+\dot{\tau}(\theta)^2}{\sinh^2(\sigma(\theta))}=\frac{1}{\epsilon^2}.
\end{align}
Here, $\epsilon$ is the cut-off parameter, and $\cdot$ denotes differentiation with respect to $\theta$. The length of the curve is $2\pi/\epsilon$ and sending $\epsilon\to 0$ sends the curve to the boundary of hyperbolic space. The boundary curve is specified by $\tau(\theta)$ alone since eq.~\eqref{eq:tdsrfde4e} fixes $\sigma(\theta)$ in terms of $\tau(\theta)$. In particular, to leading order in $\epsilon$, $\sigma(\theta)=\epsilon \dot{\tau}\theta)$. $\tau(\theta)$ can be viewed as a reparametrization of the boundary of the hyperbolic disk.

The boundary conditions for the dilaton and the scalar specify their renormalized values $\tilde{\Phi}(\theta)$ and $\tilde{y}(\theta)$ along the boundary curve:
\begin{align}\label{eq:JT bcs}
    \Phi(\sigma(\theta),\tau(\theta))&=\frac{\tilde{\Phi}(\theta)}{\epsilon}, &y(\sigma(\theta),\tau(\theta))&=\frac{\tilde{y}(\theta)}{\epsilon^{\Delta-1}}.
\end{align}
(The scalar scaling dimension $\Delta$ is related to the mass by the relation $m^2=\Delta(\Delta-1)$.) The simplest case is if the dilaton is constant on the boundary: $\tilde{\Phi}(\theta)=\tilde{\Phi}$. With these boundary conditions, in the limit $\epsilon\to 0$, one can write both the JT and matter actions in \eqref{eq:JT+matter} as boundary actions in terms of $\tilde{\Phi}$, $\tilde{y}(\theta)$ and $\tau(\theta)$. Consider first the JT action in \eqref{eq:JT action}. The bulk term vanishes due to the dilaton equation of motion, and the boundary term becomes
\begin{align}\label{eq:mik4K1iLrh}
S_{\rm JT}[\tilde{\Phi},\tau]=-\frac{\tilde{\Phi}}{8\pi G_N}\int d\theta \frac{1}{\epsilon^2}(K-1).
\end{align}
The explicit expression for the extrinsic curvature of the cut-off curve in these coordinates is:
\begin{align}
    K=\frac{\cosh{\sigma}(\dot{\sigma}^2+\dot{\tau}^2)+\sinh{\sigma}(\dot{\tau}\ddot{\sigma}-\dot{\sigma}\ddot{\tau})}{(\dot{\sigma}+\dot{\tau})^{\frac{3}{2}}}, 
\end{align}
which, since $\sigma=\epsilon\dot{\tau}+O(\epsilon^2)$, to subleading order in $\epsilon$ is:
\begin{align}\label{eq:K expansion}
    K&=1+\epsilon^2\{\tan{\frac{\tau}{2}},\theta\}+O(\epsilon^4)=1+\epsilon^2\left(\{\tau,\theta\}+\frac{1}{2}\dot{\tau}^2\right)+O(\epsilon^4).
\end{align}
Here, $\{,\}$ denotes the Schwarzian derivative:
\begin{align}\label{eq:Schwarzian derivative}
    \{f,\theta\}\equiv \frac{\dddot{f}(\theta)}{\dot{f}(\theta)}-\frac{3}{2}\frac{\ddot{f}(\theta)^2}{\dot{f}(\theta)^2}.
\end{align}
Thus, the dynamical part of the JT action in the $\epsilon\to 0$ limit is governed by the Schwarzian:
\begin{align}\label{eq:3qs9vXiyeP}
    S_{\rm Schwarzian}[\tau]&=-C\int d\theta \{\tan{\frac{\tau}{2}},\theta\}=-C\int d\theta \left(\{\tau,\theta\}+\frac{1}{2}\dot{\tau}(\theta)^2\right),
\end{align}
where $C\equiv \frac{\tilde{\Phi}}{8\pi G_N}$. As explained in \cite{Maldacena:2016upp}, the boundary representation of the JT action follows entirely from symmetry considerations. The JT action in \eqref{eq:JT action} is invariant under the $SL(2,\mathbb{R})$ isometries moving the boundary curve around in hyperbolic space. And in the derivative expansion of a local boundary Lagrangian, the leading term that is invariant under $SL(2,\mathbb{R})$ transformations acting on $\tau$ is the Schwarzian derivative.

Meanwhile, the matter action in \eqref{eq:matter action} is that of a free scalar in hyperbolic space, and its effective action is simply given by the conformal two-point function weighted by the value of the matter field along the boundary. Because the boundary condition in \eqref{eq:JT bcs} is specified along the cut-off curve instead of along $\sigma=0$, the conformal two-point function is dressed by $\tau(\theta)$. The result is:
\begin{align}\label{eq:S matter boundary action}
    S_{\rm matter}[\tilde{y},\tau]=-\frac{D}{2}\int d\theta d\theta' \frac{\dot{\tau}(\theta)^\Delta \dot{\tau}(\theta')^\Delta}{\left[2\sin\big(\frac{\tau(\theta)-\tau(\theta')}{2}\big)\right]^{2\Delta}}\tilde{y}(\theta)\tilde{y}(\theta').
\end{align}
Here, $D$ is an unimportant normalization.

Therefore, what remains of the JT gravity partition function after the dilaton, metric and matter fields are integrated out is a path integral over the boundary reparametrizations $\tau(\theta)$ weighted by the actions in \eqref{eq:3qs9vXiyeP} and \eqref{eq:S matter boundary action}:
\begin{align}\label{eq:JT+matter partition function}
    Z_{\rm JT+matter}[\tilde{y}]&=\underset{\substack{\Phi\rvert_\partial=\tilde{\Phi}\\y\rvert_\partial=\tilde{y}}}{\int} \mathcal{D}\Phi \mathcal{D}h \mathcal{D}ye^{-S_{\rm JT}[\Phi,h]-S_{\rm matter}[y,h]}&&=\underset{\mathcal{M}_L}{\int}\mathcal{D}\tau(\theta)e^{-S_{\rm Schwarzian}[\tau]-S_{\rm matter}[\tilde{y},\tau]}.
\end{align}
The reparametrization path integral integrates over all diffeomorphisms of the circle modulo $SL(2,\mathbb{R})$ transformations in order to include the contributions from all distinct geometries that can be cut out of hyperbolic space. (Two bulk geometries in AdS$_2$ cut out by two closed curves that are related by an isometry are considered identical). Thus, the domain of integration, $\mathcal{M}_L$, is again given by \eqref{eq:VBoPjO1JbC}.

\subsection{Comparing the string reparametrization mode and the Schwarzian}
We now discuss the parallels between the Schwarzian reparametrization integral determining the JT gravity partition function in \eqref{eq:JT+matter partition function} and the AdS$_2$ string reparametrization integral contributing to the string partition function in \eqref{eq:EM43CBaVED}. 

First, both the Schwarzian and string reparametrization path integrals involve integrating over reparametrizations of the circle modulo the gauge $SL(2,\mathbb{R})$ symmetries that characterize both JT gravity and the string worldsheet. It seems plausible that the measure in both path integrals should be the same, although we have not made (and, for the purposes of computing the perturbative correlators and OTOC, did not need to make) precise the definition of the measure in the string reparametrization path integral. 

An important difference between the Schwarzian and our treatment of the string reparametrization path integral is that the former is an exact representation of (the disk contribution to) the JT gravity partition function. This is because the matter action in \eqref{eq:JT+matter partition function} is quadratic, the dilaton acts as a Lagrange multiplier, and the integral over metrics reduces to an integral over possible boundary curves. By contrast, the string reparametrization path integral in \eqref{eq:EM43CBaVED} is an approximation since it captures only a part of the string path integral. In particular, it does not include certain contributions due to matter fields (such as the transverse modes in AdS that couple to longitudinal modes) or to $bc$ ghosts that come from fixing the conformal gauge. Nevertheless, as we have seen, the approximation \eqref{eq:string partition function reparametrization integral} correctly captures the four-point functions at leading order and the OTOC in the double scaled limit.

Second, the transverse mode in the analysis of the string in AdS$_2\times S^1$ plays an analogous role to that of the matter field in JT gravity. Indeed, the effective action for the transverse mode in \eqref{eq:transverse action c.o.var} takes the same form as the matter field in \eqref{eq:S matter boundary action} (when $m=0$ or $\Delta=1$), and taking variational derivatives with respect to the boundary values of the matter fields pulls down the same bilocal operators that are defined in \eqref{eq:CSl6ubNNjr}:
\begin{align}
    B_\tau(\theta_i,\theta_j)=-\frac{1}{D}\frac{\delta S_{\rm matter}}{\delta \tilde{y}(\theta_i)\delta \tilde{y}(\theta_j)}\bigg\rvert_{\Delta=1}=-\frac{\pi}{T_s}\frac{\delta S_{T}}{\delta \tilde{y}(\theta_i)\delta \tilde{y}(\theta_j)}.
\end{align}

Third, the AdS$_2$ string reparametrization action in \eqref{eq:SR = SL} is analogous to the Schwarzian action in \eqref{eq:3qs9vXiyeP}. The reparametrization action in the AdS$_2$ string originates from the dynamics of the longitudinal modes, while the Schwarzian action in JT gravity originates from the dynamics of the dilaton and metric. However, the two reparametrization actions are not the same, and differ in a number of important ways. Firstly, the Schwarzian action is local while the AdS$_2$ reparametrization action is non-local. Secondly, although both reparametrization actions are invariant under the gauge $SL(2,\mathbb{R})$ symmetry that sends $\tau(\theta)$ to $f(\tau(\theta))$--- for the AdS$_2$ string, this corresponds to worldsheet coordinate transformations that leave the worldsheet metric invariant up to Weyl rescaling; for JT gravity, this corresponds to an isometry that maps one boundary curve to another that cuts out an equivalent patch of hyperbolic space--- the AdS$_2$ reparametrization action is also invariant under the physical $SL(2,\mathbb{R})$ transformation that sends $\tau(\theta)$ to $\tau(f(\theta))$ while the Schwarzian action is not.\footnote{Recall that the Schwarzian derivative satisfies the composition rule $\{f\circ g,x\}=\{f,g(x)\}g'(x)^2+\{g,x\}$ and is zero for $SL(2,\mathbb{R})$ transformations: $\{f,x\}=0$ if $f(x)=(ax+b)/(cx+d)$ with $ad-bc=1$. It follows that $\{f\circ g,x\}=\{g,x\}$ and $\{g\circ f,x\}=\{g,f(x)\}f'(x)^2$ if $f\in SL(2,\mathbb{R})$. In other words, the Schwarzian derivative is left-invariant but not right-invariant under $SL(2,\mathbb{R})$ transformations.} Thirdly, the origins of the reparametrization mode in the two systems are different. In JT gravity, a cut-off curve near the boundary is introduced in order to regularize the divergence of the dilaton. By contrast, as we saw in section~\ref{sec:reparametrization mode for AdS string}, the reparametrization mode of the AdS$_2$ string appears as a consequence of fixing the conformal gauge, not due to regularizing a divergent action. Although it is true that the area of the AdS$_2$ string also diverges near the AdS boundary, the divergence does not lead to a boundary mode, at least in our setup where the boundary of the worldsheet is sent to the boundary of the AdS target space. We demonstrate this explicitly in appendix \ref{app: AdS2 area with cutoff}, where we compute the regularized area of the AdS$_2$ string by introducing a cut-off curve and following steps that are similar to the steps between \eqref{eq:JT action} and \eqref{eq:3qs9vXiyeP} in the derivation of the Schwarzian action from the JT action. As a consequence of symmetry, the Schwarzian again appears in the derivative expansion of the regularized area, except that in this case it is irrelevant as the cut-off curve is sent to the boundary. In particular, compare \eqref{eq:3qs9vXiyeP} and \eqref{ws-schwarzian}. 

Some works \cite{Cai:2017nwk,Banerjee:2018twd,Banerjee:2018kwy,Vegh:2019any,Gutiez:2022uof} have identified Schwarzian effective actions that appear in the dynamics of strings in AdS and have suggested that they may explain the maximal chaos of the AdS string that was noted in \cite{Maldacena:2017axo,deBoer:2017xdk,Murata:2017rbp}. As far as we can tell, the Schwarzian effective actions thus identified share the feature of the Schwarzian appearing in the derivative expansion of the area of AdS$_2$ discussed in Appendix~\ref{app: AdS2 area with cutoff} that the action is multiplied by a cut-off parameter and is irrelevant if the cut-off is removed. There may be set-ups in which there is a natural finite cut-off for the string in the bulk (e.g., one might introduce a brane near the boundary that the string ends on), in which case the Schwarzian might indeed affect the dynamics of the string. Such a modification to the string worldsheet theory would have to break the physical $SL(2,\mathbb{R})$ symmetry, because the Schwarzian does too. 

It is instructive to compare in detail the perturbative computations of the four-point functions in JT gravity and on the AdS$_2$ string. The computation on the AdS$_2$ string in section~\ref{sec:epsilon propagator and 4pt function} was guided by the similar computation in JT gravity that was presented in \cite{Maldacena:2016upp} and which we now review. We first expand the Schwarzian action in \eqref{eq:3qs9vXiyeP} about the saddle point by letting $\tau(\theta)=\theta+\epsilon(\theta)$ and taking $\epsilon$ to be small. To quadratic order in $\epsilon$,
\begin{align}\label{eq:QEqEi8nrb6}
    S_{\rm Schwarzian}[\theta+\epsilon(\theta)]=\frac{C}{2}\int d\theta \left(\ddot{\epsilon}(\theta)^2-\dot{\epsilon}(\theta)^2\right).
\end{align}
If we write $\epsilon(\theta)=\sum_{n}\epsilon_n e^{in \theta}$, the action in Fourier space is
\begin{align}
    S_{\rm Schwarzian}[\theta+\epsilon(\theta)]=2\pi C \sum_{n=2}^\infty n^2 (n^2-1)\epsilon_n \epsilon_{-n}.
\end{align}
Note that the Schwarzian dispersion relation has an extra factor of $n$ (and no absolute values, because the action is local) compared to the dispersion relation for the AdS$_2$ reparametrization mode in \eqref{eq:reparametrization quadratic action fourier}. When the $SL(2,\mathbb{R})$ zero modes (corresponding to $\epsilon_0$, $\epsilon_{\pm 1}$) are properly gauge fixed, the propagator for $\epsilon$ is found to be \cite{Maldacena:2016hyu,Maldacena:2016upp}
\begin{align}\label{eq:Schwarzian epsilon propagator}
    \braket{\epsilon(\theta)\epsilon(0)}=\frac{1}{2\pi C}\sum_{n\neq 0,\pm1}\frac{e^{in\theta}}{n^2(n^2-1)}=\frac{1}{2\pi C}\left[a+b\cos{\theta}-\frac{(|\theta|-\pi)^2}{2}+(|\theta|-\pi)\sin{|\theta|}\right].
\end{align}
Here, $a=1+\frac{\pi^2}{6}$ and $b=\frac{5}{2}$, but more generally these coefficients are gauge dependent. 

The four-point function in the Schwarzian theory to leading order in $1/C$ is again given by eq.~\eqref{eq:TLmeIfkMCn} and \eqref{eq:f7JS8ArQyv}, just as in the AdS$_2$ string reparametrization path integral. The difference is that the correlators in \eqref{eq:f7JS8ArQyv} are to be evaluated using the Schwarzian $\epsilon$ propagator in \eqref{eq:Schwarzian epsilon propagator} instead of the AdS$_2$ string propagator in \eqref{eq:sOtAmXClR9}. The cases where $V_1,V_2$ and $W_3,W_4$ are in order or alternating on the circle should be handled separately. When they are in order (i.e., $\theta_1<\theta_2<\theta_3<\theta_4$) the four-point function is \cite{Maldacena:2016upp}:
\begin{align}\label{eq:Schwarzian 4pt TO}
    \frac{\braket{V_1V_2W_3W_4}}{\braket{V_1V_2}\braket{W_3W_4}}=1+\frac{\Delta_V\Delta_W}{2\pi C}\bigg(-2+\frac{\theta_{12}}{\tan{\frac{\theta_{12}}{2}}}\bigg)\bigg(-2+\frac{\theta_{34}}{\tan{\frac{\theta_{34}}{2}}}\bigg)+O(1/C^2).
\end{align}
When they are alternating (i.e., $\theta_1<\theta_3<\theta_2<\theta_4$), the four-point function is: 
\begin{align}\label{eq:Schwarzian 4pt OTO}
    \frac{\braket{V_1V_2W_3W_4}}{\braket{V_1V_2}\braket{W_3W_4}}&=1+\frac{\Delta_V\Delta_W}{2\pi C}\bigg[\bigg(-2+\frac{\theta_{12}}{\tan{\frac{\theta_{12}}{2}}}\bigg)\bigg(-2+\frac{\theta_{34}}{\tan{\frac{\theta_{34}}{2}}}\bigg)\nonumber\\&+2\pi\frac{\sin(\frac{\theta_1-\theta_2+\theta_3-\theta_4}{2})-\sin(\frac{\theta_1+\theta_2-\theta_3-\theta_4}{2})}{\sin{\frac{\theta_{12}}{2}}\sin{\frac{\theta_{34}}{2}}}+\frac{2\pi \theta_{23}}{\tan{\frac{\theta_{12}}{2}}\tan{\frac{\theta_{34}}{2}}}\bigg].
\end{align}

We can compare the Schwarzian four-point function in eqs.~\eqref{eq:Schwarzian 4pt TO} and \eqref{eq:Schwarzian 4pt OTO} to the AdS$_2$ four-point function in \eqref{eq:four-pt function leading order}. As in the AdS$_2$ four-point function, the Schwarzian four-point function is independent of the coefficients $a$ and $b$ of the gauge dependent terms in the $\epsilon$ propagator in \eqref{eq:Schwarzian epsilon propagator}, which reflects the $SL(2,\mathbb{R})$ gauge symmetry of the Schwarzian theory at the level of the perturbative correlator. However, unlike the AdS$_2$ four-point function, the Schwarzian four-point function is not simply a function of the conformal cross ratio but rather depends on the four insertions $\theta_i$, $i=1,\ldots,4$ individually. This reflects the fact that the Schwarzian, unlike the AdS$_2$ string, does not have a physical $SL(2,\mathbb{R})$ symmetry. Nonetheless, when \eqref{eq:Schwarzian 4pt OTO} is continued to the OTOC configuration in \eqref{eq:OTOC euclidean times}, the result matches \eqref{eq:chEkwMOIlb}:
\begin{align}
    \frac{\braket{V_1V_2W_3W_4}}{\braket{V_1V_2}\braket{W_3W_4}}=1-\frac{\Delta_V\Delta_W}{2C}e^{t}+\ldots.
\end{align}

For the sake of comparing with the string boundary correlators computed from the reparametrization integral in section~\ref{sec:OTOC from the string reparametrization mode}, we have reviewed only the perturbative computation of the Schwarzian correlators. However, as is well known, the Schwarzian theory has been very successfully studied in recent years and one can go far beyond the perturbative analysis \cite{Bagrets:2017pwq,Bagrets:2016cdf,Stanford:2017thb,Mertens:2017mtv,Mertens:2018fds,Blommaert:2018oro,Iliesiu:2019xuh,Kitaev:2018wpr,Yang:2018gdb,Suh:2020lco}. Indeed, the Schwarzian theory is one-loop exact \cite{Stanford:2017thb} and its correlators can be computed exactly \cite{Mertens:2017mtv}, which also means that the double scaled OTOC can be computed rigorously \cite{Lam:2018pvp}. By contrast,  as we saw in section~\ref{sec:reparametrization mode for AdS string}, it already takes some effort to determine the reparametrization/longitudinal action on the AdS$_2$ string to quadratic order about the saddle point---compare the derivation of \eqref{eq:longitudinal action quadratic circle} in appendix~\ref{app:conformal gauge hyperbolic disk coordinates} (or of \eqref{eq:longitudinal action quadratic line} in section~\ref{sec:longitudinal modes perturbative})  with the effortlessness getting from \eqref{eq:3qs9vXiyeP} to \eqref{eq:QEqEi8nrb6}. A non-perturbative explicit expression for the string reparametrization action is not currently known.

Finally, we close this section by noting a few contexts in which non-local reparametrization actions have appeared. First, \cite{Milekhin:2021cou,Milekhin:2021sqd} recently studied a non-local reparametrization action that describes a free massive scalar in JT gravity with a specified constant value along the cut-off curve, and competes with the Schwarzian in determining the low-energy dynamics of two SYK models coupled by a particular interaction.\footnote{This action was also discussed in appendix D of \cite{Maldacena:2016upp}.} Like the Schwarzian and unlike the string reparametrization action, the non-local action has an $SL(2,\mathbb{R})$ gauge symmetry but no physical $SL(2,\mathbb{R})$ symmetry (except in the massless limit where the action becomes Diff$(S^1)$ invariant). Second, a different non-local reparametrization action with both physical and gauge $SL(2,\mathbb{R})$ symmetry, and which to quadratic order is the same as our action \eqref{eq:reparametrization quadratic action fourier}, was discussed in appendix H of \cite{Maldacena:2016hyu}, which also noted that the non-local action should describe an AdS$_2$ holographic defect. Finally, \cite{Gross:2017vhb} studied a non-local conformal variant of the SYK model in which the dynamical fields are majorana fermions rather than a reparametrization mode. It would be interesting if there were a connection between these non-local actions and the reparametrization action of the AdS$_2$ string.

\section{Discussion}\label{sec:discussion}

In this work, we studied the OTOC on the AdS$_2$ string in the double scaling limit. We first computed it to all orders as an amplitude for high energy $2\to 2$ scattering on the worldsheet, and checked it to fourth order using results from the analytic bootstrap for four-point functions in the Wilson line defect CFT. The result also matches with the OTOC of large charge correlators computed in \cite{Giombi:2021zfb,Giombi:2022anm}, in their overlapping regime of validity. We then showed how the conformal gauge analysis of the AdS$_2$ string gives rise to an effective action for the reparametrizations of the string boundary, which we used to compute the leading correction to the four-point function on the string (in particular proving agreement between conformal gauge and previously known static gauge results \cite{Giombi:2017cqn}), and the OTOC in the Lyapunov regime. We also presented a 
derivation of the OTOC in the double scaling limit using the reparametrization action that agrees with the all orders result from the scattering analysis. An important takeaway from these investigations is that the string reparametrization action shares some similarities with, but is distinct from, the Schwarzian reparametrization action in JT gravity.

The analysis of the reparametrization mode of the AdS$_2$ string presented in sections~\ref{sec:reparametrization mode for AdS string} and \ref{sec:OTOC from the string reparametrization mode} represents only the tip of the iceberg, and there are a number of open questions that would be natural to investigate in future work. It would be good to better understand: (i) the precise definition of the reparametrization path integral, (ii) how exactly the reparametrization path integral arises from the string sigma model path integral, (iii) the regime of validity of including only fluctuations of the reparametrization mode without also including fluctuations of the matter fields and ghosts, and (iv) what other observables beyond the four-point function at leading order and the OTOC in the double scaling limit can be analyzed using the reparametrization path integral. 

One direct extension of the present work would be to find an explicit exact expression for the reparametrization action, which would go beyond the implicit expression in \eqref{eq:reparametrization action implicit} and the explicit expression for the action to quadratic order about the saddle point in \eqref{eq:reparametrization action explicit}. It may be possible to make progress in this direction using Pohlmeyer reduction \cite{Pohlmeyer:1975nb} or by finding a connection between the string reparametrization action and the non-local actions discussed in \cite{Maldacena:2016upp,Milekhin:2021cou,Milekhin:2021sqd,Maldacena:2016hyu,Gross:2017vhb}. Given an exact expression for the reparametrization action, it should be possible to make the derivation of the double scaled OTOC in section~\ref{sec:double scaled OTOC from reparam integral} more rigorous, and to also compute tree-level higher-point correlators. It may be simpler to compute the higher-point correlators from the reparametrization path integral than by computing contact and exchange Witten diagrams in the static gauge (see \cite{Bliard:2022xsm} for a recent discussion of higher point contact Witten diagrams in AdS$_2$). It would be interesting to check whether the higher-point correlators at strong coupling in the Wilson line defect CFT satisfy the generalized Ward identity conjectured in \cite{Barrat:2021tpn,Barrat:2022eim} based on results for higher-point functions at weak coupling.

We studied the string restricted to an AdS$_2\times S^1$ subspace of AdS$_5\times S^5$ (or any other product of anti-de Sitter and an internal manifold, such as AdS$_4\times CP^3$, as is relevant in the analysis of ABJM \cite{Aharony:2008ug}) and studied the fluctuations of the AdS$_2$ string only along the $S^1$ direction. This allowed us to get rid of self-interactions of the transverse coordinate in conformal gauge. A natural generalization would be to allow the string to fluctuate in the other transverse directions. The generalization to the string in AdS$_2\times S^5$ (or AdS$_2\times M$ with some internal manifold $M$) is relatively straightforward. The tree level four-point amplitudes for the AdS$_2\times S^5$ case, and their agreement with the known static gauge results, are summarized in appendix~\ref{app:string in AdS2 x Sn}. 
It would also be interesting to study the fluctuations of the string along the transverse directions in AdS. As a minimal set-up, one could consider the AdS$_2$ string in AdS$_3$ in conformal gauge and study the reparametrization action. The extension of the analysis of the string in AdS$_2\times S^1$ in section~\ref{sec:reparametrization mode for AdS string} to AdS$_3$ is not trivial  because the transverse modes would be massive and would mix together with the longitudinal modes.\footnote{By contrast, the scattering analysis of the OTOC and the computation of the boundary correlators in the static gauge can handle transverse fluctuations in AdS$^5$ and $S^5$ equally well.} Perhaps one could try to extract the reparametrization action from the formalism developed in \cite{Kruczenski:2014bla} for studying the classical string in AdS$_3$.

Another generalization would be to study a string in AdS$_5\times S^5$ incident on a general curve on the boundary of AdS (i.e., not a straight line or circle), whose classical worldsheet geometry is therefore not AdS$_2$. The mixing of the longitudinal and transverse modes and the non-trivial worldsheet geometry makes the analysis of this general case more difficult. Nonetheless, it seems that, at least in principle, one should be able to write the leading order boundary correlators and the double scaled OTOC (suitably defined) in terms of a path integral over reparametrizations governed by a bilocal effective action, except the reparametrization action will now break the physical $SL(2,\mathbb{R})$ symmetry. (The $SL(2,\mathbb{R})$ gauge symmetry on the worldsheet would, of course, remain unbroken). 

One possible extension of the scattering analysis in section~\ref{sec:OTOC from shockwave S matrix} would be to study the OTOC on a string in a general AdS-Schwarzschild black hole background:
\begin{align}\label{eq:AdS Schwarzschild metric}
    ds^2=-\left(1-\frac{\mu}{r^{d-2}}+\frac{r^2}{\ell^2}\right)dt^2+\frac{dr^2}{1-\frac{\mu}{r^{d-2}}+\frac{r^2}{\ell^2}}+r^2d\Omega^2_{d-1}
\end{align}
Here, $\mu\geq 0$ determines the mass of the black hole and $d\Omega^2_{d-1}$ is the metric on $S^{d-1}$. The string extended in the $t$ and $r$ directions and sitting at a point in $S^{d-1}$ is now dual to a stationary quark in the gauge theory on $\mathbb{R}\times S^{d-1}$ at a finite temperature. One can also study the string in the hyperbolic AdS black hole (which generalizes \eqref{eq:Rindler-AdS5 metric} by replacing $-1+\frac{r^2}{\ell^2}\to -1+\frac{\mu}{r^{d-1}}+\frac{r^2}{\ell^2}$ \cite{Emparan:1999gf}; for $\mu\neq 0$ this describes the exterior of a true black hole with non-zero mass and a singularity at $r=0$), which is dual to a stationary quark in $\mathbb{R}\times H^{d-1}$ at finite temperature. Unlike in \eqref{eq:Rindler-AdS5 metric}, the temperature can now be made arbitrary by tuning $\mu\neq 0$. In the case of the string in the spherical/hyperbolic AdS black hole background, the worldsheet is no longer AdS$_2$ and the correlators (both euclidean and OTOC) are more complicated. In particular, in the scattering analysis, we expect that the scattering interaction relevant for the OTOC at late times is still given by \eqref{eq: out and in states}, but the boundary-to-bulk propagator would be more complicated than the AdS$_2$ propagator in \eqref{eq:boundary-to-bulk propagator}. Nonetheless, one expects the OTOC at leading order in $\ell_s^2$ to still take the Lyapunov form and saturate the chaos bound (see appendix C of \cite{Maldacena:2017axo}). It would be interesting to see whether the double-scaled OTOC continues to take a relatively simple form in a more general black hole background.

Another interesting direction is to study the OTOC on the field theory side, see if a similar double scaling limit exists at weak coupling, and identify the reparametrization mode. 
The Lyapunov exponent in a weakly-coupled field theory was computed in \cite{Stanford:2015owe} by resumming a class of ladder diagrams, and it might be possible to generalize some of the analyses there to the half-BPS Wilson line in $\mathcal{N}=4$ SYM. A related question is to understand better the physical meaning of the OTOC on the Wilson loop. Since the time ordering corresponds to the path ordering of the Wilson loop, a natural guess is that the OTOC is a correlation function on the Wilson loop in the presence of `zig-zags', i.e. back-tracking segments of the Wilson loop. Undestanding the precise relationship between the OTOC and zig-zags may help clarify the roles of zig-zags for the worldsheet black hole \cite{Dubovsky:2012wk}.

Finally, it would be interesting to generalize our analysis of the double-scaled OTOC for the large-charge operators to other weakly-coupled field theories. This may provide a useful alternative to \cite{Stanford:2015owe} since the correlators at large charge can be studied using semiclassics \cite{Hellerman:2015nra,Monin:2016jmo,Badel:2019oxl}, which automatically includes the effect of resumming a subclass of diagrams \cite{Arias-Tamargo:2019xld}.
\section*{Acknowledgements}

We thank Yiming Chen, Sergey Dubovsky, Akash Goel, Victor Gorbenko, Victor Ivo, Juan Maldacena, M\' ark Mezei, Baur Mukhametzhanov, Silviu Pufu, Jieru Shan, and Joaquin Turiaci for helpful discussions. The work of SG and BO is supported in part by the US NSF under Grant No.~PHY-2209997.

\appendix 

\section{String in conformal gauge in hyperbolic disk coordinates}\label{app:conformal gauge hyperbolic disk coordinates}

This appendix presents the details of the conformal gauge analysis of the classical string in AdS$_2\times S^1$ using hyperbolic disk coordinates on AdS$_2$. Concretely, the goal is to derive \eqref{eq:transverse action circle body} and \eqref{eq:longitudinal action quadratic circle body}.

We start with the string action in conformal gauge in \eqref{eq:conformal gauge action disk}, and the boundary conditions for the longitudinal and transverse modes in \eqref{eq:string bc disk coordinates}. The equations of motion for the longitudinal modes follow from \eqref{eq:longitudinal action disk}:
\begin{align}\label{eq:longitudinal eom disk}
    0&=\partial^\alpha\left(\frac{1}{\sinh^2{r}}\partial_\alpha \theta\right),&0&=\partial^\alpha \left(\frac{1}{\sinh^2{r}}\partial_\alpha r\right)+\frac{\cosh{r}}{\sinh^3{r}}(\partial^\alpha r \partial_\alpha r + \partial^\alpha \theta \partial_\alpha \theta).
\end{align}
The equations of motion for the transverse modes are given by \eqref{eq:transverse eom}. These are supplemented by the Virasoro constraint,
\begin{align}
    0&=T^L_{\alpha\beta}[r,\theta]+T^T_{\alpha\beta}[y],
\end{align}
where the contribution to the stress tensor from the transverse modes is again given by \eqref{eq:transverse stress tensor}, and the contribution from the longitudinal modes is now
\begin{align}\label{eq:longitudinal stress tensor disk}
    T^{L}_{\alpha\beta}&=\frac{\partial_\alpha r \partial_\beta r+\partial_\alpha \theta \partial_\beta \theta}{\sinh^2{r}}-\frac{1}{2}\delta_{\alpha\beta}\frac{\partial^\gamma r \partial_\gamma r + \partial^\gamma \theta \partial_\gamma \theta}{\sinh^2{r}}.
\end{align}

In analogy with \eqref{eq:8qr0R8pczj} and \eqref{eq:classical action in conformal gauge via extremization}, the classical string action can be expressed as a sum of the on-shell longitudinal and transverse actions that depend on the boundary mode $\alpha(\tau)$, which needs to be fixed either by imposing the Virasoro constraint or by extremization. This is what is expressed by \eqref{eq:classical action with Virasoro constraint body}-\eqref{eq:classical action with extremization over reparametrizations body}. Thus, if we impose the Virasoro constraint or extremize over the boundary mode only at the last step, we can study the transverse and longitudinal modes separately.

\paragraph{Transverse modes.} Consider first the transverse modes. The general solution for $y$ can be expressed using the massless scalar bulk-to-boundary propagator on the disk, which in $\sigma,\tau$ coordinates is given by
\begin{align}
    K(\sigma,\tau,\tau')&=\frac{1}{2\pi}\frac{\sinh{\sigma}}{\cosh{\sigma}-\cos(\tau-\tau')}.
\end{align}
This satisfies $(\partial_\sigma^2+\partial_\tau^2)K=0$ and $K\to \delta(\tau-\tau')$ as $\sigma \to 0$. Thus, $y(\sigma,\tau)$ is given by:
\begin{align}
    y(\sigma,\tau)&=\int_0^{2\pi} d\tau' K(\sigma,\tau,\tau')\tilde{y}(\alpha(\tau')).
\end{align}
The transverse action can be evaluated explicitly and written as a bilocal boundary integral:
\begin{align}
    S_T[\tilde{y}\circ \alpha]&=-\frac{T_s}{2}\int d\tau \left[ y(0,\tau)\partial_\sigma y(0,\tau)\right],\nonumber\\&=-\frac{T_s}{2}\lim_{\sigma \to 0}\int d\tau d\tau' \tilde{y}(\alpha(\tau))\tilde{y}(\alpha(\tau'))\partial_\sigma K(\sigma,\tau,\tau')\nonumber\\&=\frac{T_s}{4\pi}\int d\tau d\tau' \frac{\big[\tilde{y}(\alpha(\tau))-\tilde{y}(\alpha(\tau'))\big]^2}{\bigr[2\sin\big(\frac{\tau-\tau'}{2}\big)\bigr]^2}.
\end{align}
This is \eqref{eq:transverse action circle body}, as desired.

\paragraph{Longitudinal modes.} Now consider the longitudinal modes. First, we again make some general comments about the properties of the longitudinal modes before studying them perturbatively.

\textit{Symmetries.}
The longitudinal action is invariant under both a physical and a gauge $SL(2,\mathbb{R})$ symmetry. We can represent a generic $SL(2,\mathbb{R})$ transformation as 
\begin{align}
    f(x)=e^{i\lambda}\frac{x-a}{1-x\bar{a}},
\end{align}
where $\lambda \in \mathbb{R}$ and $|a|<1$. This is an $SL(2,\mathbb{R})$ transformation on the circle if $|x|=1$ and on the unit disk if $x$ is complex with $|x|<1$.

The physical $SL(2,\mathbb{R})$ symmetry acts as an AdS$_2$ isometry on the longitudinal modes and trivially on the transverse mode:
\begin{align}
    e^{-r(\sigma,\tau)+i\theta(\sigma,\tau)}~~&\to ~~ e^{-\bar{r}(\sigma,\tau)+i\bar{\theta}(\sigma,\tau)}=f(e^{-r(\sigma,\tau)+i\theta(\sigma,\tau)}),\\
    y(\sigma,\tau)~~&\to ~~\bar{y}(\sigma,\tau)=y(\sigma,\tau).
\end{align}
Consistency with the boundary condition in \eqref{eq:string bc disk coordinates} means the transformation acts on the boundary reparametrization and the boundary curve as as
\begin{align}
    e^{i\theta(\tau)}~~&\to ~~e^{i\bar{\theta}(\tau)}=f(e^{i\theta(\tau)}),\label{eq:jR9ruCAwrT}\\
    \tilde{y}(\alpha)~~&\to ~~\bar{\tilde{y}}(\alpha)=\tilde{y}(\bar{\alpha}),
\end{align}
where $e^{i\bar{\alpha}}=f(e^{i\alpha})$. The longitudinal and transverse actions and stress tensors are invariant under this transformation, both off-shell and on-shell. This $SL(2,\mathbb{R})$ transformation actually moves the string in AdS$_2\times S^1$ and the curve on the boundary, and is therefore physical.

Meanwhile, the gauge $SL(2,\mathbb{R})$ symmetry acts on the longitudinal modes as 
\begin{align}
     r(\sigma,\tau)-i\theta(\sigma,\tau)~~&\to~~\bar{r}(\sigma,\tau)-i\bar{\theta}(\sigma,\tau)=r(\bar{\sigma},\bar{\tau})-i\theta(\bar{\sigma},\bar{\tau}),\\
     y(\sigma,\tau)~~&\to ~~ \bar{y}(\sigma,\tau)=y(\bar{\sigma},\bar{\tau}),
\end{align}
where $e^{-\bar{\sigma}(\sigma,\tau)+i\bar{\tau}(\sigma,\tau)}=f(e^{-\sigma+i\tau})$. Consistency with the boundary condition in \eqref{eq:string bc disk coordinates} means it also transforms the boundary reparametrization but not the boundary curve:
\begin{align}
    \alpha(\tau)~~&\to ~~\bar{\alpha}(\tau)=\alpha(\bar{\tau}),\label{eq:s10Y7PjWQZ}\\
    \tilde{y}(\alpha)~~&\to ~~\bar{\tilde{y}}(\alpha)=\tilde{y}(\alpha),
\end{align}
where $e^{i\bar{\tau}}=f(e^{i\tau})$. The longitudinal and transverse actions and stress tensors are invariant under this transformation, both off-shell and on-shell. This $SL(2,\mathbb{R})$ tranformation simply relabels the worldsheet coordinates without actually moving the string in target space or the curve on the boundary, and is therefore gauge.

\textit{Behavior near the boundary.} Next, we show that the longitudinal action is well behaved at the boundary. It is again useful to study the general form of the solutions to the equations of motion in \eqref{eq:longitudinal eom disk} using series expansions near the boundary. In this case, we expand  $\theta(\sigma,\tau)$ and $r(\sigma,\tau)$ in powers of $\sigma$:
\begin{align}\label{eq:theta r series expansion}
    \theta(\sigma,\tau)&=\alpha(\tau)+\sum_{n=1}^\infty a_n(\tau)\sigma^n, &r(\sigma,\tau)&=\sum_{n=1}^\infty b_n(\tau)\sigma^n.
\end{align}
We can solve for $a_n$ and $b_n$ recursively by substituting the above expansions into \eqref{eq:longitudinal eom disk} and setting the coefficient of each power $\sigma^n$ to zero. The expansion to order $\sigma^3$ is given by
\begin{align}
    \theta(\sigma,\tau)&=\alpha(\tau)-\frac{1}{2}\ddot{\alpha}(\tau)\sigma^2+\frac{1}{3}g(\tau)\sigma^3+\ldots\label{eq:theta series expansion}\\
    r(\sigma,\tau)&=\dot{\alpha}(\tau)\sigma+\frac{1}{3}\bigg[h(\tau)-\frac{1}{2}\dddot{\alpha}(\tau)\bigg]\sigma^3+\ldots.\label{eq:r series expansion}
\end{align}
Higher terms in \eqref{eq:theta r series expansion}, $a_n(\tau)$ and $b_n(\tau)$ for $n>3$, are fixed in terms of $\alpha(\tau)$, $g(\tau)$ and $h(\tau)$. In principle, $g(\tau)$ and $h(\tau)$ are also fixed in terms of $\alpha(\tau)$ by requiring that $\theta(\sigma,\tau)$ and $r(\sigma,\tau)$ be regular in the middle of the hyperbolic disk at $\sigma=\infty$, but this is hard to implement because the series do not converge there. Note that \eqref{eq:theta series expansion} and \eqref{eq:r series expansion} take the same form as the expansions in \eqref{eq:z series expansion} and \eqref{eq:x series expansion} for $x$ and $z$ on the hyperbolic half-plane. This is only true for the lowest order terms, however, and the series on the disk and on the half-plane differ starting at $O(\sigma^5)$.

The series in \eqref{eq:theta series expansion} and \eqref{eq:r series expansion} allow us to check explicitly that the longitudinal action in \eqref{eq:longitudinal action disk} is well behaved near the boundary and finite. The steps are essentially identical to the ones leading to \eqref{eq:longitudinal lagrangian near s=0} because $\sinh^{-2}{r}=r^{-2}+O(r^0)$. Furthermore, we could repeat the argument we used in the hyperbolic half-plane coordinates to show that extremizing over the reparametrization $\alpha(\tau)$ is indeed equivalent to the Virasoro constraint, as claimed in \eqref{eq:classical action with Virasoro constraint body}-\eqref{eq:classical action with extremization over reparametrizations body}.

\textit{Longitudinal modes without transverse modes.} The analysis of the longitudinal modes is again very simple when the transverse modes are turned off. In this case, we take $\tilde{y}=0$, which implies $y=T^T=S_T=0$. If we impose the Virasoro constraint before solving the longitudinal equations of motion, it implies $T^L=0$ and therefore, from \eqref{eq:longitudinal stress tensor disk} that $\dot{r}r'+\dot{\theta}\theta'=0$ and $\dot{r}^2-{r'}^2+\dot{\theta}^2-{\theta'}^2=0$. These have two sets of solutions: $\dot{r}=\pm \theta'$ and $r'=\mp\theta'$, which also automatically satisfy the equations of motion in \eqref{eq:longitudinal eom disk}. The two solutions correspond to $-r+i\theta$ being a holomorphic or antiholomorphic function of $-\sigma+i\tau$ (up to a branch cut because $\theta$ and $\tau$ are angular coordinates) or, equivalently, to $e^{-r+i\theta}$ being a holomorphic or antiholomorphic function of $e^{-\sigma+i\tau}$. Now we can again invoke the result from complex analysis that the biholomorphic bijections of the unit disk in the complex plane are the $SL(2,\mathbb{R})$ transformations. This means that the general solutions for $\theta(\sigma,\tau)$ and $r(\sigma,\tau)$ that is consistent with the Virasoro constraint and the equations of motion is:
\begin{align}\label{eq:longitudinal modes w/o transverse modes}
    e^{-r(\sigma,\tau)+i\theta(\sigma,\tau)}=e^{i\lambda}\frac{e^{-\sigma \pm i \tau}-a}{1-e^{-\sigma \pm i \tau}\bar{a}},
\end{align}
where $\lambda\in\mathbb{R}$ and $a\in \mathbb{C}$ with $|a|<1$. Restricted to the unit circle, this means that the general form of the boundary parametrization $\alpha(\tau)$ is given by:
\begin{align}\label{eq:longitudinal bdy mode w/o transverse modes}
    e^{i\alpha(\tau)}=e^{i\lambda}\frac{e^{i\pm\tau}-a}{1-\bar{a}e^{\pm i\tau}}. 
\end{align}

\textit{Perturbative analysis of the longitudinal modes.} 
Finally, we again study the longitudinal action to first non-trivial order in perturbation theory, treating the transverse fluctuations as being small. We want to compute the on-shell longitudinal action to quadratic order in small fluctuations on the boundary about the solutions given in \eqref{eq:longitudinal modes w/o transverse modes}-\eqref{eq:longitudinal bdy mode w/o transverse modes}.

In particular, we expand around the solution given by $\theta(\sigma,\tau)=\tau$, $r(\sigma,\tau)=\sigma$ and $\alpha(\tau)=\tau$. This is a convenient choice of $SL(2,\mathbb{R})$ gauge. Thus, we expand
\begin{align}\label{eq:perturbative expansion theta, r}
    \alpha(\tau)&=\tau+\epsilon(\tau), &\theta(\sigma,\tau)&=\tau+\eta(\sigma,\tau), &r(\sigma,\tau)&=\sigma+\rho(\sigma,\tau),
\end{align}
and treat $\epsilon$, $\eta$ and $\rho$ as small perturbations. The boundary conditions for $\eta$ and $\rho$ are:
\begin{align}
    \eta(0,\tau)&=\epsilon(\tau), &\rho(0,\tau)=0.
\end{align}
The fluctuation fields are periodic in $\tau$: $\epsilon(\tau+2\pi)=\epsilon(\tau)$, $\eta(\sigma,\tau+2\pi)=\eta(\sigma,\tau)$, $\rho(\sigma,\tau+2\pi)=\rho(\sigma,\tau)$. 

Substituting \eqref{eq:perturbative expansion theta, r} into \eqref{eq:longitudinal eom disk} and expanding, we find the linear order equations of motion are: 
\begin{align}
    0&=\sinh{\sigma}(\ddot{\eta}+\eta'')-2\cosh{\sigma}(\dot{\rho}+\eta'),&
    0&=\sinh{\sigma}(\ddot{\rho}+\rho'')+2\cosh{\sigma}(\dot{\eta}-\rho').\label{eq:longitudinal linear eom disk}
\end{align}
Substituting \eqref{eq:perturbative expansion theta, r} into \eqref{eq:longitudinal action disk}, we find that the longitudinal action expanded to quadratic order is:
\begin{align}
    S_{L}&=T_s\int  \frac{d\sigma d\tau}{\sinh^2{\sigma}}\biggr[\dot{\eta}+\rho'-2\rho\coth{\sigma}-2\rho\rho'\coth{\sigma}+\rho^2 \left(2+\frac{3}{\sinh^2{\sigma}}\right)\nonumber\\&\hspace{7cm}-2\rho \dot{\eta}\coth{\sigma}+\frac{1}{2}\big(\partial_\alpha \rho \partial^\alpha \rho + \partial_\alpha \eta \partial^\alpha \eta\big)\biggr]+\ldots\nonumber\\&=T_s\int  d\sigma d\tau\biggr[\partial_\tau\left(\frac{\eta}{\sinh^2{\sigma}}\right)+\partial_\sigma\left(\frac{\rho}{\sinh^2{\sigma}}\right)-\partial_\sigma\left(\frac{\rho^2\cosh{\sigma}}{\sinh^3{\sigma}}\right)\label{eq:ng3erNO1On}\\&\hspace{7cm}-\partial_\tau\left(\frac{\eta\rho\cosh{\sigma}}{\sinh^3{\sigma}}\right)+\partial_\alpha\left(\frac{\rho\partial^\alpha \rho+\eta\partial^\alpha \eta}{2\sinh^2{\sigma}}\right)\biggr]+\ldots\nonumber
\end{align}
We used \eqref{eq:longitudinal linear eom disk} to get to the second line.

All the terms in \eqref{eq:ng3erNO1On} are total derivatives, but we again need to be careful about their behavior near $\sigma=0$. Given the expansions of $\theta$ and $r$, and therefore of $\eta$ and $\rho$, in \eqref{eq:theta series expansion}-\eqref{eq:r series expansion}, we see that all the singular terms near $\sigma=0$ cancel, and the only finite contribution to the action is:
\begin{align}
    S_{L}&=-\frac{T_s}{2}\int d\tau \eta(0,\tau)g(\tau) =-\frac{T_s}{4}\int d\tau \eta(0,\tau)\eta'''(0,\tau).\label{eq:zml11ueZAp}
\end{align}
 This action has corrections that are of third order in $\eta$ and $\rho$.

Once we solve \eqref{eq:longitudinal linear eom disk} for $\eta(\sigma,\tau)$ to linear order in $\epsilon$, we can determine the on-shell longitudinal action to quadratic order in $\epsilon$. The general solution to the linear equations of motion can be written as boundary-to-bulk integrals:
\begin{align}\label{eq:longitudinal bdy-to-bulk integrals disk}
    \eta(\sigma,\tau)&=\int d\tau' K_\theta(\sigma,\tau,\tau')\epsilon(\tau),&
    \rho(\sigma,\tau)&=\int d\tau' K_r(\sigma,\tau,\tau')\epsilon(\tau'),
\end{align}
where the two boundary-to-bulk propagators $K_\theta$ and $K_r$ are given explicitly by:
\begin{align}
    K_\theta(\sigma,\tau,\tau')&=\frac{1}{\pi}\frac{(\cos(\tau-\tau')\cosh{\sigma}-1)\sinh^3{\sigma}}{(\cosh{\sigma}-\cos(\tau-\tau'))^3},&
    K_r(\sigma,\tau,\tau')&=-\frac{1}{\pi}\frac{\sin(\tau-\tau')\sinh^4{\sigma}}{(\cosh{\sigma}-\cos(\tau-\tau'))^3}.\label{eq:longitudinal bdy-to-blk propagators disk}
\end{align}
It is easy to check that $K_\theta(\sigma,\tau,\tau')$ and $K_r(\sigma,\tau,\tau')$ solve \eqref{eq:longitudinal linear eom disk}, become sharply peaked at $\tau=\tau'$ as $\sigma\to 0$ and satisfy $\int d\tau' K_\theta(\sigma,\tau,\tau')=1$ and $\int d\tau' K_r(\sigma,\tau,\tau')=0$ for any $\sigma$. These properties and \eqref{eq:longitudinal bdy-to-bulk integrals disk} imply that $\eta(\sigma,t)\to \epsilon(\tau)$ and $\rho(\sigma,\tau)\to 0$ as $\sigma\to 0$, as desired.

Combining \eqref{eq:zml11ueZAp} with \eqref{eq:longitudinal bdy-to-bulk integrals disk}, we arrive at the following expression for the longitudinal action to quadratic order in $\epsilon$:
\begin{align}\label{eq:pIHDFOuXGC}
    S_L[\tau+\epsilon(\tau)]=-\frac{T_s}{4}\lim_{\sigma\to 0}\int d\tau \int d\tau' \partial_\sigma^3 K_\theta(\sigma,\tau,\tau')\epsilon(\tau)\epsilon(\tau').
\end{align}
To put this into a manifestly finite form, we use the following slightly formal argument. First, we note that $\partial_\sigma^3 K_\theta(0,\tau,\tau')=-\frac{24}{\pi}[2\sin{\frac{\tau-\tau'}{2}}]^{-4}$ and take the limit in \eqref{eq:pIHDFOuXGC} inside the integral to get
\begin{align}
S_L=\frac{6}{\pi}\int d\tau d\tau' \frac{\epsilon(\tau)\epsilon(\tau')}{[2\sin{\frac{\tau-\tau'}{2}}]^4}.
\end{align}
We can make sense of this expression using analytic regularization. Then, noting that $\sin^{-4}\frac{\tau-\tau'}{2}=\frac{2}{3}\left[1-\partial_\tau\partial_{\tau'}\right]\sin^{-2}{\frac{\tau-\tau'}{2}}$, using integration by parts to transfer the derivatives to $\epsilon(\tau)$ and $\epsilon(\tau')$ and invoking the identity $\int d\tau \sin^{-2}{\frac{\tau-\tau'}{2}}=0$ to replace $\epsilon(\tau)\epsilon(\tau')\to -\frac{1}{2}(\epsilon(\tau)-\epsilon(\tau'))^2$ and $\dot{\epsilon}(\tau)\dot{\epsilon}(\tau')\to -\frac{1}{2}(\dot{\epsilon}(\tau)-\dot{\epsilon}(\tau'))^2$, we arrive at the following manifestly finite expression for the longitudinal action to quadratic order:
\begin{align}\label{eq:longitudinal action quadratic circle}
    S_L[\tau+\epsilon(\tau)]&=\frac{T_s}{2\pi}\int d\tau d\tau' \frac{(\dot{\epsilon}(\tau)-\dot{\epsilon}(\tau'))^2-(\epsilon(\tau)-\epsilon(\tau'))^2}{[2\sin\left(\frac{\tau-\tau'}{2}\right)]^2}.
\end{align}
This is \eqref{eq:longitudinal action quadratic circle body}, as desired.

\section{Regularized area of hyperbolic space with a cut-off}\label{app: AdS2 area with cutoff}

This appendix reviews the computation of the regularized area of the AdS$_2$ string with a cut-off. It is similar to the derivation of the Schwarzian action from the JT action in section~\ref{sec:Schwarzian from JT gravity}, except the absence of a dilaton means the Schwarzian is irrelevant and the regularized area is cut-off independent.

The area of the AdS$_2$ string can be regularized by cutting off the worldsheet along a curve $B$ near the boundary and subtracting its length:\cite{Maldacena:1998im,Rey:1998ik,Drukker:1999zq}
\begin{align}\label{eq:regularized area}
    A_{\rm ws}=\int_{\mathcal{M}}d^2\sigma \sqrt{h}-\int_{B}d\theta \sqrt{h_{\theta\theta}}.
\end{align}
Here, $\sigma^\alpha=(\sigma,\tau)$ and $h_{\alpha\beta}$ are the coordinates and metric on the worldsheet, $\theta$ and $h_{\theta\theta}$ are the coordinate and induced metric on the boundary. Subtracting the length of the curve can be interpreted as a renormalization of the mass of the boundary particle that the string is dual to.

The Gauss-Bonnet theorem relates the Euler characteristic to the bulk and boundary integrals of the Ricci and extrinsic curvatures:
\begin{align}
2\pi \chi_E= \frac{1}{2}\int_{\mathcal{M}} d^2\sigma \sqrt{h}R+\int_{B} d\theta \sqrt{h_{\theta\theta}}K.
\end{align}
Thus, since $R=-2$ in AdS$_2$, the regularized area in \eqref{eq:regularized area} can be written as:\footnote{See e.g. the discussions in \cite{Kruczenski:2008zk,Kitaev:2018wpr,Yang:2018gdb}.}
\begin{align}\label{eq:regularized area }
    A_{\rm ws}=-2\pi \chi_E + \int_B d\theta \sqrt{h_{\theta\theta}}(K-1).
\end{align}
The second term above is the same as the boundary term in the JT action in \eqref{eq:JT action} without the dilaton. If we use the same coordinates as in \eqref{sec:Schwarzian from JT gravity}, parametrize the curve as $(\sigma(\theta),\tau(\theta))$, and let $\theta$ be the renormalized length coordinate, then we can use \eqref{eq:tdsrfde4e} and \eqref{eq:K expansion} and the fact that $\chi_E=1$ for a disk to write the regularized area as
\begin{align}
    A_{\rm ws}=-2\pi +\epsilon\int d\theta \{\tan{\frac{\tau}{2}},\theta\}+O(\epsilon^3).
\label{ws-schwarzian}
\end{align}
If we push the cut-off curve to the boundary by sending $\epsilon\to 0$, the Schwarzian vanishes and the regularized area is simply $A_{\rm ws}=-2\pi$. This is to be contrasted with \eqref{eq:mik4K1iLrh}, where the divergence of the dilaton at the boundary keeps the Schwarzian term relevant in JT gravity. The supergravity result for the circular Wilson loop is therefore $\langle \mathcal{W}\rangle=e^{-T_s A_{\rm ws}}=e^{\sqrt{\lambda}}$, which matches the result from supersymmetric localization \cite{Erickson:2000af,Drukker:2000rr,Pestun:2007rz}.

The analysis can also be repeated on the hyperbolic plane. The regularized area is again given by \eqref{eq:regularized area }. Working with Poincar\'e coordinates $\sigma^\alpha = (s,t)$ on the half-plane, parametrizing the cut-off curve as $(s(x),t(x))$ where $\theta\to x$ is the renormalized length coordinate such that $h_{xx}=\frac{1}{\epsilon^2}$, and setting the effective Euler characteristic of the plane to be $\chi_E=0$, one finds
\begin{align}
    A_{\rm ws}=\epsilon \int_{-\infty}^\infty dx \{t,x\}+O(\epsilon^3).
\end{align}
Taking $\epsilon\to 0$ yields $A_{\rm ws}=0$, as claimed in \eqref{eq:regularized are hyperbolic plane}. The supergravity result for the Wilson line is therefore $\langle \mathcal{W}\rangle=e^{-T_sA_{\rm ws}}=1$, which is the exact result from localization.

\section{Equivalence of the classical string in static and conformal gauge}\label{sec:static = conformal gauge correlators}

In section~\ref{sec:reparametrization mode for AdS string}, we used the conformal gauge to study the classical string action and found that it can be expressed as the sum of the longitudinal and transverse actions subject to either the Virasoro constraint as in \eqref{eq:8qr0R8pczj} or, equivalently, extremization over the boundary reparametrizations as in \eqref{eq:classical action in conformal gauge via extremization}. In this appendix, we show how imposing the Virasoro constraint fixes the boundary reparametrization and determines the classical action in closed form to fourth order in the perturbation of the boundary curve about the straight line. We show that the result agrees with the static gauge analysis in section~\ref{sec:static gauge analysis}. 

The explicit expression for the classical string action computed in the static gauge is given in \eqref{eq:fourth order classical action}. Another way to write it is:
\begin{align}\label{eq:classical action TTbar}
    S_{\rm cl}[\tilde{y}]&=\frac{T_s}{4\pi}\int dt_1 dt_2 \frac{(\tilde{y}(t_1)-\tilde{y}(t_2))^2}{(t_1-t_2)^2}-\frac{T_s}{2}\int d^2\sigma s^2 ([T_{ts}^T(s,t)]^2+[T_{tt}^T(s,t)]^2)+O(\tilde{y}^6).
\end{align}
Here, $T_{ts}^T$ and $T_{tt}^T$ are the leading order on-shell transverse stress tensors (i.e., they are quadratic in $\tilde{y}$). Eq.~\eqref{eq:classical action TTbar} follows from \eqref{eq:classical action 4th order} and \eqref{eq:L_n} combined with the definition of the stress tensor in \eqref{eq:complex stress tensors}. Another way to write the quartic term is $T_{tt}^2+T_{ts}^2=T^{T}\bar{T}^T$, where $T^T$ and $\bar{T}^T$ are the holomorphic and antiholomorphic stress tensors defined in \eqref{eq:transverse stress tensor}.

We will now rederive \eqref{eq:classical action TTbar} starting from the conformal gauge result \eqref{eq:classical action in conformal gauge via extremization} and the expressions for the longitudinal and transverse actions in \eqref{eq:longitudinal action quadratic line} and \eqref{eq:transverse action on-shell explicit}. We write $\alpha(t)=t+\epsilon(t)$ and work perturbatively to quadratic order in $\epsilon$ (i.e., to quartic order in $\tilde{y}$). To this order, the longitudinal and transverse actions are given by
\begin{align}
    S_L[t+\epsilon]&=\frac{T_s}{2\pi}\int dt dt' \frac{(\dot{\epsilon}(t)-\dot{\epsilon}(t'))^2}{(t-t')^2}\label{eq:KofxqDqsYd},\\
    S_T[t+\epsilon]&=\frac{T_s}{4\pi}\int dt dt' \frac{(\tilde{y}(t)-\tilde{y}(t'))^2}{(t-t')^2}+\frac{T_s}{2\pi}\int dt dt' \frac{(\tilde{y}(t)-\tilde{y}(t'))(\dot{\tilde{y}}(t)\epsilon(t)-\dot{\tilde{y}}(t')\epsilon(t'))}{(t-t')^2}.\label{eq:CkF3v9PEP3}
\end{align}
We want to derive and solve the Euler-Lagrange equation for $\epsilon$. If $\epsilon(t)\to \epsilon(t)+\delta \epsilon(t)$ in \eqref{eq:KofxqDqsYd}, the variation of the longitudinal action is:
\begin{align}
    \delta S_L &= \frac{T_s}{\pi}\int dt dt' \frac{(\dot{\epsilon}(t)-\dot{\epsilon}(t')(\delta\dot{\epsilon}(t)-\delta\dot{\epsilon}(t'))}{(t-t')^2}.
\end{align}
To isolate $\delta \epsilon(t)$, it is convenient to introduce a regulator by letting $(t-t')^2+s^2$ and take $s\to 0$ at the end. Integrating by parts leads to
\begin{align}
    \delta S_L &= -\lim_{s\to 0}\frac{2T_s}{\pi}\int dt \delta \epsilon(t)\left[ \frac{d}{dt}\int dt' \frac{\dot{\epsilon}(t)-\dot{\epsilon}(t')}{(t-t')^2+s^2}\right] &&=-\lim_{s\to 0}\frac{2T_s}{\pi}\int dt\text{ }\delta \epsilon(t) \int dt' \frac{\ddot{\epsilon}(t)-\ddot{\epsilon}(t')}{(t-t')+s^2},\nonumber\\&=-\frac{2T_s}{\pi}\int dt \delta\epsilon(t) \fint dt' \frac{\ddot{\epsilon}(t)-\ddot{\epsilon}(t')}{(t-t')^2}.\label{eq:NKOvZa0zsh}
\end{align}
The second line is expressed using the slash notation for the Cauchy principal value. Similarly, the variation of the transverse action is given by:
\begin{align}\label{eq:dRysSieZ3j}
    \delta S_T=\frac{T_s}{\pi}\int dt \delta \epsilon(t) \dot{\tilde{y}}(t)\fint dt'\frac{\tilde{y}(t)-\tilde{y}(t')}{(t-t')^2}.
\end{align}
Setting $0=\delta S_L+\delta S_T$ leads to the Euler-Lagrange equation for $\epsilon$:
\begin{align}\label{eq:epsilon EL eqn}
    0&=\fint dt'\left[\frac{\ddot{\epsilon}(t)-\ddot{\epsilon}(t')}{(t-t')^2}-\frac{\dot{\tilde{y}}(t)}{2}\frac{\tilde{y}(t)-\tilde{y}(t')}{(t-t')^2}\right].
\end{align}

From~\eqref{eq:wQbrprkLwp} and \eqref{eq:Bo74FAqNmH}, we know the variations of the actions are related to the stress tensors on the boundary. \eqref{eq:NKOvZa0zsh} and \eqref{eq:dRysSieZ3j} determine the longitudinal and transverse stress tensors at leading order to be:
\begin{align}\label{eq:stress tensor CPV}
    T_{ts}^L(0,t)&=\frac{2}{\pi}\fint dt' \frac{\ddot{\epsilon}(t)-\ddot{\epsilon}(t')}{(t-t')^2},&T_{ts}^T(0,t)&=-\frac{\dot{\tilde{y}}(t)}{\pi}\fint dt' \frac{\tilde{y}(t)-\tilde{y}(t')}{(t-t')^2}.
\end{align}
Thus, \eqref{eq:epsilon EL eqn} is equivalent to $0=T_{ts}^L(0,t)+T_{ts}^T(0,t)$, in accordance with section~\ref{sec:extremization and Virasoro}. 

Before attempting to solve for $\epsilon(t)$, we note that the action in \eqref{eq:KofxqDqsYd}-\eqref{eq:CkF3v9PEP3} can also be rewritten in terms of the Cauchy principal values:
\begin{align}
    S_L[t+\epsilon(t)]&=-\frac{T_s}{\pi}\int dt \epsilon(t)\fint dt' \frac{\ddot{\epsilon}(t)-\ddot{\epsilon}(t')}{(t-t')^2},\\
    S_T[t+\epsilon(t)]&=\frac{T_s}{4\pi}\int dt dt' \frac{(\tilde{y}(t)-\tilde{y}(t'))^2}{(t-t')^2}+\frac{T_s}{\pi}\int dt \epsilon(t)\dot{\tilde{y}}(t)\fint dt'\frac{\tilde{y}(t)-\tilde{y}(t')}{(t-t')^2}. 
\end{align}
This, combined with the sum of the stress tensors in \eqref{eq:stress tensor CPV} being zero when $\epsilon$ satisfies its equation of motion, means that the longitudinal action and the second term in the transverse action are the same up to a factor of $-\frac{1}{2}$. Thus, the classical action becomes
\begin{align}\label{eq:qI39t6o0gs}
    S_{\rm cl}[\tilde{y}]=\frac{T_s}{4\pi}\int dt dt' \frac{(\tilde{y}(t)-\tilde{y}(t'))^2}{(t-t')^2}-\frac{T_s}{2\pi}\int dt \epsilon(t) T_{ts}^T(0,t),
\end{align}
where $\epsilon(t)$ is given in terms of $\tilde{y}$ by \eqref{eq:epsilon EL eqn}.

We will now solve for $\epsilon$ in terms of $T_{ts}^T$ and show that \eqref{eq:qI39t6o0gs} is equivalent to \eqref{eq:classical action TTbar}. First, we rewrite \eqref{eq:qI39t6o0gs} as 
\begin{align}
    \fint dt' \frac{\ddot{\epsilon}(t)-\ddot{\epsilon}(t')}{(t-t')^2}=-\frac{\pi}{2}T_{ts}^T(0,t).
\end{align}
This non-local equation can be solved in Fourier space. Taking the Fourier transform yields
\begin{align}\label{eq:epsilon and Tts}
    \epsilon(\omega)|\omega|^3=\frac{1}{2}T_{ts}^T(\omega),
\end{align}
where we write $\epsilon(t)=\int \frac{d\omega}{2\pi} e^{-i\omega t}\epsilon(\omega)$ and $T_{ts}^T(0,t)=\int \frac{d\omega}{2\pi}e^{-i\omega t}T_{ts}^T(\omega)$. Thus, if we write the second term in \eqref{eq:qI39t6o0gs} in Fourier space, the action becomes
\begin{align}\label{eq:X5YHmfK7eZ}
    S_{\rm cl}[\tilde{y}]=\frac{T_s}{4\pi}\int dt dt' \frac{(\tilde{y}(t)-\tilde{y}(t'))^2}{(t-t')^2}-\frac{T_s}{8\pi^2}\int d\omega \frac{T_{ts}^T(\omega) T_{ts}^T(-\omega)}{|\omega|^3}.
\end{align}

The above expression derived from the conformal gauge results is equivalent to the static gauge result in \eqref{eq:classical action TTbar}. This is easiest to see if \eqref{eq:classical action TTbar} is expressed in Fourier space. First, because the transverse stress tensor is holomorphic, $T_{ts}^T(s,t)$ and $T_{tt}^T(s,t)$ in the bulk are determined by $T_{ts}^T(0,t)$ on the boundary. The explicit relation is given in \eqref{eq:kramersKronig stress tensor 2}. Thus, the quartic term in \eqref{eq:classical action TTbar} is
\begin{align}
    S_{\rm cl}^{(4)}[\tilde{y}]=-\frac{T_s}{2\pi^2}\int ds dt \int dt' dt'' \frac{s^4+s^2(t-t')(t-t'')}{[s^2+(t-t')^2][s^2+(t-t'')^2]}T_{st}(0,t')T_{st}(0,t'').
\end{align}
It is tempting to interchange the two integrals over the boundary with the integral over the bulk, and express the quartic contribution to the action as a term bilocal in the transverse stress tensor. However, the integral over $s$ and $t$ does not converge. Instead, we write $T_{st}(0,t')$ and $T_{st}(0,t'')$ in their Fourier representation, and evaluate the integrals over $t$ and $t'$ instead. The result is\footnote{We used $\int dt \frac{1}{t^2+s^2}e^{-i \omega t}=\frac{\pi}{s}e^{-|\omega|s}$ and $\int dt \frac{t}{t^2+s^2}e^{-i \omega t}=-i\pi\text{sgn}(\omega)e^{-|\omega|s}$}
\begin{align}
    S_{\rm cl}^{(4)}[\tilde{y}]=-\frac{T_s}{2}\int \frac{d \omega}{2\pi} \frac{d\nu}{2\pi} \int ds dt s^2(1-\text{sgn}(\omega\nu))e^{-|\omega|s-|\nu|s-i\omega t-i\nu t}T_{ts}^T(\omega)T_{ts}^T(\nu).
\end{align}
Evaluating the $t$, $\nu$ and $s$ integrals (in that order), yields
\begin{align}
    S_{\rm cl}^{(4)}[\tilde{y}]=-\frac{T_s}{8\pi}\int d\omega \frac{T_{ts}^T(\omega)T_{ts}^T(-\omega)}{|\omega|^3}.
\end{align}
Thus the static gauge expression for the classical string action in \eqref{eq:classical action TTbar} matches the conformal gauge expression in \eqref{eq:X5YHmfK7eZ}.

For completeness, we also write down the real space representation of the relation between the reparametrization $\epsilon$ and the transverse stress tensor $T_{ts}^T$ in \eqref{eq:epsilon and Tts}, and of the action in \eqref{eq:X5YHmfK7eZ}. These involve the Fourier transform of $1/|\omega|^3$, which we encountered in the computation of the $\epsilon$ propagator on the line in \eqref{eq:BiyOKdFweq}. The Fourier transform needs to be defined with a regularization, and the result is given in \eqref{eq:Pdq5ZKUGWh}. We again ignore the constant and quadratic terms in \eqref{eq:Pdq5ZKUGWh} to write $\int \frac{d\omega}{2\pi}e^{-i\omega t}/|\omega|^3\sim \frac{1}{4\pi}t^2\log(t^2)$\footnote{We can ignore the constant and quadratic terms in \eqref{eq:Pdq5ZKUGWh} when computing the $\epsilon$ propagator in \eqref{eq:BiyOKdFweq} because they are gauge dependent. We can ignore those terms in \eqref{eq:1WQrI5e2oG} and \eqref{eq:NGlrvSnEDx} if the stress tensor satisfies $\int dt T_{st}^T(0,t)t^n=0$ for $n=0,1,2$. In Fourier space, this is equivalent to $\frac{d^n}{d\omega^n}T_{st}(\omega=0)=0$ for $n=0,1,2$, which is necessary for $\epsilon(\omega)$ to be well behaved according to \eqref{eq:epsilon and Tts}.} and therefore find that $\epsilon(t)$ in terms of $T_{ts}^T(0,t)$ is:
\begin{align}\label{eq:1WQrI5e2oG}
    \epsilon(t)=\frac{1}{8\pi}\int dt' (t-t')^2\log[(t-t')^2]T_{ts}^T(0,t').
\end{align}
Likewise, the action in \eqref{eq:X5YHmfK7eZ} in real space becomes 
\begin{align}\label{eq:NGlrvSnEDx}
    S_{\rm cl}[\tilde{y}]=\frac{T_s}{4\pi}\int dt dt' \frac{(\tilde{y}(t)-\tilde{y}(t'))^2}{(t-t')^2}-\frac{T_s}{16\pi}\int dt dt'  T_{ts}^T(0,t)T_{ts}^T(0,t')(t-t')^2\log[(t-t')^2].
\end{align}

\section{Tree level 4-point function of \texorpdfstring{$S^5$}{S5} fluctuations in conformal and static gauge}\label{app:string in AdS2 x Sn}
In this appendix, we compute the leading connected four-point functions of $S^{5}$ fluctuations using the conformal gauge and reproduce the result computed in \cite{Giombi:2017cqn} using the static gauge.

Let us first review the difference between the conformal gauge and the static gauge actions for the $S^{5}$ fluctuations. In the conformal gauge, it reads
\beq
\begin{aligned}
S_{\rm conformal}&=\frac{T_s}{2}\int d^{2}\sigma \, \frac{\partial_{\alpha}y^{m}\partial^{\alpha}y^{m}}{(1+\frac{1}{4}y^2)^2}\\&=T_s\int d^{2}\sigma \, \left[\frac{1}{2}\partial_{\alpha}y^{m}\partial^{\alpha}y^{m}-\frac{1}{4}(y^{n}y^{n})(\partial_{\alpha}y^{m}\partial^{\alpha}y^{m})+O(y^{6})\right]\period
\end{aligned}
\eeq
Here, $y^m$, $m=1,\ldots,5$, are stereographic coordinates on $S^5$, $\sigma^\alpha=(s,t)$ are worldsheet coordinates, and the worldsheet indices are contracted with $\delta^{\alpha\beta}$. On the other hand, the action in the static gauge expanded up to quartic order is
\beq
\begin{aligned}
S_{\rm static}&=T_s\int d^{2}\sigma \, \sqrt{g}\left[L_2 +L_4+O(y^{6})\right]\comma\\
L_2&=\frac{1}{2}g^{\alpha\beta}\partial_{\alpha}y^{m}\partial_{\beta}y^{m}\comma\\
L_4&=-\frac{1}{4}(y^{n}y^{n})(g^{\alpha\beta}\partial_{\alpha}y^{m}\partial_{\beta}y^{m})+\frac{1}{8}(g^{\alpha\beta}\partial_{\alpha}y^{m}\partial_{\beta}y^{m})^2-\frac{1}{4}(g^{\alpha\beta}\partial_{\alpha}y^{m}\partial_{\beta}y^{n})(g^{\gamma\delta}\partial_{\gamma}y^{m}\partial_{\delta} y^{n})\period
\end{aligned}
\eeq
Here, $g_{\alpha\beta}=\frac{1}{s^2}\delta_{\alpha\beta}$ is the AdS$_2$ metric on the worldsheet. As shown above, the quartic interaction in the conformal gauge contains only two derivatives  while the quartic interaction in the static gauge includes both the two-derivative interaction and the four-derivative interactions. Thus, to show the equivalence of the two, all we need to do is to check that the reparametrization mode in the conformal gauge reproduces the contribution from the four-derivative interactions.

Thanks to the SO(5) symmetry, the connected four-point function takes the following general form\footnote{Note that $T_s=\frac{\sqrt{\lambda}}{2\pi}$.}:
\beq
\langle y^{m_1}(x_1)y^{m_2}(x_2)y^{m_3}(x_3)y^{m_4}(x_4)\rangle_{\rm conn}=\frac{T_s}{\pi^4}\left[\delta^{m_1m_2}\delta^{m_3m_4}G_{1}+\delta^{m_1m_3}\delta^{m_2m_4}G_{2}+\delta^{m_1m_4}\delta^{m_2m_3}G_{3}\right]\period\nonumber
\eeq
Since all the three terms on the right hand side are related by the permutation of indices, we focus on $G_{1}$ in what follows. The result for $G_{1}$ was computed in the static gauge in  (4.11) of \cite{Giombi:2017cqn} and is given by
\beq
\begin{aligned}
G_{1}=&2D_{1111}-2x_{34}^2 D_{1122}-2x_{12}^2D_{2211}\\
&+D_{1111}-2x_{13}^2D_{2121}-2x_{14}^2D_{2112}-2x_{23}^2D_{1221}-2x_{24}^2D_{1212}+2x_{34}^2 D_{1122}+2x_{12}^2D_{2211}\\
&+4(x_{13}^2x_{24}^2+x_{14}^2x_{23}^2-x_{12}^2x_{34}^2)D_{2222}\comma
\end{aligned}
\eeq
where the first line denotes the contribution from the two derivative term while the second and the third lines are contributions from the four derivative terms, and $D_{\Delta_1\Delta_2\Delta_3\Delta_4}$ is the $D$-function given by the following contact Witten diagram,
\beq
D_{\Delta_1\Delta_2\Delta_3\Delta_4}=\int \frac{dsdt}{s^{2}}\tilde{K}_{\Delta_1}(s,t;x_1)\tilde{K}_{\Delta_2}(s,t;x_2)\tilde{K}_{\Delta_3}(s,t;x_3)\tilde{K}_{\Delta_4}(s,t;x_4)\comma
\eeq
with $\tilde{K}(s,t;t^{\prime})\equiv \frac{s}{s^2+(t-t^{\prime})^2}$. For integer $\Delta$'s, the $D$-functions can be evaluated explicitly\footnote{For readers' convenience, here we display two $D$-functions needed for evaluating the contribution from the two derivative interaction:
\beq
\begin{aligned}
D_{1111}&=\frac{\pi}{2}\frac{1}{x_{13}^2x_{24}^2}\left[\frac{\log |\chi|}{\chi-1}-\frac{\log|1-\chi|}{\chi}\right]\comma\\
D_{2211}&=\frac{x_{34}^2}{x_{13}^4x_{24}^2}\left[-\frac{\pi (\chi+2)\log |1-\chi|}{8\chi^3}+\frac{\pi}{8\chi^2(1-\chi)}+\frac{\pi\log |\chi|}{8(\chi-1)^2}\right]\period
\end{aligned}
\eeq
$D_{1122}$ can be obtained from $D_{2211}$ by the permutation of indices $1\leftrightarrow 3$ and $2\leftrightarrow 4$.
} in terms of polylogarithms (see e.g. appendix A of \cite{Bliard:2022xsm}). As a result, we obtain
\begin{align}
\frac{T_s}{\pi^4}G_1&=\frac{T_s^2}{\pi^2}\frac{1}{x_{12}^2x_{34}^2}\left(F_{2}+F_4\right)\comma\label{eq:apcfinal1}\\
F_2&=\frac{1}{2\pi T_s}\left[\frac{\chi^2}{\chi-1}+\chi^2\log |1-\chi|-\frac{\chi^2(2-2\chi+\chi^2)\log |\chi|}{(\chi-1)^2}\right]\comma\\
F_4&=-\frac{1}{4\pi T_s}\left[4+\frac{2-\chi}{\chi}\log \left((1-\chi)^2\right)\right]\period
\end{align}
Here we factored out the result for the disconnected diagram $T_s^2/(\pi^2x_{12}^2x_{34}^2 )$ on the right hand side of \eqref{eq:apcfinal1}, and $F_2$ and $F_4$ are contributions from the two derivative and the four derivative interactions respectively. 

As we can see, the result for $F_4$ coincides with the $s$-channel disconnected diagram dressed by the reparametrization mode in the conformal gauge, \eqref{eq:four-pt function leading order}, upon setting $\Delta_V=\Delta_W=1$. This confirms the equivalence of the two formulations at this order. At higher orders in perturbation theory in the conformal gauge, one would also need to dress the connected diagrams by the reparametrization mode in order to reproduce the results in the static gauge. For example, at the next order, one needs to include the two-derivative contact Witten diagram dressed by the reparametrization mode.

\section{Large charge OTOC}\label{app:heavy-light OTOC}
In this appendix, we derive the result for the large charge OTOCs on the Wilson line that was stated without proof in \eqref{eq:PhPhJJ OTOC}. Our starting point is some results from \cite{Giombi:2022anm} for the following four-point functions on the Wilson line introduced in section~\ref{sec:WL defect correlators}:
\begin{align}
    \frac{\braket{\Phi(x_1)\Phi(x_2)Z^J(x_3)\bar{Z}^J(x_4)}}{\braket{\Phi(x_1)\Phi(x_2)}\braket{Z^J(x_3)\bar{Z}^J(x_4)}}&=G_1(\chi),&\frac{\braket{\mathbb{D}(x_1)\mathbb{D}(x_2)Z^J(x_3)\bar{Z}^J(x_4)}}{\braket{\mathbb{D}(x_1)\mathbb{D}(x_2)}\braket{Z^J(x_3)\bar{Z}^J(x_4)}}&=G_4(\chi).\label{eq:large charge 4-pt functions}
\end{align}
Here, $\Phi=\Phi^1$, $Z= \Phi^4+i\Phi^5$ and $\bar{Z}=\Phi^4-i\Phi^5$, where $\Phi^1$, $\Phi^4$ and $\Phi^5$ are three of the scalars orthogonal to the scalar $\Phi^6$ that is coupled to Wilson line; $Z^J,\bar{Z}^J$ are chiral primaries of rank $J$; $\mathbb{D}=\mathbb{D}_1$ is one of the three displacement operators; and $\chi$ is the conformal cross-ratio defined in \eqref{eq:cross ratio line}. In eq.~\eqref{eq:PhPhJJ OTOC}, we were schematic by writing $\Phi^J$ for $Z^J$ and $\bar{Z}^J$, and also only listed the four-point function with light scalars although the analysis of the four-point function with displacement operators is essentially identical.

In \cite{Giombi:2022anm}, we computed the correlators in \eqref{eq:large charge 4-pt functions} in the double scaling limit $J,\lambda\to \infty$ with $\mathcal{J}\equiv \frac{J}{\sqrt{\lambda}}$ held fixed\footnote{This definition of $\mathcal{J}$ differs from the one in \cite{Giombi:2022anm} by a factor of $\frac{1}{4\pi}$.} by studying the classical string with angular momentum $J$ in $S^5$ that is dual to the Wilson line with $Z^J$ and $\bar{Z}^J$ inserted \cite{Drukker:2006xg,Miwa:2006vd,Gromov:2012eu,Giombi:2021zfb}. In particular, the four-point functions can be viewed as two-point functions of $\Phi$ and $\mathbb{D}$ in the large charge background created by $Z^J$ and $\bar{Z}^J$. Via AdS/CFT, they are equal to the boundary-to-boundary propagators for the fluctuations on the classical string in the transverse directions in $S^5$ and AdS$_5$.  The induced geometry on the classical string is non-trivial and depends on $\mathcal{J}$.

We need the following integral representations for the four-point functions in \eqref{eq:large charge 4-pt functions} \cite{Giombi:2022anm} :
\begin{align}
    G_1(\chi)&=\frac{\chi^2}{\chi-1}\int_{-\infty}^\infty dk \frac{k\sqrt{1-c^2+k^2}\text{exp}\left(-ik\log(\chi-1)\right)}{2\sqrt{1+k^2}\sinh\left(\pi\int_0^k\tilde{\rho}(\ell) d\ell \right)},\label{eq:G1 integral rep}\\
    G_4(\chi)&=\frac{\chi^4}{(1-\chi)^2}\int_{-\infty}^\infty dk \frac{k\sqrt{1+k^2}\sqrt{1-c^2+k^2}\text{exp}\left(-ik\log(\chi-1)\right)}{12\sinh\left(\pi \int_0^k \tilde{\rho}(\ell)d\ell\right)}.\label{eq:G4 integral rep}
\end{align}
Here, the parameter $c\in [0,1)$ is related to $\mathcal{J}\in [0,\infty)$ by $\pi \mathcal{J}=\ellK-\ellE$, 
where $\ellK\equiv \ellK(c^2)$ and $\ellE\equiv \ellE(c^2)$ are the complete elliptic integrals of the first and second kind. $\mathcal{J}$ is a monotonically increasing function of $c$, such that $c=0$ when $\mathcal{J}= 0$ and $c^2\to 1$ as $\mathcal{J}\to \infty$. Furthermore, $\tilde{\rho}(\ell)$ contains information about the density of excitation energies on the string and is given by:
\begin{align}\label{eq:rho(ik)}
    \tilde{\rho}(k)=\frac{2}{\pi}\frac{\ellK k^2+\ellE}{\sqrt{1+k^2}\sqrt{1-c^2+k^2}},
\end{align}
The expressions in \eqref{eq:G1 integral rep} and \eqref{eq:G4 integral rep} are valid for $\chi>1$.

To compute the OTOCs from \eqref{eq:G1 integral rep} and \eqref{eq:G4 integral rep}, we analytically continue along the path in \eqref{eq:OTOC cross ratio}. If we keep $c$ fixed and finite as we vary $t$, then at early times the OTOC does not have a period of exponential decay (because there is no small parameter) while at late times OTOC exhibits the standard behavior $G_1\sim \# e^{-2t}$ and $G_4\sim \# e^{-4t}$ as $t\to \infty$. The more interesting case is to take $c$ to be small, which sets up a parametric separation between the dissipation time $t_d\sim \beta\sim 1$ and the scrambling time $t_s\sim -\beta \ln{c}\sim -\ln{c}$ and gives rise to a period of exponential decay. More precisely, we will evaluate \eqref{eq:G1 integral rep} and \eqref{eq:G4 integral rep} in the standard OTOC configuration in the double scaling limit $t\to \infty$ and $c\to 0$ with $c^2 e^{t}$ fixed. This is equivalent to taking $t\to \infty$ and $\mathcal{J}\to 0$ with $\mathcal{J}e^{t}$ held fixed. 

At late times, $-i\log(\chi(t)-1)= \pi-4e^{-t}+O(e^{-3t})$, which means the numerators in the integrands in \eqref{eq:G1 integral rep} and \eqref{eq:G4 integral rep} grow exponentially as functions of $k$ only slightly slower than $e^{\pi k}$. Meanwhile, at large $k$, the energy density in \eqref{eq:rho(ik)} asymptotically approaches $\tilde{\rho}(k)\sim \frac{2\ellK}{\pi}$ and therefore $\int_0^k \tilde{\rho}(\ell)d\ell\sim \frac{2\ellK}{\pi}k$. At small $c^2$, $\ellK=\frac{\pi}{2}+\frac{\pi c^2}{8}+O(c^4)$, so the denominators in the integrands in \eqref{eq:G1 integral rep} and \eqref{eq:G4 integral rep} grow exponentially as functions of $k$ only slightly faster than $e^{\pi k}$ at large $k$. These observations imply that, at large $t$ and small $c$, the integrands in \eqref{eq:G1 integral rep} and \eqref{eq:G4 integral rep} decay exponentially at large values of $k$ but do so only slowly and therefore develop long tails extending towards infinity that dominate the integrals. Thus, only the large $k$ behavior of the integrand is relevant, and the OTOC in the limit is given by
\begin{align}\label{eq:PhPhJJ OTOC 2}
    G_1(\chi(t))&=\lim_{\substack{t\to \infty\\c\to 0}}\big(4e^{-t}\big)^2\int^\infty dk\text{ }k \frac{\text{exp}\big(\big(\pi -4e^{-t}+\ldots\big)k\big)}{\text{exp}\big(\big(\pi +\frac{\pi c^2}{4}+\ldots\big)k\big)}=\frac{1}{\big(1+\frac{\pi c^2}{16}e^{t}\big)^2}.
\end{align}
In agreement with \eqref{eq:PhPhJJ OTOC}. Likewise,
\begin{align}\label{eq:DDJJ OTOC 2}
    G_4(\chi(t))&=\lim_{\substack{t\to \infty\\c\to 0}}\big(4e^{-t}\big)^4\int^\infty dk\text{ }\frac{k^3}{6} \frac{\text{exp}\big(\big(\pi -4e^{-t}+\ldots\big)k\big)}{\text{exp}\big(\big(\pi +\frac{\pi c^2}{4}+\ldots\big)k\big)}=\frac{1}{\big(1+\frac{\pi c^2}{16}e^{t}\big)^4}.
\end{align}
Given that $c^2=4\mathcal{J}+\ldots$ as $\mathcal{J}\to 0$, \eqref{eq:PhPhJJ OTOC 2} leads to \eqref{eq:PhPhJJ OTOC}, as we wanted to show. It therefore agrees with \eqref{eq:general large charge OTOC} for the case $\Delta_V=\Delta_{Z^J}=J$ and $\Delta_W=\Delta_\Phi=1$. Similarly, \eqref{eq:DDJJ OTOC 2} agrees with \eqref{eq:general large charge OTOC} for the case $\Delta_W=\Delta_{\mathbb{D}}=2$.

\bibliographystyle{ssg}
\bibliography{mybib}

\end{document}